\documentclass[aps,prf,twocolumn,groupedaddress,amsmath,amssymb,longbibliography]{revtex4-1}
\usepackage{graphicx}  
\usepackage{dcolumn}   
\usepackage{bm}        
\usepackage{verbatim}   

\usepackage{newtxtext,newtxmath}
\usepackage{amsmath}
\usepackage{color}

\usepackage{braket}
\usepackage{pstricks}
\usepackage{epsfig}

\newcommand*{\Ek}{{\rm E}}

\newcommand*{\Pec}{{\rm Pe}}

\newcommand*{\Ray}{{\rm Ra}}
\newcommand*{\Rayt}{{\rm \widetilde{Ra}}}
\newcommand*{\Rayc}{{\rm Ra_c}}
\newcommand{\Pra}{{\rm Pr}}

\newcommand{\kv}{\ensuremath{\mathbf{k}}}
\newcommand{\rv}{\ensuremath{\mathbf{r}}}
\newcommand{\ve}{{\mathbf{v}}}
\newcommand{\lp}{\ensuremath{\left(}}
\newcommand{\rp}{\ensuremath{\right)}}

\begin{document}

\title{Weak branch and multimodal convection in rapidly rotating
  spheres at low Prandtl number}

\affiliation{Department of Magnetohydrodynamics, Helmholtz-Zentrum Dresden-Rossendorf, Bautzner Landstra\ss e 400, D-01328 Dresden, Germany}
\affiliation{D\'epartement de Math\'ematiques et Applications, UMR-8553, \'Ecole Normale Sup\'erieure, CNRS,PSL University, 75005 Paris, France}
\affiliation{Department of Fluid Mechanics, Universitat Polit\`ecnica de Catalunya-BarcelonaTech, Barcelona 08019, Spain}

\author{F.~Garcia}
\affiliation{Serra H\'unter Fellow, Department of Fluid Mechanics, Universitat Polit\`ecnica de Catalunya-BarcelonaTech, Barcelona 08019, Spain}

\author{F.~Stefani}
\affiliation{Department of Magnetohydrodynamics, Helmholtz-Zentrum Dresden-Rossendorf, Bautzner Landstra\ss e 400, D-01328 Dresden, Germany}

\author{E. Dormy}
\affiliation{D\'epartement de Math\'ematiques et Applications, UMR-8553, \'Ecole Normale Sup\'erieure, CNRS,PSL University, 75005 Paris, France}

%
%
%
\vskip 0.25cm


\date{\today}

\begin{abstract}
The focus of this study is to investigate primary and secondary
bifurcations to weakly nonlinear flows (weak branch) in convective
rotating spheres in a regime where only strongly nonlinear oscillatory
sub- and super-critical flows (strong branch) were previously found in
[E. J. Kaplan, N. Schaeffer, J. Vidal, and P. Cardin,
  Phys. Rev. Lett. {\bf{119}}, 094501 (2017)]. The relevant regime
corresponds to low Prandtl and Ekman numbers, indicating a
predominance of Coriolis forces and thermal diffusion in the system.
We provide the bifurcation diagrams for rotating waves (RWs) computed
by means of continuation methods and the corresponding stability
analysis of these periodic flows to detect secondary bifurcations
giving rise to quasiperiodic modulated rotating waves
(MRWs). Additional direct numerical simulations (DNS) are performed
for the analysis of these quasiperiodic flows for which Poincar\'e
sections and kinetic energy spectra are presented. The diffusion time
scales are investigated as well. Our study reveals very large initial
transients (more than 30 diffusion time units) for the nonlinear
saturation of solutions on the weak branch, either RWs or MRWs, when
DNS are employed. In addition, we demonstrate that MRWs have
multimodal nature involving resonant triads. The modes can be located
in the bulk of the fluid or attached to the outer sphere and exhibit
multicellular structures. The different resonant modes forming the
nonlinear quasiperiodic flows can be predicted with the stability
analysis of RWs, close to the Hopf bifurcation point, by analyzing the
leading unstable Floquet eigenmode.
\end{abstract}

\maketitle

\section{Introduction}
\label{sec:intro}

Present knowledge of many geophysical and astrophysical phenomena has
been acquired with the support of computer simulations of thermal
rotating convection in spherical geometry. This is especially the case
for the geodynamo \cite{GlRo95a,SJNF17}, for gas giant atmospheres
\cite{HGW15,GCW20}, and the Sun \cite{Rud89,BMT04} since flow
measurements in these environments are extremely difficult. In the
specific case of fluid planetary cores, including the Earth,
convective motions are thought to be driven by thermal and
compositional gradients \cite{Jon07} and are responsible for the
generation of magnetic fields \cite{GLPGS02,MoDo19}. In this context,
the dynamics is strongly influenced by rotation which constrains the
flow to form convective columns aligned with the axis of rotation
(e.\,g. \cite{DoSo07}). This quasi-geostrophic structure may prevail
even in turbulent regimes \cite{JRGK12,GCS19}.

Usually, a spherical shell is considered to model the existence of an
inner core (as in \cite{CACDGGGHJKMSTTWZ01}) but simulations in a full
sphere have been also performed for the modelling of ancient cores (as
in \cite{Mar_et_al15}). One of the simplest models, which has been
widely used, is the Boussinesq approximation of the Navier-Stokes and
energy equations in a rotating frame of reference \cite{Cha81}. If a
full sphere is considered the governing equations depend on three
parameters -the Prandtl ($\Pra$), Ekman ($\Ek$), and Rayleigh ($\Ray$)
numbers- which account for the physics of the problem. Concretely,
$\Pra$ measures the ratio of viscous (momentum) diffusivity to thermal
diffusivity, $\Ek$ the relevance of viscous over Coriolis forces,
while in the present study $\Ray$ is associated with an internal
heating source.

When the sphere is rapidly rotating (small $\Ek$) the fluid is at rest
up to a certain value of $\Ray$, and this value and the type of
emerging convective flow depends strongly on $\Pra$. For $\Pr>0.1$ the
onset of convection takes place in the form of quasi-geostrophic
columns, with spiral morphology, steadily drifting in the azimuthal
direction. These solutions are called rotating waves (RW) in the
context of symmetry theory \cite{Ran82,GLM00}. The spiral modes,
predicted by linear studies \cite{Zha92,DSJJC04} are nonaxisymmetric
(i.\,e. depend on the azimuthal coordinate) and equatorially
symmetric. For smaller values of $\Pra$ the topology of the linear
nonaxisymmetric modes is more diverse. The modes can be equatorially
symmetric or antisymmetric. The former are either trapped
(\cite{Zha93}) on the equatorial region, or multicellular and attached
to the outer boundary (\cite{NGS08}), while the latter are located at
high latitudes (\cite{GSN08,GCW18}).  In addition, a purely
axisymmetric mode can be preferred if $\Pra$ is sufficiently small
\cite{SGN16b,ZLK17}.

While the dynamics of nonlinear flows in the regime of large $\Pra$
has been investigated for several decades
(e.\,g. \cite{ABW97,SiBu03,OrDo14,GWA16,SJNF17} among many others) the
regime of small $\Pra$ has been less studied. This has, however,
started to change during the last decade
(e.\,g. \cite{GSDN15,HoSc17,KSVC17,LKZ18,ABGHV18,GCW19,Lin21}) because
low Prandtl numbers are more relevant for planetary and stellar
interiors \cite{Mas91}. When $\Pr$ is small enough, strong oscillatory
flows with multimodal nature, in which the interaction of certain
modes with different spatial localizations play a relevant role in the
dynamics, may appear right after the onset
(\cite{HoSc17,ABGHV18}). For instance, a flow consisting of convective
structures, either attached to the boundary or located in the
interior, has been observed in a recent experiment \cite{VHA21} with
liquid gallium ($\Pra=0.026$) inside a cylindrical vessel. In these
experiments, in agreement with \cite{HoSc17,ABGHV18}, steady
convective columns (i.e RWs) have not been found to exist at the
onset.

For a full rotating sphere, as in the present study, low $\Pra$
convection can be sub-critical and strongly energetic if $\Ek$ is
sufficiently small \cite{KSVC17}. However, in this regime weakly
energetic nonlinear flows (weak branch of \cite{KSVC17}), which
include non-axisymmetric RWs (steadily drifting flows in the azimuthal
direction), have not been found although they were predicted by the
linear theory \cite{EZK92}. The situation is different in the case of
low $\Pra$ and stress-free boundaries \cite{SaNe19}, because the first
convective instability is axisymmetric, i.e periodic torsional
oscillations develop at the onset. That study revealed a rich
dynamical regime including bifurcations to quasiperiodic flows and
solutions in which the amplitude is slowly increasing and rapidly
decaying, repeatedly. This repeated behavior was interpreted in terms
of heteroclinic chains connecting unstable states close to the onset
of torsional oscillations. The complex nature of low $\Pra$ flows,
described by several thermal-inertial modes with different symmetries,
is also demonstrated in \cite{LKZ18}, for the case of liquid
gallium. Moreover, triadic resonances involving convective and
inertial modes have been analyzed very recently in \cite{Lin21} for
$\Pra\le 0.01$. The above mentioned studies, and the results presented
here, are based on numerical simulations with parameters quite remote
from those of real planets. However, these studies model fundamental
features of planetary cores such as rapid rotation, spherical
geometry, or second order viscosity and thermal diffusivity effects,
and thus help to shed light onto flow instabilities occuring in
planetary interiors.

In the present study we compute RWs by means of continuation methods
(\cite{Kel77,DoTu00,SaNe16}) in a regime where they have not yet been
found. We select the parameters according to \cite{KSVC17} and
investigate the stability of the RWs demonstrating their
existence. These RWs consist of a single multicellular mode with fixed
azimuthal symmetry as described in \cite{NGS08}. By performing several
direct numerical simulations, we show the difficulty of obtaining
RWs at this regime since very long initial transients are required to
saturate the solutions. We investigate further bifurcations to
modulated rotating waves \cite{Ran82,GLM00,GNS16}, which are
quasiperiodic flows. These MRWs are multimodal, consisting of several
modes with different azimuthal symmetries and time scales, and we
demonstrate that this multimodal character can be indeed predicted
from the stability analysis of the RWs. The unstable eigenfunction
(Floquet mode) at the bifurcation reveals the main mode structure of
the multimodal MRWs, which include wall-attached and interior modes as
seen in recent numerical and experimental studies
\cite{HoSc17,ABGHV18,VHA21}. Finally, in agreement with \cite{Lin21},
triadic resonances have been found and interpreted in terms of MRWs as
done in \cite{GGS21} in the case of the magnetized spherical Couette
flow.  The outline of the paper is the following: First, the model
equations, numerical methods and parameters, are detailed in
\S\ \ref{sec:model}. The description of the main results obtained for
the RWs is undertaken in \S\ \ref{sec:rw} while the analysis of
quasiperiodic flows (MRWs) is left to \S\ \ref{sec:t_ev}. Finally, the
paper concludes in \S\ \ref{sec:conc} with a brief summary.

\section{The model}
\label{sec:model}

Boussinesq thermal convection in a self-gravitating, internally
heated, and rotating spherical shell, defined by the inner and outer
radius $r_i$ and $r_o$, is considered as in \cite{SiBu03}. To compare
with the full sphere results of \cite{KSVC17} we set
$\eta=r_i/r_o=0.01$. The effect of considering a very small inner
sphere in the modelling of Boussinesq rotating thermal convection
within a full sphere was considered in \cite{Mar_et_al15} where
several codes have been benchmarked. They have found errors below
0.4\% and 4\% for the volume-averaged kinetic energy and the main time
scale of a purely hydrodynamic RW close to the onset of convection,
which is just the same type of solutions considered in our study.

The physical properties of the fluid -- thermal diffusivity $\kappa$,
thermal expansion coefficient $\alpha$, and dynamic viscosity $\mu$ --
are constant and the density is assumed to vary linearly with the
temperature, $\rho=\rho_0(1-\alpha(T-T_0))$, just in the gravitational
term ${\bm g}=-\gamma \rv$ ($\gamma$ is constant and $\rv$ the
position vector). The system rotates with uniform angular velocity
${\bm{\Omega}}=\Omega {\kv}$ about the vertical axis ${\kv}$.

\subsection{Governing equations and numerical method}
\label{sec:gov_eq}

The Navier-Stokes and energy equations are derived in the rotating
frame of reference and expressed in terms of velocity ($\ve$) field
and temperature ($\Theta$) perturbation of the conductive state. They
are
\begin{align}
&\nabla\cdot\ve=0,\label{eq:cont}\\
&\partial_t\ve+\ve\cdot\nabla\ve+2\Ek^{-1}\kv\times\ve = 
-\nabla p^*+\nabla^2\ve+\Theta\rv,\label{eq:mom}\\
&\Pr\lp\partial_t \Theta+\ve\cdot\nabla \Theta\rp= \nabla^2
\Theta+\Ray~ \rv\cdot\ve.  \label{eq:ener}
\end{align}
No-slip boundary conditions $v_r=v_{\theta}=v_{\varphi}=0$,
where $(r,\theta,\varphi)$ are the radial, colatitudinal, and azimuthal
coordinates, are considered for the velocity field and the temperature
is fixed at the thermally conducting boundaries.  The
characteristic scales are $d=r_o-r_i$ for the distance,
$\nu^2/\gamma\alpha d^4$ for the temperature, and $d^2/\nu$ for the
time. The non-dimensional parameters -the aspect ratio ($\eta$), the
Rayleigh ($\Ray$), Prandtl ($\Pr$), and Ekman ($\Ek$) numbers- are
defined as
\begin{equation}
  \eta=\frac{r_i}{r_o},\quad
  \Ray=\frac{q\gamma\alpha d^6}{3c_p\kappa^2\nu},\quad
  \Ek=\frac{\nu}{\Omega d^2},\quad
  \Pr=\frac{\nu}{\kappa},
\label{eq:param}
\end{equation}
where $c_p$ is the specific heat at constant pressure and $q$ is the
rate of heat due to internal sources per unit mass. In these units the
conduction state is $\ve=0$ and $T_c(r)=T_0-(\Ray/2\Pr)r^2$.

The toroidal-poloidal formulation
(\cite{Cha81}) expresses a divergence-free velocity field in terms of
toroidal, $\Psi$, and poloidal, $\Phi$, potentials
\begin{equation}
  \ve=\nabla\times\lp\Psi\rv\rp+\nabla\times\nabla\times\lp\Phi\rv\rp,
\label{eq:pot}  
\end{equation}
and a pseudo-spectral method (see \cite{GNGS10}), in which a
Gauss--Lobatto mesh of $N_r$ radial collocation points\citep{SGN16} is
used in the radial direction and spherical harmonics are used for the
angular coordinates, is employed. The unknowns of the governing
equations \ref{eq:cont}-\ref{eq:ener} are then
\begin{eqnarray}
  \Psi(t,r,\theta,\varphi)=\sum_{l=0}^{L_{\text{max}}}\sum_{m=-l}^{l}{\Psi_{l}^{m}(r,t)Y_l^{m}(\theta,\varphi)},\label{eq:serie_psi}\\
  \Phi(t,r,\theta,\varphi)=\sum_{l=0}^{L_{\text{max}}}\sum_{m=-l}^{l}{\Phi_{l}^{m}(r,t)Y_l^{m}(\theta,\varphi)},\label{eq:serie_phi}\\
    \Theta(t,r,\theta,\varphi)=\sum_{l=0}^{L_{\text{max}}}\sum_{m=-l}^{l}{\Theta_{l}^{m}(r,t)Y_l^{m}(\theta,\varphi)},\label{eq:serie_the}
\end{eqnarray}
with $\Psi_l^{-m}=\overline{\Psi_l^{m}}$,
$\Phi_l^{-m}=\overline{\Phi_l^{m}}$, $\Psi_0^0=\Phi_0^0=0$ to uniquely
determine the two potentials, and
$Y_l^{m}(\theta,\varphi)=P_l^m(\cos\theta) e^{im\varphi}$, where
$P_l^m$ is the normalized associated Legendre functions of degree $l$
and order $m$ up to $L_{\text{max}}$.

The code is parallelized in the spectral $(m,l)$ as well as the
physical $(r,\theta,\varphi)$ space using OpenMP directives. The
computation of the nonlinear term relies on the pseudo-spectral
transform method~\citep{Ors70} which requires fast Fourier and
Legendre transforms. These are implemented using the optimized
libraries FFTW3 \cite{FrJo05} and dgemm \cite{GoGe08}.  The time
integration is based on high order implicit-explicit backward
differentiation formulas IMEX--BDF \cite{GNGS10}. The nonlinear terms
are integrated explicitly, to avoid implicit solution of nonlinear
systems but the Coriolis term is considered fully implicit to allow
larger time steps during the time integration \citep{GNGS10}.

\subsection{Computation of rotating waves}
\label{sec:co_meth}

Rotating waves (RW) in spherical systems are periodic solutions for
which the time and azimuthal coordinates are coupled, i.\,e. their
time dependence is described by a steady drift in the azimuthal
direction with uniform rotation frequency. This type of solution is
common in spherical systems since these are invariant by azimuthal
rotations ({\bf{SO}}$(2)$) and reflections with respect to the
equatorial plane ({\bf{Z}}$_2$). Generally, in {\bf{SO}}$(2)$
symmetric systems, non-axisymmetric RWs, which can be stable or
unstable, bifurcate after the axisymmetric base state becomes unstable
(primary Hopf bifurcation~\cite{EZK92,CrKn91}).

The computation of RW and the study of their stability helps to
understand the origin and structure of secondary flows,
i.\,e. modulated rotating waves (MRW), which are quasiperiodic and
oscillatory solutions found near the onset of convection
(e.\,g. \cite{Ran82,CoMa92,GLM00}). The symmetry properties of flows
occurring near the onset can thus be understood in terms of
bifurcation theory \cite{CrKn91}. The study of periodic and
quasiperiodic unstable flows is important since these types of
solutions act as organizing centers for the global dynamics
\cite{KUL12}. Moreover, the analysis of unstable RW provides useful
insights into the appearance of turbulent flows \cite{Hof_et_al04}.

In this section we outline the method to compute RWs which are indeed
the simplest time dependent solutions belonging to the weak
branches studied in \cite{KSVC17} and, more generally, in rotating
thermal convection in spherical geometry. Concretely, we use
continuation methods (e.\,g. \cite{Kel77,Doe86,DoTu00}) of periodic
orbits since RWs are periodic flows.  We refer the reader
to~\cite{SNGS04b}, or the comprehensive tutorial~\cite{SaNe16}, for a
full description of continuation methods in large-scale dissipative
systems such as the considered in our study. Continuation methods have
been already applied for thermal convection in rotating spherical
shells in \cite{SGN13,GNS16} and \cite{GCW19}, so only few details are
provided here.

For fixed $\Pr$ and $\Ek$ we want to study the dependence of RWs,
having $m_0$-fold azimuthal symmetry and rotating in the azimuthal
direction with frequency $\omega$, with respect to the control
parameter $p=\Ray$. Pseudo-arclength continuation methods obtain the
branch of periodic solutions
$x(s)=(u(s),\tau(s),p(s))\in\mathbb{R}^{n+2}$, where $u$ is the
rotating wave, $\tau=2\pi/(m_0\omega)$ is the rotation period, and $s$
is the arclength parameter. We note that the vector
$u\in\mathbb{R}^{n}$ contains the spherical harmonic amplitudes, at
the radial collocation points, of the scalar potentials and the
temperature perturbation. The dimension of the vector is
$n=(3L_{\text{max}}^2+6L_{\text{max}}+1)(N_r-1)$.

The pseudo-arclength methods require the condition
\begin{equation}
h(u,\tau,p)\equiv\langle w,x-x^0 \rangle=0,
\end{equation}
where $x^0=(u^0,\tau^0,p^0)$ and $w=(w_u,w_\tau,w_p)$ are the
predicted point and the tangent to the curve of solutions,
respectively, obtained by extrapolation of the previous points along
the curve. We note that $\langle .,. \rangle$ stands for the inner
product in $\mathbb{R}^{n+2}$. To find a single solution
$x=(u,\tau,p)$ on the branch we solve the system:
\begin{equation}
H(u,\tau,p)= \left(
\begin{array}{c}
u-\phi(\tau,u,p)\\
g(u)\\
h(u,\tau,p)\\
\end{array}
\right)
=0,
\label{eq:H_eq}
\end{equation}
where $\phi(\tau,u,p)$ is a solution of
Eqs. (\ref{eq:cont}-\ref{eq:ener}) at time $\tau=2\pi/(m_0\omega)$ and
initial condition $u$ for fixed $p$. The additional constraint
$g(u)=0$ is imposed to fix the azimuthal phase of the RW with respect
to the rotating reference frame. See eq. (9) of \cite{SGN13} for
further details on the definition of $g(u)$.

Newton-Krylov methods are employed to solve the large non-linear
system defined by Eq.~(\ref{eq:H_eq}). Krylov methods are used since
they only require the action of the Jacobian
$D_{(u,\tau,p)}H(u,\tau,p)$ on a given vector, and not its explicit
computation, which due the spatial resolutions used in our study would
be prohibitive. For the evaluation of the Jacobian a time integration
of a system obtained from the Navier-Stokes and energy equations must
be performed. We note that periodic rotating waves can also be
obtained efficiently by Newton-Krylov continuation methods but as
steady solutions of the equations written in a reference frame which
is rotating with the wave, see for instance~\cite{SGN13,FSTG13,FTGS15}
for thermal convection or dynamo problems in spherical geometries
or~\cite{TLW19} for the pipe flow.

Floquet theory (e.\,g. \cite{JoSm07}) is applied to study the
stability of RWs so the dominant eigenvalue of the map $\delta
u\longrightarrow D_u \phi(\tau,u,p)\delta u= v(\tau)$, where $v(\tau)$
is the solution of the first variational equation (see \cite{GNS16}
and \cite{GCW19} for further details), must be estimated. Arnoldi
methods (ARPACK \cite{LSY98}) are used to compute eigenvalues of
larger modulus corresponding to the dominant Floquet multiplier
$\lambda=|\lambda|e^{i\text{Arg} \lambda}$. When $|\lambda|>1$ the RW
is unstable. The Floquet multiplier with $|\lambda|=1$ and
eigenfunction $v_1=\partial_t u$, associated to the invariance under
azimuthal rotations, is deflated by redefining the map $\delta u
\longrightarrow v(\tau)-\braket{v(\tau),v_1}v_1$. The azimuthal
symmetry, $m_1$, of the leading eigenfunction should be a factor of
the azimuthal symmetry, $m_0$, of the RW. We note that this eigenvalue
problem requires the time integration of an ODE system of dimension
$2n$ over one rotation period, which is an extensive computational
task. Because the periodic orbit is a RW there is a more efficient
alternative to this procedure (see \cite{SGN13,FSTG13,Tuc15,FTGS15})
which consists of studying the stability as a fixed point of a vector
field. However, this method requires to apply shift-invert techniques
to the eigenvalue solver. Numerical tests performed in \cite{SGN13}
found the Floquet analysis method more robust than the steady state
method, but less efficient.

\subsection{Parameters for the study of the weak branch}
\label{sec:par}

\begin{table*}[t!] 
  \begin{center}
    \begin{tabular}{lccccccccc}
\vspace{0.1cm}           
Set&$N_r$ & $L_{\text{max}}$ & $\Ek$            & $\Pra$   & $m_c$  & $\Rayc$              & $\omega_c\Ek$  & $\Rayc^*$            & $\omega_c^*\Ek$      \\
\hline\\[-8.pt]
$P_1$&$50$  & $160$        & $3\times 10^{-6}$ & $0.03$  & $12$   & $2.3392\times 10^7$    & $-0.042852$    & $2.336\times 10^7$    & $-0.04275$     \\
    &$60$  & $156$        & $3\times 10^{-6}$ & $0.03$  & $12$   & $2.3365\times 10^7$    & $-0.042861$    & $2.336\times 10^7$    & $-0.04275$     \\
\hline\\[-8.pt]
$P_2$&$50$  & $154$        & $10^{-6}$         & $0.01$  & $11$   & $5.7165\times 10^7$    & $-0.039801$    & $5.475\times 10^7$    & $-0.03895$    \\
     &$80$  & $154$        & $10^{-6}$         & $0.01$  & $11$   & $5.4949\times 10^7$    & $-0.039022$    & $5.475\times 10^7$    & $-0.03895$    \\
     &$80$  & $192$        & $10^{-6}$         & $0.01$  & $11$   & $5.4949\times 10^7$    & $-0.039022$    & $5.475\times 10^7$    & $-0.03895$    \\
\hline\\[-8.pt]
$P_3$& $80$  & $192$        & $3\times 10^{-7}$ & $0.003$ & $12$   & $1.2609\times 10^8$    & $-0.042899$    & $1.255\times 10^8$    & $-0.04287$     \\
     &$100$  & $192$       & $3\times 10^{-7}$ & $0.003$ & $12$   & $1.2536\times 10^8$    & $-0.042913$    & $1.255\times 10^8$    & $-0.04287$     \\
     &$100$  & $192$       & $3\times 10^{-7}$ & $0.003$ & $11$   & $1.2584\times 10^8$    & $-0.041916$    &                       &      \\
\hline
  \end{tabular}
    \caption{Number of radial collocation points $N_r$, spherical
      harmonic truncation parameter $L_{\text{max}}$, and critical
      Rayleigh numbers $\Rayc$, azimuthal wave numbers $m_c$ and
      critical frequencies $\omega_c$ for the onset of convection for
      the three sets of parameters $P_i$, $i=1,2,3$, considered. For
      the set $P_3$ the critical parameters for the nonpreferred
      $m=11$ eigenfunction are also listed. Values marked with the *
      symbol are taken from~\cite{KSVC17}.}
  \label{tab:lin}
  \end{center}
\end{table*}

Several combinations of the parameters given in Eq. (\ref{eq:param})
are considered to explore the appearance of solutions belonging to the
weak branch. This branch bifurcates supercritically from the
conductive state and the flow is localized away from the interior of
the sphere and characterized by the predominance of diffusion rather
than advection transport. In contrast, for solutions belonging to the
strong branch advection dominates and there is a strong thermal
anomaly and noticeable zonal flow near the sphere's origin. At
moderate rotation rates the strong branch is found at usually larger
forcing than that required for the weak branch, but in rapidly
rotating spheres at low $\Pra$ the strong branch can be subcritical
\cite{KSVC17}. The regimes selected in our study are characterized by
low $\Pr$ and $\Ek$ in accordance with the study of \cite{KSVC17} in a
full sphere. Because our formulation of the problem is different than
that used in \cite{KSVC17}, we describe the results in terms of their
definition. The relation between the Rayleigh number of
Eq. (\ref{eq:param}) and the Rayleigh number $\Ray^{\text{K}}$ defined
in \cite{KSVC17} is $\Ray=2(1-\eta)^6\Ray^{\text{K}}$. For the sake of
simplicity we use $\Ray=\Ray^{\text{K}}$ from now on.  In addition,
following \cite{KSVC17} the diffusion time scale
$\tau_{\kappa}=d^2/\kappa$ is used for analyzing the results giving
rise to the dimensionless time $t_{\kappa}=t_{\nu}/\Pra$, where
$t_{\nu}$ is the dimensionless viscous time employed in our numerical
code.

Following \cite{KSVC17} three different pairs $P_i=(\Pr_i,\Ek_i)$,
$i=1,2,3$, are considered. They are $P_1=(0.03,3\times 10^{-6})$,
$P_2=(0.01,10^{-6})$, and $P_3=(0.003,3\times 10^{-7})$.  The critical
Rayleigh numbers, azimuthal wave numbers and critical frequencies for
the onset of convection for the three different sets are listed in
Table~\ref{tab:lin}. The frequencies are normalized by
$\omega_c\Ek=\omega_c^{\text{dim}}/\Omega$, where
$\omega_c^{\text{dim}}=\omega_c \nu/d^2$ is the dimensional
frequency. In this table the number of radial collocation points
($N_r$) and spherical harmonic truncation parameter ($L_{\text{max}}$)
used for the computations are listed as well. For the sets $P_1$ and
$P_2$ the critical mode flow patterns can be described (\cite{Zha92})
as a set of columns, with a single convective cell, which are parallel
to the axial direction, spiral in the azimuthal direction, and are
located in the interior of the shell. However, the columns become
multicellular and attached to the outer sphere in the case of the set
$P_3$. The onset of multicellular modes has been already studied in
\cite{NGS08} for the case of a thick rotating spherical shell.

The motivation for the choice of these three sets is described in the
following. The DNS of \cite{KSVC17} showed the existence of the weak
branch for the sets $P_1$ and $P_2$, i.e where the onset of convection
is in the form of spiraling modes, but the weak branch was not found
for the set $P_3$, where the onset of convection is multicellular and
equatorially attached. In the present study we show that RWs,
solutions belonging to the weak branch, can also be found for the set
$P_3$ if continuation methods are employed. Considering the sets $P_1$
and $P_2$ allows us to check our results and to investigate why it is
difficult to obtain the weak branch by means of DNS for the set
$P_3$. The focus of the present study is then on the set $P_3$ since
the weak branch for this set has not yet been described. By performing
additional DNS we also study quasiperiodic flows, bifurcating from
RWs, that also belong to the weak branch regime.

\begin{figure}[b!]
\begin{center}
  \includegraphics[scale=1.4]{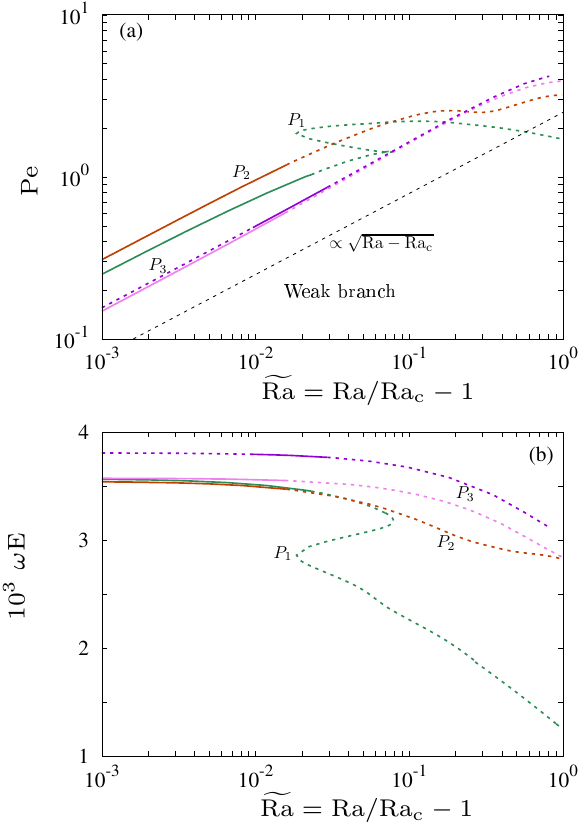}
\end{center}  
\caption{Bifurcation diagrams of rotating waves for the 3 sets of
  parameters: $P_1$ with $(\Pr,\Ek)=(0.03,3\times 10^{-6})$, $P_2$
  with $(\Pr,\Ek)=(0.01,10^{-6})$, and $P_3$ with
  $(\Pr,\Ek)=(0.003,3\times 10^{-7})$. For the set $P_3$ two different
  branches with $m_0=11$ (violet) and with $m_0=12$ (magenta) are shown.
  (a) P\'eclet number $\Pec$ versus $\Rayt=\Ray/\Rayc-1$. The dashed line
  indicates the $\sqrt{\Ray-\Rayc}$ scaling predicted
  in~\cite{EZK92}.  (b) Scaled rotating frequency $\omega \Ek$ versus
  $\Rayt$. Solid/dashed lines mark stable/unstable rotating
  waves.}
\label{fig:bif_diag}   
\end{figure}

We note that both, $\Pr$ and $\Ek$ roughly decrease by a factor of $3$
from $P_1$ to $P_2$, and from $P_2$ to $P_3$, so there is an increase
of computational complexity from $P_1$ to $P_3$. As is clear from
Table~\ref{tab:lin} the critical frequencies $\omega_c$ increase since
the product $\omega_c \Ek$ remains very similar for all the cases. In
addition, as the Prandtl and Ekman numbers are decreased the marginal
stability curves for the onset of convection corresponding to a single
azimuthal wave number approach each other (e.\,g. \cite{NGS08,SGN16b})
meaning that multitudes of radial and colatitudinal structures are
unstable just after the onset.

The global quantity analyzed, in correspondence with \cite{KSVC17}, is
the Peclet number $\Pec=r_oU/\kappa$, which in terms of the
dimensionless volume-averaged kinetic energy $K$ becomes
$\Pec=(1-\eta)^{-1}\Pr~(2K)^{1/2}$. According to \cite{KSVC17} the
Peclet number helps to identify if a solution belongs to the weak
branch or not depending on whether $\Pec<10$ or not. This threshold
separates flows dominated by diffusion (weak branch) to flows
dominated by advection (strong branch). The frequencies of the RWs,
$\omega \Ek$, or the volume-averaged kinetic energy $K_m$, computed by
considering only the azimuthal wave number $m$ in the spherical
harmonics expansion of the toroidal and poloidal potentials
(Eqs. (\ref{eq:serie_psi}) and (\ref{eq:serie_phi})), are also
considered as global data. Regarding local data, the time series of
the temperature perturbation, picked up at some points inside the
fluid, and the time series of the real part of the poloidal amplitudes
of Eq. (\ref{eq:serie_phi}) of different modes $(m,l)$ in the middle
of the sphere, are considered.

\section{Rotating waves}
\label{sec:rw}

\begin{figure}[t!]
  \begin{center}
    \includegraphics[scale=1.2]{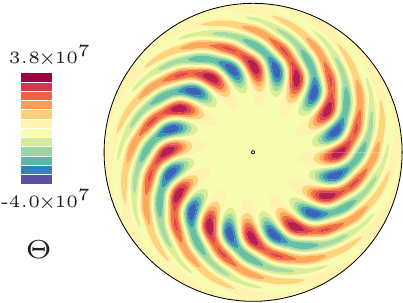}
    \includegraphics[scale=1.2]{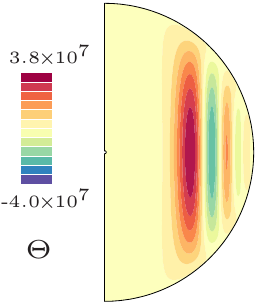}\\[2.mm]
    \includegraphics[scale=1.2]{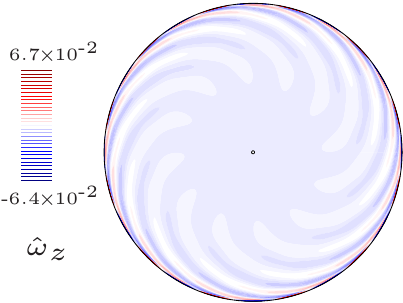}
    \includegraphics[scale=1.2]{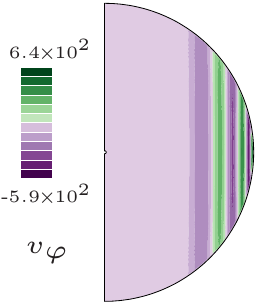}    
\end{center}  
\caption{Rotating wave with $m_0=12$, in the case of the set $P_3$
  ($\Ek=3\times 10^{-7}$,$\Pra=0.003$) and $\Ray=1.2634\times 10^8$
  ($\Rayt=7.9\times 10^{-3}$). Top row: Contour plots for the
  temperature perturbation $\Theta$ on an equatorial and meridional
  section. Bottom row: Contour plots for the vertical vorticity
  $\hat{\omega}_z$ on an equatorial section and for the azimuthal
  velocity $v_{\varphi}$ on a meridional section.}
\label{fig:rw1_cp_E3.d-7}   
\end{figure}
\begin{figure*}[t!]
  \begin{center}
    \includegraphics[scale=0.85]{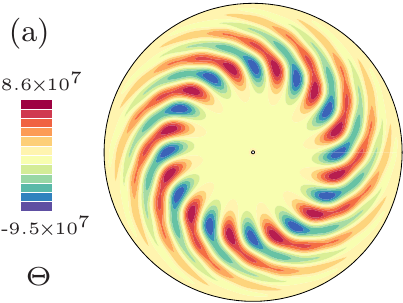}
    \includegraphics[scale=0.85]{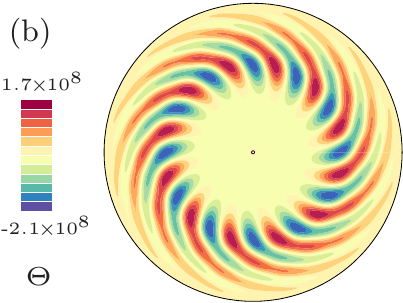}
    \includegraphics[scale=0.85]{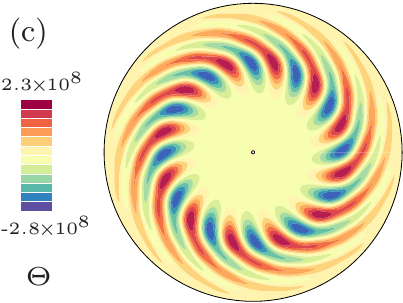}
    \includegraphics[scale=0.85]{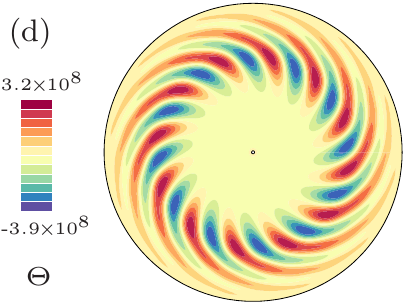}
    \includegraphics[scale=0.85]{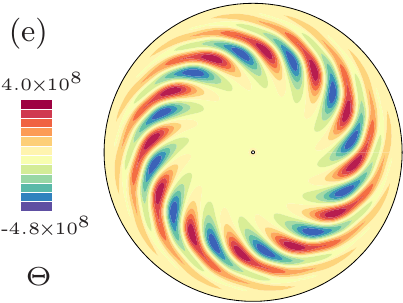}\\
    \includegraphics[scale=0.85]{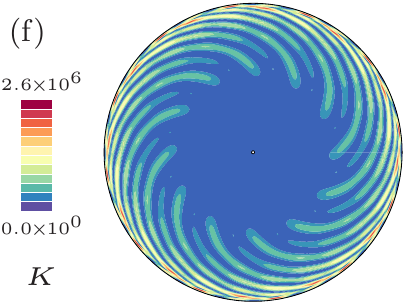}
    \includegraphics[scale=0.85]{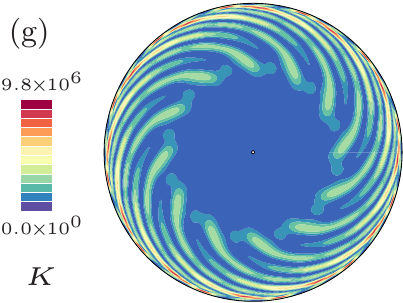}
    \includegraphics[scale=0.85]{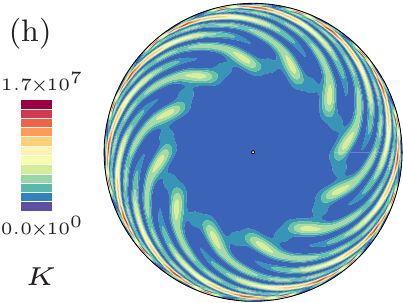}
    \includegraphics[scale=0.85]{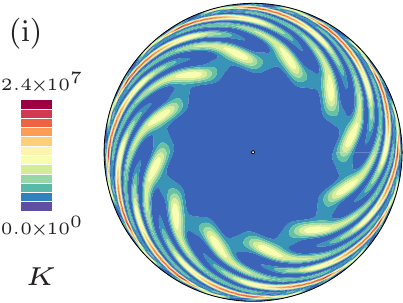}
    \includegraphics[scale=0.85]{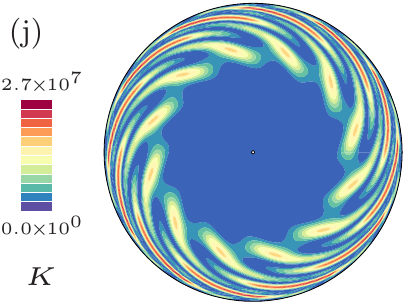}
  \end{center}
\caption{Rotating waves with $m_0=12$, in the case of the set $P_3$
  ($\Ek=3\times 10^{-7}$,$\Pra=0.003$). Contour plots for the
  temperature perturbation $\Theta$ (top row) and for the kinetic
  energy density $K$ (bottom row) on an equatorial section. The
  Rayleigh numbers are $\Ray=1.3030\times 10^8$ (a,f),
  $\Ray=1.4511\times 10^8$ (b,g), $\Ray=1.5932\times 10^8$ (c,h),
  $\Ray=1.9650\times 10^8$ (d,i), and $\Ray=2.5067\times 10^8$
  (e,j). This corresponds to $\Rayt=3.9\times 10^{-2}$,
  $\Rayt=1.6\times 10^{-1}$, $\Rayt=2.7\times 10^{-1}$,
  $\Rayt=5.7\times 10^{-1}$, and $\Rayt=10^{0}$.}
\label{fig:rw2_cp_E3.d-7}   
\end{figure*}

By means of the continuation method described in \S\ \ref{sec:co_meth}
the bifurcation diagrams for RWs corresponding to the three sets,
$P_1=(\Pr,\Ek)=(0.03,3\times 10^{-6})$,
$P_2=(\Pr,\Ek)=(0.01,10^{-6})$, and $P_3=(\Pr,\Ek)=(0.003,3\times
10^{-7})$, are obtained. For each set $P_i$ the azimuthal symmetry of
the RWs correspond to that at the onset of convection given in Table
\ref{tab:lin}. Concretely, $m_0=12$ for $P_1$, $m_0=11$ for $P_2$, and
$m_0=12$ for $P_3$. Figure~\ref{fig:bif_diag} displays the Peclet number
$\Pec$ and the normalized rotation frequencies $\omega\Ek$ versus the
parameter $\Rayt=\Ray/\Rayc-1$, which measures the departure from the
onset. Stable (resp. unstable) RWs are denoted by solid (resp. dashed)
lines.

Figure~\ref{fig:bif_diag}(a) is the same as Figure~1 of \cite{KSVC17}
but note that in \cite{KSVC17} two additional sets, one at
$\Ek=10^{-5}$ and the other at $\Ek=10^{-7}$, were displayed. To
compare both figures one must take into account that slight deviations
of the value of $\Rayc$ imply important deviations in $\Pec$ for
values of $\Rayt$ close to $10^{-2}$. For instance, if for the set
$P_2$ we use $\Rayc=5.475\times 10^7$, given in \cite{KSVC17}, instead
of our computed $\Rayc=5.4949\times 10^7$, given in Table
\ref{tab:lin}, we would obtain a value of $\Pec\approx 0.7$ (in
agreement with \cite{KSVC17}) instead of $\Pec\approx 1$ marked in
Figure~\ref{fig:bif_diag}(a). Note that for the set $P_1$ our results
agree with those of \cite{KSVC17} since the critical Rayleigh numbers
for this set have the same three first significant figures (see Table
\ref{tab:lin}).

In contrast to \cite{KSVC17}, we have found stable the branch of RWs
with azimuthal symmetry $m_0=12$ (weak branch) bifurcating from the
onset in the case of the set $P_3$. Certainly, these solutions can be
found up to a critical value of the Rayleigh number marking the
interval of stability of the branch. This interval is comparable to
those of the weak branches bifurcating from the onset for the sets
$P_1$ and $P_2$. Aside the branch of RWs with azimuthal symmetry
$m_0=12$, we have computed a branch of RWs with azimuthal symmetry
$m_0=11$. This branch is born unstable as it corresponds to the second
preferred eigenfunction at the onset of convection but becomes stable
very close to the onset. As it will be shown in the next sections, to
assess the stability of these RWs, or to compute them using DNS, is a
computationally challenging task.

Figure \ref{fig:bif_diag}(a) evidences that all the branches follow
the $\sqrt{\Ray-\Rayc}$ scaling, since there's a Hopf bifurcation
breaking the axisymmetry of the basic state (\cite{EZK92}). This
scaling is only valid close to the bifurcation point. Notice how the
scaling is valid in a larger interval as we go from set $P_1$ to the
set $P_3$ indicating that the validity of the scaling depends on the
other parameters ($\Pra,\Ek$). In addition, very close to the onset
the branches become more steep. This is only clear for the branch with
$m_0=11$ corresponding to the set $P_3$ but also occurs for the other
branches. Notice that for larger values of $\Rayt$ the Peclet number
departs from the predicted scaling and in the case of the set $P_1$
two saddle-node bifurcations occur (the folds of the curve).

In figure~\ref{fig:bif_diag}(b) the dependence of the rotation
frequencies on the Rayleigh number is analyzed by displaying $\omega
\Ek$ versus $\Rayt$. We recall that the rotation frequency ($\omega$)
of a RW with azimuthal symmetry $m_0$ is related to the critical
frequency at the onset ($\omega_c$) by $\omega=-\omega_c/m_0$. This is
clear when comparing the values of Fig. \ref{fig:bif_diag}(b) at
$\Rayt=10^{-3}$ with Table \ref{tab:lin}. Note that the frequencies
$\omega$ remain nearly constant among the three different sets up to
$\Rayt=2\times 10^{-2}$. From this point the frequency decreases
significantly in the case of $P_1$. In addition, the frequencies of
the branches bifurcating from the onset are almost equal for the three
sets.

The flow and temperature patterns for a stable RW with azimuthal
symmetry $m_0=12$ corresponding to the case $P_3$ at
$\Ray=1.2634\times 10^8$ ($\Rayt=7.9\times 10^{-3}$) are investigated
in figure \ref{fig:rw1_cp_E3.d-7} and correspond to the patterns of a
multicellular mode described in \cite{NGS08}. The first row displays,
from left to right, the contour plots of the temperature perturbation
on the equatorial plane and on a meridional section. On the second
row, the contour plots for the vertical vorticity $\hat{\omega}_z$
(normalized by the planetary vorticity $\hat{\omega}_z=\omega_z\Ek/2$)
on an equatorial plane, and the contour plots for the azimuthal
velocity $v_{\varphi}$ on a meridional section, are shown. The
meridional sections cut relative maxima of the fields. All the fields
shown in Fig. \ref{fig:rw1_cp_E3.d-7} were already shown in Figs. 2
and S2 of \cite{KSVC17}, but for a solution corresponding to the case
$P_2$ at $\Rayt=10^{-2}$ for which the flow patterns are very
similar. The flow is strongly geostrophic displaying convective
columns aligned with the rotation axis (see meridional sections). In
addition, azimuthal velocity and vertical vorticity tend to be
attached to the outer sphere and multicelullar spiral arms are clearly
seen on the equatorial section for the temperature
perturbation. Additional contour plots in the case of a tricelullar
mode can be found in Figure 4 of \cite{NGS08}.

The main effect of increasing the Rayleigh number is to slightly
displace the hot fluid cells (i.e the maximum of temperature
perturbation) towards the outer sphere, in the cylindrical radial
direction (see the first row of Fig. \ref{fig:rw2_cp_E3.d-7}), whereas
the flow regions with maximum kinetic energy are progressively moved
inwards, towards the inner sphere, although they still remain located
close to the outer sphere (see the second row of
Fig. \ref{fig:rw2_cp_E3.d-7}). In addition, for $\Rayt>2.\times
10^{-1}$ (three rightmost plots), fluid motions start to develop near
the middle of the shell developing a ring of vortices which displays a
characteristic polygonal structure.

\subsection{Stability of rotating waves}
\label{sec:rw_stab}

By means of the method described in Sec.  \ref{sec:co_meth} the
stability of RWs for each set of parameters is analyzed. We have found
that for all the three sets the RWs become unstable due to Hopf
bifurcations giving rise to modulated rotating waves (MRW). This
scenario, which has been already described in \cite{SGN13,GNS16} for
thermal convection in rotating spherical shells, is typical in
{\bf{SO}}$(2)$ symmetric systems \citep{Ran82,GLM00,CrKn91}.

\begin{table*}[t!] 
  \begin{center}
    \begin{tabular}{lccccccccc}
\vspace{0.1cm}           
$N_r$ & $L_{\text{max}}$ & $\Ek$            & $\Pra$  & $\Ray$               & $\Rayt$      & $K$                 & $\omega$             &$|\lambda|$ &$m_0$\\
\hline\\[-3.mm]
$60$  & $156$          & $3\times 10^{-6}$ & $0.03$  & $2.38589\times 10^7$ & $2.0\times 10^{-2}$   & $1.0981\times 10^3$ & $1.1314\times 10^3$ &$0.98915$ & $12$\\
$70$  & $192$          & $3\times 10^{-6}$ & $0.03$  & $2.38589\times 10^7$ & $2.0\times 10^{-2}$   & $1.1036\times 10^3$ & $1.1312\times 10^3$ &$0.98923$ & $12$\\
\hline\\[-3.mm]
$80$  & $154$          & $10^{-6}$         & $0.01$  & $5.53968\times 10^7$ & $8.2\times 10^{-3}$   & $7.4015\times 10^3$ & $3.4380\times 10^3$ &$0.97097$ & $11$\\
$100$ & $198$          & $10^{-6}$         & $0.01$  & $5.53968\times 10^7$ & $8.2\times 10^{-3}$   & $7.5082\times 10^3$ & $3.4375\times 10^3$ &$0.97033$ & $11$\\
\hline\\[-3.mm]
$100$ & $192$         & $3\times 10^{-7}$ & $0.003$ & $1.27467\times 10^8$ & $1.7\times 10^{-2}$   & $4.2477\times 10^4$ & $1.1603\times 10^4$ &$1.00214$ & $12$\\
$120$ & $192$         & $3\times 10^{-7}$ & $0.003$ & $1.27467\times 10^8$ & $1.7\times 10^{-2}$   & $4.2723\times 10^4$ & $1.1603\times 10^4$ &$1.00218$ & $12$\\
\hline
  \end{tabular}
    \caption{Number of radial collocation points $N_r$, spherical
      harmonic truncation parameter $L_{\text{max}}$, Ekman $\Ek$,
      Prandtl $\Pra$, and Rayleigh $\Ray$ numbers, volume-averaged
      kinetic energy $K$, rotating frequency $\omega$, and modulus of
      the dominant Floquet multiplier $|\lambda|$, for rotating waves
      with azimuthal symmetry $m_0$. The value $\Rayt=\Ray/\Rayc-1$,
      where $\Rayc$ is the critical Rayleigh number for the onset of
      convection, can be used to locate the rotating wave on the
      bifurcation diagram of Fig. \ref{fig:bif_diag}(a).}
  \label{tab:val_rw}
  \end{center}
\end{table*}

\begin{figure}[h!]
\begin{center}
  \includegraphics[scale=1.5]{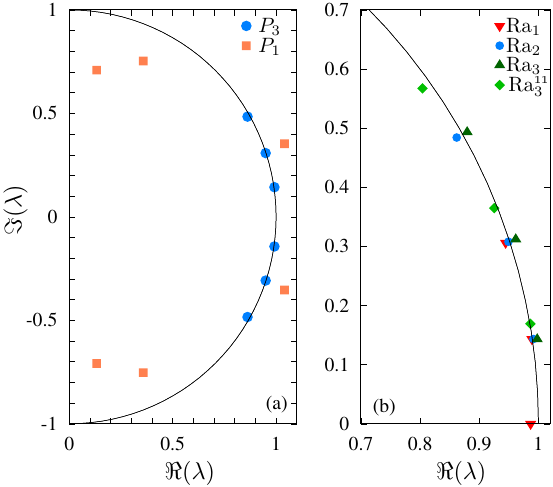}
\end{center}  
\caption{(a) Real and imaginary part of the leading Floquet
  multipliers $\lambda$ corresponding to the first unstable RWs with
  azimuthal symmetry $m_0=12$ for the set $P_1=(\Pr,\Ek)=(0.03,3\times
  10^{-6})$ (squares, orange online) and for the set
  $P_3=(\Pr,\Ek)=(0.003,3\times 10^{-7})$ (circles, blue online). The
  Rayleigh numbers are $\Ray=2.43043\times 10^7$ ($\Rayt=4\times
  10^{-2}$) and $\Ray=1.27467\times 10^8$ ($\Rayt=1.7\times 10^{-2}$),
  respectively. (b) Leading Floquet multipliers for the last stable
  (triangles down, red online) and first (circles, blue online) and
  second unstable (triangles up, dark-green online) RWs with azimuthal
  symmetry $m_0=12$ for the set $P_3$ at $\Ray_1=1.26344\times 10^8$,
  $\Ray_2=1.27467\times 10^8$, and $\Ray_3=1.3030\times 10^8$,
  respectively. The first unstable RW with azimuthal symmetry $m_0=11$
  at $\Ray_3^{11}=1.3011\times 10^8$ is also shown (diamonds, green
  online). The conjugate Floquet multipliers are not shown in (b) and
  the solid line marks the unit circle.}
\label{fig:arp}   
\end{figure}

\begin{figure}[h!]
\begin{center}
\includegraphics[scale=1.5]{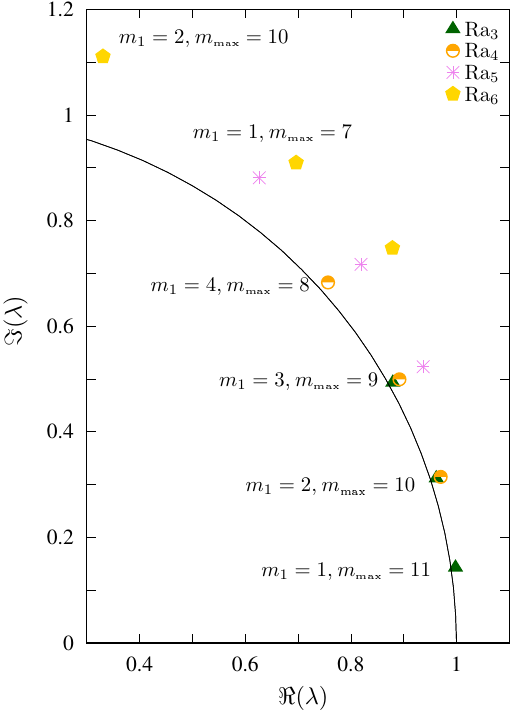}
\end{center}  
\caption{Leading Floquet multipliers for unstable RWs with azimuthal
  symmetry $m_0=12$ and the set $P_3$ ($\Ek=3\times
  10^{-7}$,$\Pra=0.003$) at $\Ray_3=1.3030\times 10^8$ (triangles,
  dark-green online), at $\Ray_4=1.3228\times 10^8$ (circles, orange
  online), $\Ray_5=1.3919\times 10^8$ (asterisk, violet online), and
  $\Ray_6=1.4511\times 10^8$ (pentagon, yellow online). The labels
  indicate the azimuthal symmetry $m_1$ of the eigenfunction and its
  most energetic wave number $m_{\text{max}}$. The conjugate Floquet
  multipliers are not shown and the solid line marks the unit circle.}
\label{fig:arp2}   
\end{figure}

\begin{figure}[t!]
  \begin{center}
    \includegraphics[scale=1.2]{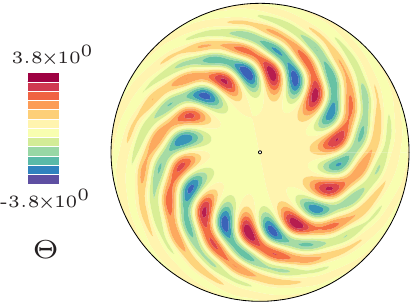}
    \includegraphics[scale=1.2]{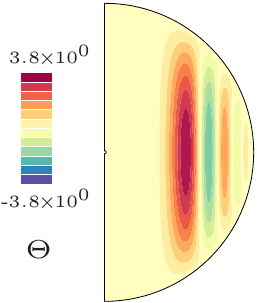}\\[2.mm]
    \includegraphics[scale=1.2]{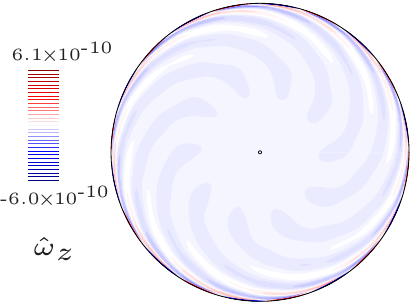}
    \includegraphics[scale=1.2]{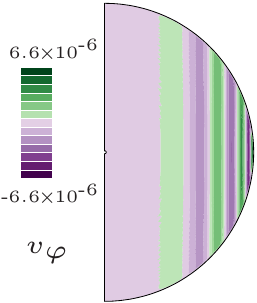}\\[2.mm]
    \includegraphics[scale=1.2]{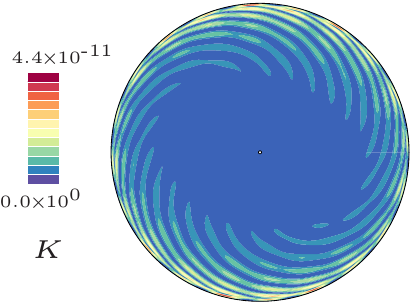}
    \includegraphics[scale=1.2]{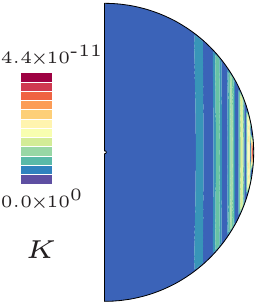}\\[2.mm]    
  \end{center}
\caption{Leading eigenfunction of a RW with $m_0=11$, in the case of
  the set $P_3$ ($\Ek=3\times 10^{-7}$,$\Pra=0.003$), and
  $\Ray=1.3011\times 10^8$. Top row: Contour plots for the temperature
  perturbation $\Theta$ on an equatorial and meridional
  sections. Middle row: Contour plots for the vertical vorticity
  $\hat{\omega}_z$ on an equatorial section and for the azimuthal
  velocity $v_{\varphi}$ on a meridional section. Bottom row: Contour
  plots for the kinetic energy density $K$ on an equatorial and
  meridional sections. The azimuthal symmetry and most energetic wave
  number are $m_1=1$ and $m_{\text{max}}=10$, respectively.}
\label{fig:cp_eig1}   
\end{figure}
\begin{figure*}[t!]
  \begin{center}
    \includegraphics[scale=0.85]{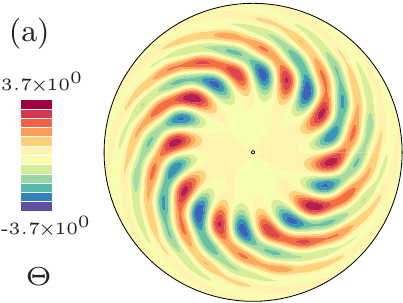}
    \includegraphics[scale=0.85]{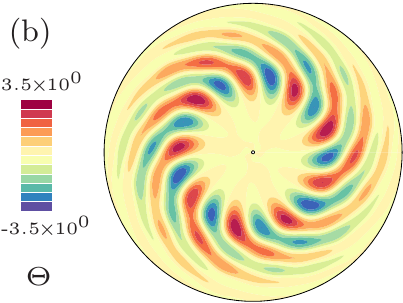}
    \includegraphics[scale=0.85]{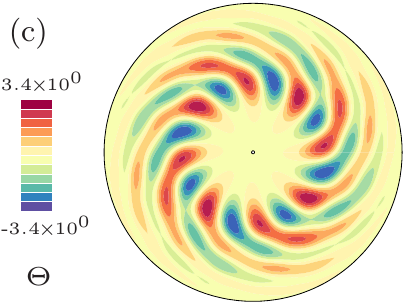}
    \includegraphics[scale=0.85]{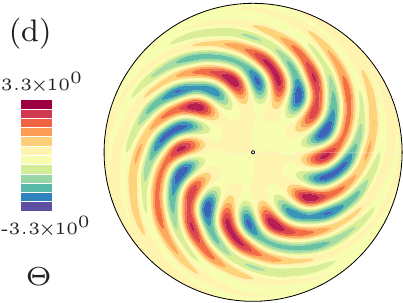}
    \includegraphics[scale=0.85]{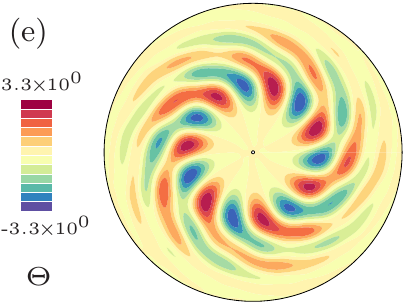}\\
    \includegraphics[scale=0.85]{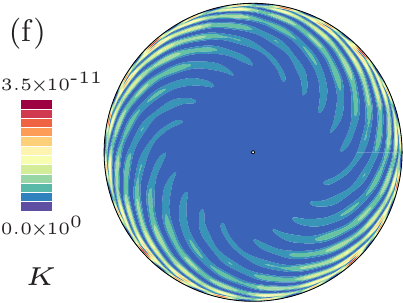}
    \includegraphics[scale=0.85]{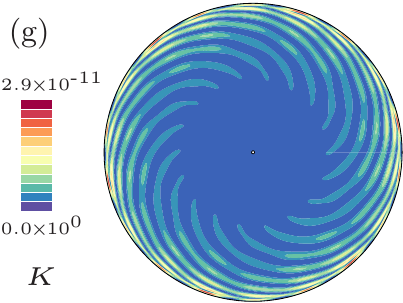}
    \includegraphics[scale=0.85]{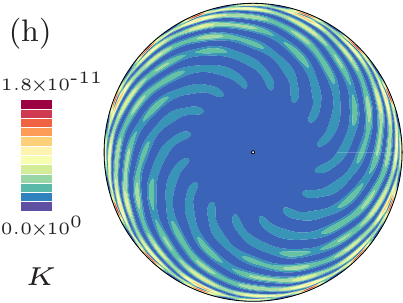}
    \includegraphics[scale=0.85]{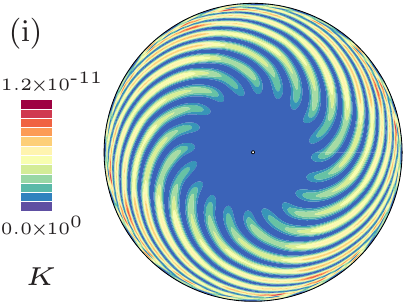}
    \includegraphics[scale=0.85]{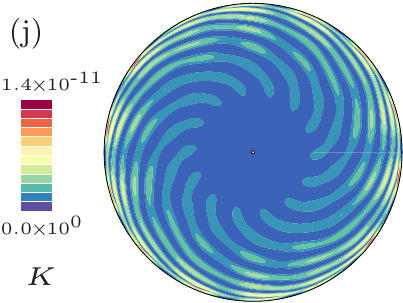}            
  \end{center}
\caption{Leading eigenfunctions of rotating waves with $m_0=12$, in
  the case of the set $P_3$ ($\Ek=3\times
  10^{-7}$,$\Pra=0.003$). Contour plots for the temperature
  perturbation $\Theta$ (top row) and for the kinetic energy density
  $K$ (bottom row) on an equatorial section.  The Rayleigh numbers are
  $\Ray=1.3030\times 10^8$ (a,f), $\Ray=1.3228\times 10^8$ (b,g),
  $\Ray=1.3919\times 10^8$ (c,h), and $\Ray=1.4511\times 10^8$
  (d,e,i,j). For the latter $\Ray$, (d,i) and (e,j) correspond to the
  1st and 3rd leading eigenfunctions, respectively. Their respective
  azimuthal symmetry and most energetic wave number are $m_1=2$ and
  $m_{\text{max}}=10$, $m_1=3$ and $m_{\text{max}}=9$, $m_1=4$ and
  $m_{\text{max}}=8$, $m_1=2$ and $m_{\text{max}}=10$, and $m_1=1$ and
  $m_{\text{max}}=7$.}
\label{fig:cp_eig2}   
\end{figure*}

Specifically, the Hopf bifurcations occur at $\Ray=2.3903\times 10^7$
($\Rayt=2.2\times 10^{-2}$) for the set $P_1=(\Pr,\Ek)=(0.03,3\times
10^{-6})$, at $\Ray=5.5823\times 10^7$ ($\Rayt=1.6\times 10^{-2}$) for
the set $P_2=(\Pr,\Ek)=(0.01,10^{-6})$. For the set
$P_3=(\Pr,\Ek)=(0.003,3\times 10^{-7})$, RWs with $m_0=12$ become
unstable at $\Ray=1.2731\times 10^8$ ($\Rayt=1.6\times 10^{-2}$) and
RWs with $m_0=11$ become unstable at $\Ray=1.2952\times 10^8$
($\Rayt=2.9\times 10^{-2}$). The values of $\Ray$ marking the
bifurcation point have been obtained by linear interpolation between
the last stable and the first unstable available RWs, which have the
pairs $(\Ray_1,|\lambda_1|)$ and $(\Ray_2,|\lambda_2|)$, with
$|\lambda_1|<1$ and $|\lambda_2|>1$, where $\lambda_i$ is the dominant
Floquet multiplier. The values $(\Ray_i,|\lambda_i|)$, with
$|\lambda_i|$ closest to unity, the volume-averaged kinetic energy
$K$, and the rotation frequency $\omega$ of the RWs, are listed in
Table \ref{tab:val_rw} for the three sets of parameters and different
resolutions to look for spatial discretization errors. We have found
that the radial resolution is critical to correctly assess the
stability of the waves. In the case of the set $P_3$ all the RWs have
been found unstable if $N_r=80$ is employed.

Figure \ref{fig:arp}(a) displays the six leading Floquet multipliers
for two unstable RWs corresponding to the sets $P_1$ (squares) and
$P_3$ (circles) at $\Ray=2.43043\times 10^7$ ($\Rayt=4\times 10^{-2}$)
and at $\Ray=1.27467\times 10^8$ ($\Rayt=1.7\times 10^{-2}$),
respectively. For both cases any leading Floquet multiplier has its
corresponding complex conjugate and for the set $P_3$ the leading
Floquet multipliers are arranged near the unit circle. This is
specially true for Rayleigh numbers close to the critical Rayleigh
number determining the onset of unstable RWs, either for the branch of
RWs with azimuthal symmetry $m_0=12$ or for the branch of RWs with
azimuthal symmetry $m_0=11$, see Fig \ref{fig:arp}(b). In this figure,
the solution at $\Ray_3=1.3030\times 10^8$ ($\Rayt_3=3.9\times
10^{-2}$), corresponding to the branch with azimuthal symmetry
$m_0=12$ and the set $P_3$, has the Floquet multipliers more clustered
near the unit circle than the solution at a similar $\Rayt=4\times
10^{-2}$ ($\Ray=2.43043\times 10^7$) for the set $P_1$ (shown in
Fig. \ref{fig:arp}(a)). This means that
$|\lambda|\equiv|\lambda(\Rayt)|$ is steeper in the case of the set
$P_1$ and quite flat for the set $P_3$, at least near the onset of
convection ($\Rayt<0.1$).  For this reason, the stability analysis for
the RWs shown in Fig \ref{fig:arp}(b) is computationally challenging
because of the convergence of the eigenvalue solver
(\cite{Saa92}). Before starting the Arnoldi iteration procedure
(\cite{LSY98}), more than 400 power method iterations have been
performed to the initial guess to filter out the components associated
to non-leading Floquet multipliers.

\begin{figure*}[t!]
\begin{center}
  \includegraphics[scale=1.5]{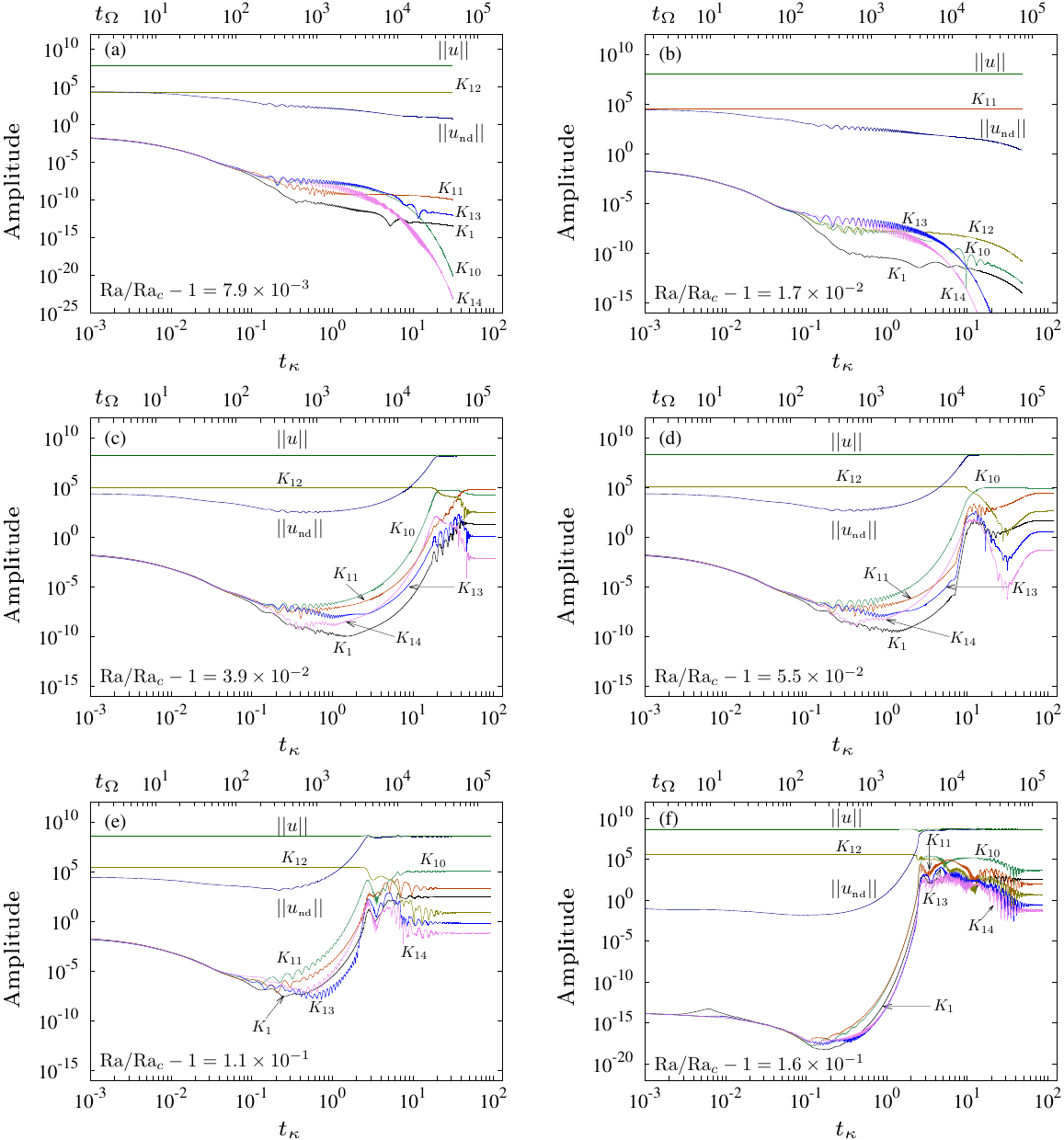}
\end{center}  
\caption{Time integration with initial conditions obtained by adding a
  random perturbation to the RWs for the set $P_3$ ($\Ek=3\times
  10^{-7}$,$\Pra=0.003$). Norm of the amplitudes of the potential
  scalars and the temperature perturbation $||u||$, and the norm
  $||u_{\text{\tiny{nd}}}||$ for only $m\ne 12k,~k\in\mathbb{Z}$ (in
  (a,c,d,e,f)) and $m\ne 11k,~k\in\mathbb{Z}$ (in (b)), versus
  diffusion time (also rotation time on top horizontal axis). The
  volume averaged kinetic energies for each wave number
  $m=1,10,11,12,13,14$ are displayed as well. The Rayleigh numbers are
  (a) $\Ray=1.2634\times 10^8$, (b) $\Ray=1.2747\times 10^8$, (c)
  $\Ray=1.3030\times 10^8$, (d) $\Ray=1.3228\times 10^8$, (e)
  $\Ray=1.3919\times 10^8$, and (f) $\Ray=1.4511\times 10^8$.}
\label{fig:K_tk}   
\end{figure*}

Figure \ref{fig:arp2} displays the leading Floquet multipliers for
several Rayleigh numbers up to $\Rayt=1.6\times 10^{-1}$ which are
still located near the unit circle. This means that any perturbation
applied to the RWs grows very slowly and gives rise to very long
transients if DNS are employed. This will be illustrated later in
Sec.~\ref{sec:t_ev}. The study of the symmetry of the unstable
eigenfunctions, when coupled with the symmetry of the RWs, allows to
infer the spatial structure of MRWs which bifurcate from the branch of
RWs (e.g. \cite{SGN13,GNS16}). This is because close to the
bifurcation point a MRW denoted by $u_2$ can be approximated by
$u_2\approx u_0+\epsilon u_1$, where $u_0$ is the parent RW, $u_1$ is
the leading Floquet mode, and $\epsilon$ is a small value. As the
azimuthal symmetry of the RWs is $m_0=12$ only Floquet eigenfunctions
with azimuthal symmetry $m_1\in\{1,2,3,4,6,12\}$ are possible ($m_1$
should be a factor of $m_0$) since the RWs and their eigenfunctions
are coupled in the variational equations (e.g. \cite{GDSN14}). In
addition, at the bifurcation point, the azimuthal symmetry, $m_2$, of
the MRW should be equal to $m_1$. The azimuthal symmetry $m_1$ and
most energetic wave number $m_{\text{max}}$ of the corresponding
eigenfunctions are labeled on each multiplier shown in
Fig. \ref{fig:arp2}. The figure shows that all the eigenvalues have
azimuthal symmetry which is not $m_1=12$, meaning that the
bifurcations broke the azimuthal symmetry giving rise to the
excitation of low azimuthal wave numbers $m_1\in\{1,2,3,4\}$. Figure
\ref{fig:arp2} also helps to visualize the increase of the real and
imaginary parts of a given Floquet multiplier (described by the
azimuthal symmetry and $m_{\text{max}}$) with the Rayleigh number.

The patterns of the temperature perturbation, axial vorticity,
azimuthal velocity, and kinetic energy for the leading eigenfunction
of a RW with azimuthal symmetry $m_0=11$ corresponding to the case $P_3$
at $\Ray=1.3011\times 10^8$ ($\Rayt=3.8\times 10^{-2}$) are displayed
in Fig. \ref{fig:cp_eig1}. The corresponding Floquet multiplier is
shown in Fig. \ref{fig:arp}(b) (diamond at the bottom) and it is
located just outside the unit circle, i.e. a Hopf bifurcation has
occurred. The azimuthal symmetry of the eigenfunction is $m_1=1$ and
the most energetic wave number is $m_{\text{max}}=10$ so the $m_0=11$
azimuthal symmetry of the parent RW is broken and MRWs with azimuthal
symmetry $m_1=1$ develop. These MRWs are studied later on
Sec.~\ref{sec:t_ev}. As described for the RWs, the eigenfunction's
velocity field is aligned in the axial direction and attached to the
outer sphere. There are 10 hot (cold) cells with larger magnitude for
the temperature perturbation since $m_{\text{max}}=10$ but they have
slightly different shapes due to the $m_1=1$ azimuthal symmetry. The
latter symmetry is best displayed in the equatorial section of the
kinetic energy contour plots where the spiraling arms form an oval
structure in the interior of the sphere. The interior structures of
this eigenfunction will be further studied in Sec. \ref{sec:tri} and
compared with the topology of nonlinear flows (MRWs) near the
bifurcation point.

Figure \ref{fig:cp_eig2} displays the topology of the eigenfunctions
corresponding to some Floquet multipliers (shown in
Fig. \ref{fig:arp2}) with different azimuthal symmetries at different
Rayleigh numbers. The temperature and velocity patterns are
multicellular and look very similar to those analyzed in
Fig. \ref{fig:cp_eig1}. By increasing $\Ray$ the main difference is
that the temperature cells as well as kinetic energy vortices tend to
move to the interior of the fluid as was the case for RWs (see
Fig. \ref{fig:rw2_cp_E3.d-7}). Similarly to the case of the leading
eigenfunction at $\Ray=1.3011\times 10^8$ the spiralling arms for the
leading eigenfunction at $\Ray=1.3030\times 10^8$ (leftmost plot of
the kinetic energy density, Figure \ref{fig:cp_eig2}) form a regular
pattern, a square in this case, in the interior of the sphere.

\section{Time evolutions for $\Ek=3\times 10^{-7}$ and $\Pra=0.003$}
\label{sec:t_ev}

The aim of this section is to investigate oscillatory flows for the
set $P_3$ obtained for $\Ray$ larger than that required for the
stability of RWs. The analysis is conducted by performing DNS with
selected initial conditions, at different $\Ray$, along the branch of
RWs already studied in Sec. \ref{sec:rw}. For each initial condition a
random perturbation (of order $10^{-6}$) to all spherical harmonic
amplitudes is added and the system is integrated around $100$
diffusion time units, which is more than one order of magnitude larger
than the typical final times of the DNS presented
in~\cite{KSVC17}. This is particularly challenging since the dimension
of the system is of order $10^7$ ($N_r=100$ and $L_{\text{max}}=192$
are used) and a time step of $1.7\times 10^{-5}$ diffusion time units
is employed.

\begin{figure}[h!]
\begin{center}
  \includegraphics[scale=1.8]{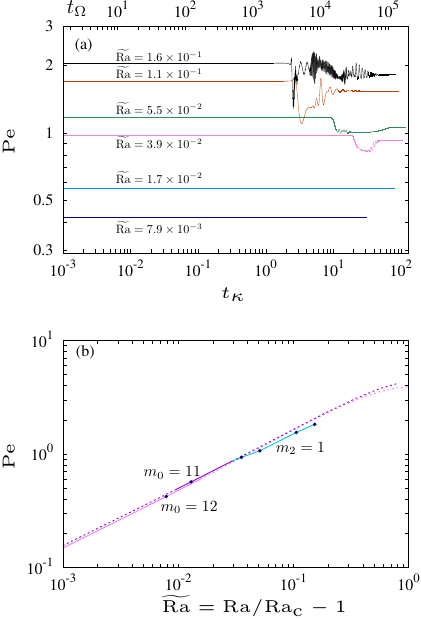}
\end{center}  
\caption{(a) Peclet number versus diffusion time (also rotation time
  on top horizontal axis), in the case of the set $P_3$ ($\Ek=3\times
  10^{-7}$,$\Pra=0.003$) for the same solutions as shown in
  Fig.~\ref{fig:K_tk}. The Rayleigh numbers, increasing from bottom to
  top, are $\Ray=1.2634\times 10^8$, $\Ray=1.2747\times 10^8$,
  $\Ray=1.3030\times 10^8$, $\Ray=1.3228\times 10^8$,
  $\Ray=1.3919\times 10^8$, and $\Ray=1.4511\times 10^8$. (b)
  Bifurcation diagrams of the time averaged Peclet number
  corresponding to the branches of RWs with azimuthal symmetry
  $m_0=12$ and $m_0=11$ and of MRWs with azimuthal symmetry
  $m_2=1$. The points correspond to the curves shown in (a).}
\label{fig:ev_Pe_tk}   
\end{figure}

\begin{figure}[h!]
\begin{center}
  \includegraphics[scale=1.8]{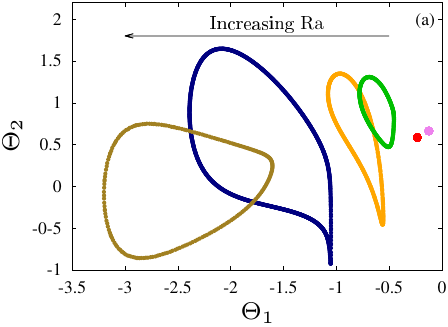}
\end{center}  
\caption{(a) Poincar\'e section defined by
  $0=\Theta(r,\varphi,\theta)$ with
  $(r,\varphi,\theta)=(0.51,0,5\pi/8)$. The temperatures
  $\Theta_1=\Theta(0.16,0,5\pi/8)$ and
  $\Theta_2=\Theta(0.86,0,5\pi/8)$ are displayed on the horizontal and
  vertical axis, respectively. We recall that $\eta=0.01$ implies
  $r_i=0.0101$ and $r_o=1.0101$ and that solutions belong to the set
  $P_3$ ($\Ek=3\times 10^{-7}$,$\Pra=0.003$). The Rayleigh numbers of
  each section increase from right to left in the figures (see the
  arrow). They are $\Ray=1.2634\times 10^8$, $\Ray=1.2747\times 10^8$,
  $\Ray=1.3030\times 10^8$, $\Ray=1.3228\times 10^8$,
  $\Ray=1.3919\times 10^8$, and $\Ray=1.4511\times 10^8$,
  corresponding to panels (a),(b),(c),(d),(e) and (f), respectively,
  of Fig.~\ref{fig:K_tk}. }
\label{fig:poinc}   
\end{figure}

\begin{figure*}[t!]
\begin{center}
  \includegraphics[scale=1.8]{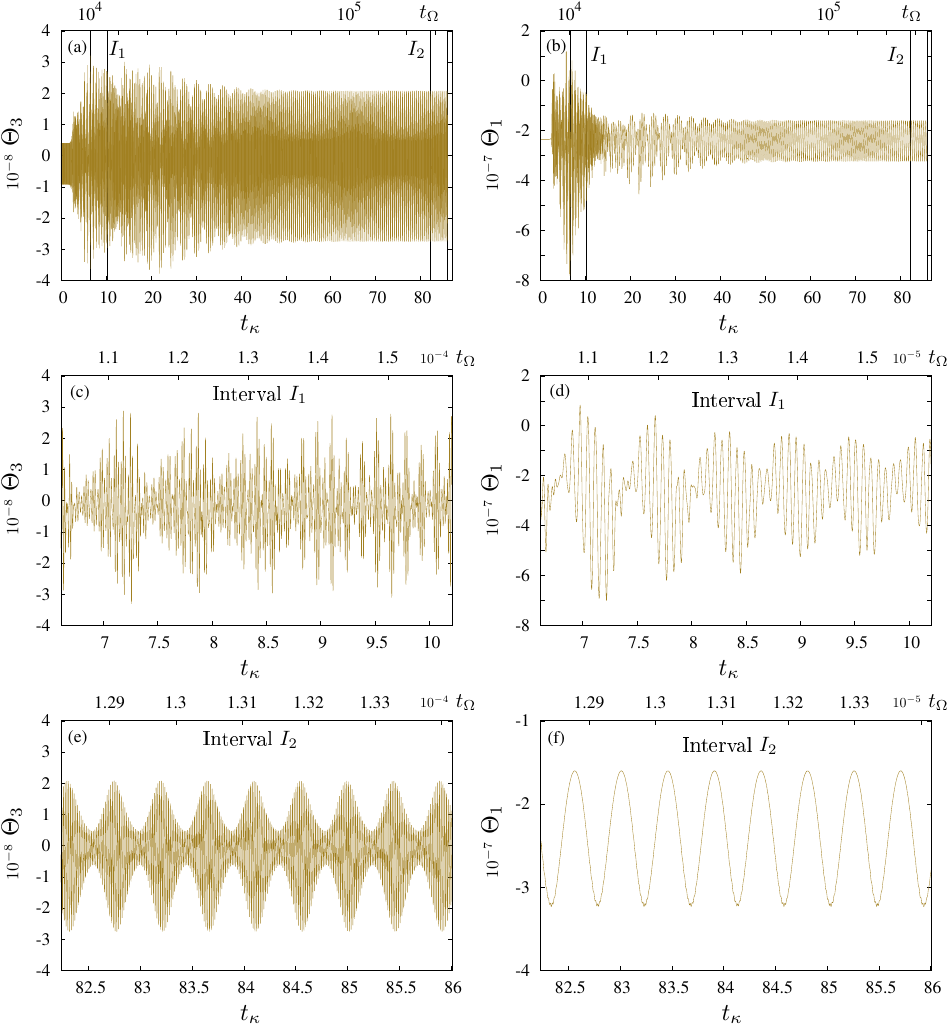}
\end{center}  
\caption{(a) Temperature $\Theta_3=\Theta(r,\varphi,\theta)$, with
  $(r,\varphi,\theta)=(0.51,0,5\pi/8)$, versus diffusion time. (b)
  Same as (a) but for the temperature
  $\Theta_1=\Theta(0.16,0,5\pi/8)$. The Rayleigh number is
  $\Ray=1.4511\times 10^8$ corresponding to panel (f) of
  Fig.~\ref{fig:K_tk}. The solution belongs to the set $P_3$
  ($\Ek=3\times 10^{-7}$,$\Pra=0.003$). The time averages of the
  kinetic energy spectra of Fig. \ref{fig:ener_spec}(a) and
  Fig. \ref{fig:ener_spec}(b) are taken over the time intervals $I_1$
  and $I_2$, respectively. (c) and (e) correspond to details of (a) in
  the intervals $I_1$ and $I_2$, respectively. (d) and (f) correspond
  to details of (b) in the intervals $I_1$ and $I_2$, respectively.}
\label{fig:ts_temp}   
\end{figure*}

\begin{figure}[b!]
\begin{center}
  \includegraphics[scale=1.5]{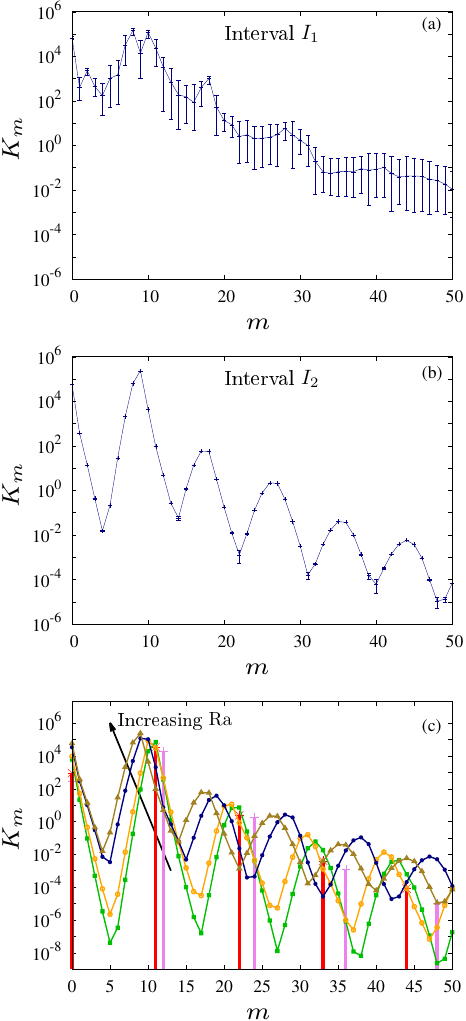}
\end{center}  
\caption{Time and volume averaged kinetic energy spectra $K_m$ versus
  the azimuthal wave number $m$.  In (a) and (b) the Rayleigh number
  is $\Ray=1.4511\times 10^8$ and the time average is taken over the
  interval $I_1$ and $I_2$, respectively, which are shown in
  Fig. \ref{fig:ts_temp}. In (c) several Rayleigh numbers are shown
  which increase from right to left in the figure (see the
  arrow). They are $\Ray=1.2634\times 10^8$, $\Ray=1.2747\times 10^8$,
  $\Ray=1.3030\times 10^8$, $\Ray=1.3228\times 10^8$,
  $\Ray=1.3919\times 10^8$, and $\Ray=1.4511\times 10^8$,
  corresponding to panels (a),(b),(c),(d),(e) and (f), respectively,
  of Fig.~\ref{fig:K_tk}. All these solutions belong to the set $P_3$
  ($\Ek=3\times 10^{-7}$,$\Pra=0.003$).  For each $m$, the maximum and
  minimum values of $K_{m}$ over the time interval are shown with
  error bars. The ime average, maximum and minimum values are taken
  over the last 5 diffusion time units of each time series.}
\label{fig:ener_spec}   
\end{figure}

Figure \ref{fig:K_tk} illustrates the procedure by displaying the
volume-averaged kinetic energy $K_m$ for each azimuthal wave number
$m\in\{1,10,11,12,13,14\}$ versus time in diffusion units (also in
rotation units) for the DNS corresponding to different $\Ray$. In each
panel the norm $||u||$, of the vector containing the amplitudes of the
scalar potentials and the temperature perturbation, and the norm
$||u||_{\text{nd}}$ -when only the azimuthal wave numbers which are
not multiples of $m_0=12$ (or $m_0=11$ for panel (b)) are considered-
are plotted as well. The initial condition corresponds to a stable RW
with azimuthal symmetry $m_0=12$ in Fig. \ref{fig:K_tk}(a), and to a
stable RW with azimuthal symmetry $m_0=11$ in Fig. \ref{fig:K_tk}(b),
and thus the added random perturbation (affecting all the spherical
harmonics of the RWs) is damped but on a very large time scale, see
the curve of $||u||_{\text{nd}}$ containing the norm of vector
containing the spherical harmonic amplitudes of the azimuthal wave
numbers which are not multiple of $m_0$. In agreement with the results
presented in Sec. \ref{sec:rw_stab} the azimuthal wave numbers for
which $K_m$ decreases more slowly correspond to the azimuthal symmetry
$m_1$ and the most energetic wave number $m_{\text{max}}$ of the
leading eigenfunction, because the associated eigenvalues are very
close to the unit circle (see Fig. \ref{fig:arp}(b)). The slowly
damped modes are $m=1$ and $m=11$ for the RW with azimuthal symmetry
$m_0=12$ (Fig. \ref{fig:K_tk}(a)) and $m=1$ and $m=10$
(Fig. \ref{fig:K_tk}(b)) for the RW with azimuthal symmetry
$m_0=11$. Notice that in Fig. \ref{fig:K_tk}(a) the mode $m=13$ is
slowly damped as well because of the coupling of the azimuthal
symmetry of the RW $m_0=12$ and the azimuthal symmetry of the
eigenfunction $m_1=1$. The same occurs in Fig. \ref{fig:K_tk}(b) for
the mode $m=12$.

For the same arguments as described above (i.e. the eigenvalues of the
eigenfunctions are clustered around the unit circle) the perturbations
added to unstable RWs grow very slowly and the stable attractor is
reached on a very large time scale. This is displayed in
Fig. \ref{fig:K_tk}(c,d,e,f) where at least 30 diffusion times (or
$5\times 10^4$ planetary rotations) are needed to saturate the
flow. In all the cases after a sharp increase of the unstable modes
(after around 2-10 diffusion times) a strongly oscillatory transient
lasts more than 30 diffusion times. The final attractor is a MRW,
i.e. a quasiperiodic flow with two incommensurable frequencies, which
has a certain spatio-temporal symmetry. Notice that for MRW the value
of $||u||_{\text{nd}}$ is almost equal to $||u||$ since the spherical
harmonics amplitudes corresponding to the azimuthal wave numbers
$m=12k,~k\in\mathbb{Z}$ are significantly smaller when compared with
other azimuthal wave numbers (for instance $m=10$).  The systematic
computation of MRW has been performed in \cite{GNS16} for the same
problem as described here but for spherical shells. These types of
oscillatory flows are still in the weak branch regime since their
Peclet numbers are of order one. This is illustrated in
Fig. \ref{fig:ev_Pe_tk}(a) where the time series of the Peclet number
$\Pec$ are displayed for the same solutions as analyzed in
Fig. \ref{fig:K_tk}. Figure \ref{fig:ev_Pe_tk}(b) corresponds to the
bifurcation diagrams of the time-averaged $\Pec$ for these MRWs
including also the branches of RWs already displayed in
Fig. \ref{fig:bif_diag}.

\begin{figure}[t!]
  \begin{center}
    \includegraphics[scale=1.2]{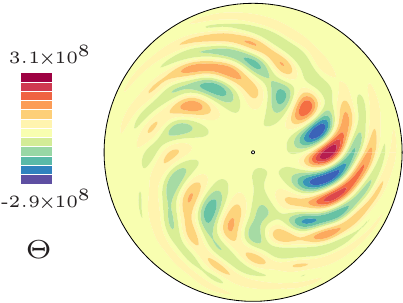}
    \includegraphics[scale=1.2]{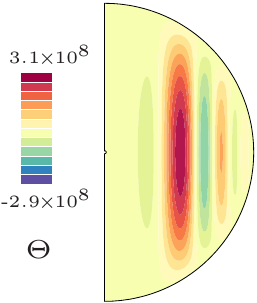}\\[2.mm]
    \includegraphics[scale=1.2]{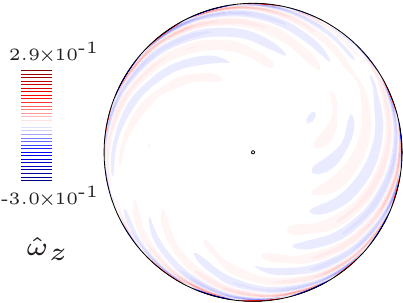}
    \includegraphics[scale=1.2]{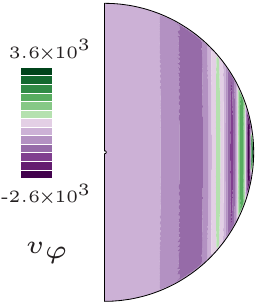}    
\end{center}  
  \caption{Solution bifurcating from rotating waves with $m_0=12$, in
    the case of the set $P_3$ ($\Ek=3\times 10^{-7}$,$\Pra=0.003$) at
    $\Ray=1.4511\times 10^8$. The snapshot is taken in the transient
    phase, at the end of the time interval $I_1$ shown in
    Fig. \ref{fig:ts_temp}(a). Top row: Contour plots for the
    temperature perturbation $\Theta$ on an equatorial and meridional
    section. Bottom row: Contour plots for the vertical vorticity
    $\hat{\omega}_z$ on an equatorial section and for the azimuthal
    velocity $v_{\varphi}$ on a meridional section.}
\label{fig:te1_cp_E3.d-7}   
\end{figure}

\begin{figure}[h!]
  \begin{center}
    \includegraphics[scale=1.2]{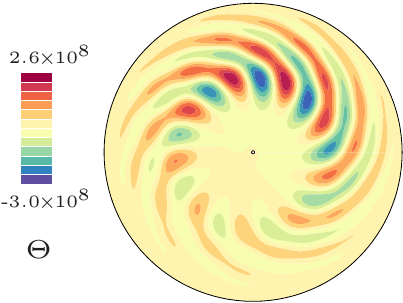}
    \includegraphics[scale=1.2]{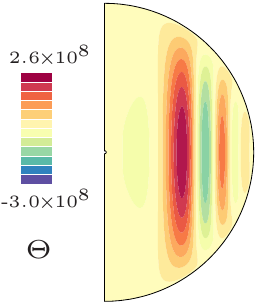}\\[2.mm]
    \includegraphics[scale=1.2]{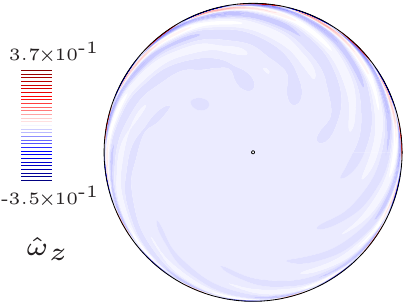}
    \includegraphics[scale=1.2]{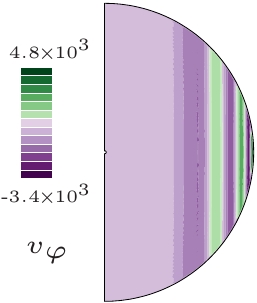}    
  \end{center}
  \caption{As Fig. \ref{fig:te1_cp_E3.d-7} but with the snapshot taken
    in the saturated phase, at the end of the time interval $I_2$
    shown in Fig. \ref{fig:ts_temp}(a).}
\label{fig:te2_cp_E3.d-7}   
\end{figure}

To demonstrate the quasiperiodic nature of MRWs, Poincar\'e sections,
extracted from the time series of temperature perturbation, are
displayed in Fig. \ref{fig:poinc} for the same solutions as analyzed
in Fig. \ref{fig:K_tk} (points in Fig. \ref{fig:ev_Pe_tk}(b)). The
Poincar\'e section of a RW (periodic flow) is a point, whereas it
corresponds to a closed curve in the case of MRWs (quasiperiodic
flow). Increasing the Rayleigh number up to $\Ray=1.3919\times 10^8$
results in larger oscillations of the temperature perturbation since
the curves enclose a larger area. Notice that for $\Ray=1.4511\times
10^8$ the curve spreads over a smaller interval in the vertical axis
than in the case of $\Ray=1.3919\times 10^8$ so the oscillations of
temperature close to the outer boundary become smaller.

To further investigate the nature of the temperature fluctuations, the
time series of $\Theta$ are displayed in Fig. \ref{fig:ts_temp}(a,b)
at two different points close to the equatorial plane (see figure
caption), one in the middle of the sphere (panel (a)) and the other
close to the inner boundary (panel (b)). The time series are for the
MRW at $\Ray=1.4511\times 10^8$ corresponding to panel (f) of
Fig.~\ref{fig:K_tk}. In Fig. \ref{fig:ts_temp}(a,b) the long initial
transients (around 50 diffusion times) required to saturate this
solution (see discussion of Fig. \ref{fig:K_tk}) are clearly
visible. Figure \ref{fig:ts_temp}(c,d) displays a detail of
Fig. \ref{fig:ts_temp}(a) (i.e. $\Theta$ in the middle of the sphere)
in two different time intervals, one during the transient phase
(interval $I_1$), and the other during the saturated phase (interval
$I_2$) of the solution. Figure \ref{fig:ts_temp}(e,f) is as
Fig. \ref{fig:ts_temp}(c,d) but displays the details of
Fig. \ref{fig:ts_temp}(b) (i.e. $\Theta$ close to the center of the
sphere). The comparison between the different panels summarizes
several facts. First, the oscillations have different main time scales
depending on whether $\Theta$ is measured in the middle of the shell
(small and large scales, clearly quasiperiodic) or close to the center
of the sphere (mainly large scales and periodic). Second,  the
long transients (interval $I_1$) exhibit intermittent-like
structures. Finally, for the long transients an intermediate time
scale is additionally present for $\Theta$ picked up close to the
center of the sphere.

The mode structure of the long initial transients and the saturated
MRW at $\Ray=1.4511\times 10^8$ is significantly different. This is
demonstrated in Fig. \ref{fig:ener_spec} displaying the time averaged
kinetic energy spectra versus the azimuthal wave number $m$ over the
interval $I_1$ (panel (a)) and over the time interval $I_2$ (panel
(b)). The figure also displays (with error bars) the amplitude of the
kinetic energy oscillations. The transients are characterized by
strong time oscillations of all the modes. In addition the flow is
bimodal, in the sense that the azimuthal wave numbers $m=8$ and $m=10$
have maximum energy. Also, low wave numbers $m<6$ have a similar and
noticeable (larger than $10^2$) magnitude. In contrast, the kinetic
energy spectra of the saturated MRW have a single maximum (at $m=9$)
and the time dependence of $K_m$ is only noticeable for the modes at
the relative minima of the spectrum. In addition, only the low wave
number $m=1$ has magnitude larger than $10^2$. The other MRWs analyzed
in the previous figures have similar kinetic energy spectra as shown
in Fig. \ref{fig:ener_spec}(c). In this figure RWs have nonzero
kinetic energy only in the wave numbers of the form
$km_0,~k\in\mathbb{Z}$ ($m_0$ is the azimuthal symmetry of the RW)
whereas MRWs, all of them with azimuthal symmetry $m_2=1$, have
nonzero kinetic energy in all the modes. As the Rayleigh number is
increased $m_{\text{max}}$ decreases (from $m=12$ at the smallest
$\Ray$ down to $m=9$ at the largest $\Ray$). Moreover, the relative
difference between dominant modes (relative maxima) and non-dominant
modes (relative minima) decreases. Notice that for the low wave
numbers (specially $m=5$) the kinetic energy $K_m$ sharply increases
with $\Ray$. Scalloped spectra like those of
Fig. \ref{fig:ener_spec}(b,c), already studied for the case of the
spherical Couette flow in \cite{MaTu87a}, are a consequence of a
periodic spatial structure modulated by an envelope as is shown in the
following paragraph.

The flow and temperature spatial structures during the transient as
well as the saturated phase for the DNS at $\Ray=1.4511\times 10^8$
can be visualized in Fig. \ref{fig:te1_cp_E3.d-7} and
Fig. \ref{fig:te2_cp_E3.d-7}, respectively. In both cases the flow is
strongly geostrophic; the temperature perturbation exhibit
multicellular patterns, and the maximum azimuthal velocity is located
close to the outer sphere as described for the RWs in
Sec. \ref{sec:rw}. In contrast to RWs, the shapshots presented in
Fig. \ref{fig:te1_cp_E3.d-7} and Fig. \ref{fig:te2_cp_E3.d-7} have a
clear asymmetry (a modulation by an envelope as described in
\cite{MaTu87a}) between temperature cells because of the excitation of
low wave numbers (see Fig. \ref{fig:ener_spec}) predicted by the
stability analysis conducted in Sec. \ref{sec:rw_stab}. The main
difference between the contour plots of the transient flow and the
saturated phase (Fig. \ref{fig:te1_cp_E3.d-7} and
Fig. \ref{fig:te2_cp_E3.d-7}, respectively) is that, for the former,
the azimuthal asymmetry of the $m=10$ structure is more irregular
(i.\,e. modulated by several low wave numbers) whereas for the latter
the azimuthal modulation is mainly due to $m=1$. A further description
of the flow and temperature patterns for a MRW, in terms of its
different mode components, will be provided in the next section,
Sec. \ref{sec:tri}.

\subsection{Triadic resonances}
\label{sec:tri}

In this section we add further evidence to the recent study of
\cite{GGS21} in which triadic resonances occurring in spherical
systems have been interpreted in terms of MRWs. Triadic resonances in
the spherical Couette problem have been comprehensively studied in
\cite{BTHW18} and are characterized by the existence of azimuthal wave
numbers $m_i$, $m_j$ and $m_k$ with main time dependencies provided by
the frequencies $\omega_i$, $\omega_j$, and $\omega_k$, respectively,
for which the relations $m_i=m_j\pm m_k$ and
$\omega_i=\omega_j\pm\omega_k$ hold. Triadic resonances as described
in \cite{BTHW18} have been also analyzed in \cite{Lin21} for the same
problem as studied here but for $\Ek\ge 10^{-6}$. As will be shown in
the following, multiple resonances between several modes can be
identified from the DNS of the MRWs previously studied. The resonant
modes are excited by the Hopf bifurcations giving rise to MRWs
(see \cite{GGS21}).

Following the same procedure as in \cite{GGS21} the time series of the
real part of the poloidal amplitudes of Eq. (\ref{eq:serie_phi}),
$\Re{\Phi^m_l(r_o/2)}$ for several $m\in \{1,...,20\}$ and $l\in
\{1,...,40\}$, are considered to investigate the time scales of the
flow and triadic resonances among the different modes $(m,l)$. We have
considered two different MRWs at $\Ray=1.3030\times 10^8$ and at
$\Ray=1.3919\times 10^8$. The 1st MRW is very close to the bifurcation
point from the branch of RWs with $m_0=11$ (see the left point on the
$m_2=1$ branch of Fig. \ref{fig:ev_Pe_tk}(b)) whereas the 2nd MRW is far
away (2nd rightmost point on the $m_2=1$ branch of
Fig. \ref{fig:ev_Pe_tk}(b))).

\begin{figure}[t!]
\begin{center}
  \includegraphics[scale=1.5]{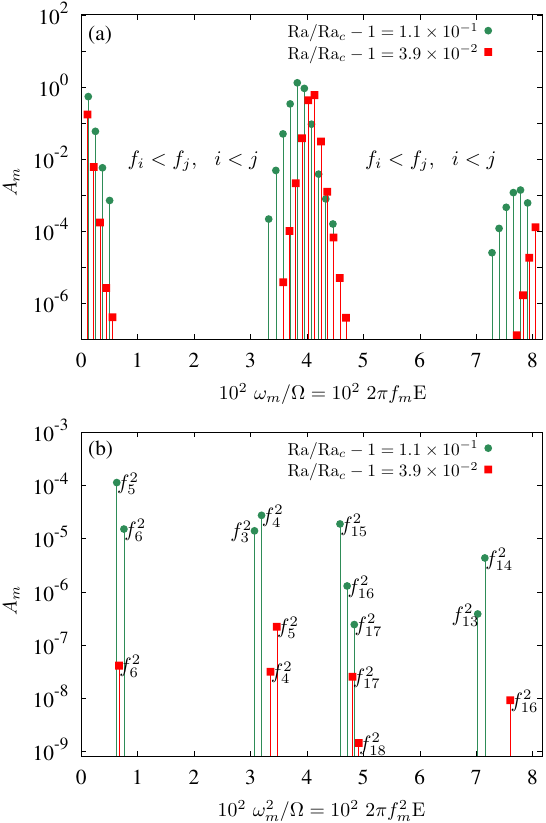}
\end{center}  
\caption{Frequency analysis for two solutions at $\Ray=1.3030\times
  10^8$ (squares) and at $\Ray=1.3919\times 10^8$ (circles), in the
  case of the set $P_3$ ($\Ek=3\times 10^{-7}$,$\Pra=0.003$). (a)
  Leading frequencies, $f_m$, and amplitudes, $A_m$, of the time
  series of poloidal component, $\Re{\Phi^m_m(r_d)}$, of the different
  modes $(m,m)$, $1\le m\le 20$. (b) as in (a) but for the 2nd leading
  frequencies $f^2_m$.}
\label{fig:sdds}   
\end{figure}

An accurate frequency analysis, based on Laskar's algorithm
\cite{Las93}, has been applied to each of the time series to determine
the fundamental frequencies. We note that for a time series of a large
scale magnetohydrodynamic periodic flow Laskar's algorithm detects the
main frequency up to a relative error of order $10^{-5}$ (see
discussion in Sec. 3.1 of \cite{GSGS21}). Because the flow is
equatorially symmetric (see meridional sections of
Fig. \ref{fig:te2_cp_E3.d-7}) we have considered the modes with
$(m,l)=(m,m)$ which are equatorially symmetric for the poloidal
potential. Other equatorially symmetric modes, such as $(m,m+2)$, are
not considered since their time dependence is analogous to that of the
mode $(m,m)$ (see \cite{GGS21}). The frequencies, normalized by the
global rotation of the sphere $\Omega$, are $\omega_m/\Omega=2\pi
f_m\Ek$, where $f_m$ is the main peak in the dimensionless frequency
spectrum. They are plotted in Fig. \ref{fig:sdds}(a) for each mode
$(m,l)=(m,m)$ and $1\le m\le 20$. Figure \ref{fig:sdds}(b) displays
the frequencies $\omega^2_m/\Omega$ obtained from the second largest
peak in the frequency spectrum (so $\omega^2_m$ is not the square of
$\omega_m$).

As in \cite{GGS21}, the leading frequencies of the modes $(i,i)$ and
$(j,j)$ verify $\omega_i<\omega_j$ if $i<j$, so the frequencies are
ordered following the azimuthal wave number ordering. A characteristic
feature seen in Fig. \ref{fig:sdds}(a) is that there exists a
particular distribution of the frequencies, in the sense that there
are separated blocks of clustered frequencies. The three blocks of
Fig. \ref{fig:sdds}(a) are $\omega_m/\Omega<0.006$,
$0.032<\omega_m/\Omega<0.048$, and $\omega_m/\Omega>0.072$, which are
associated to slow, moderate, and fast modes that correspond to small,
moderate and large azimuthal wave numbers. For the MRW at
$\Ray=1.3030\times 10^8$ these are $m\le 5$, $6\le m\le 16$, and $m\ge
17$, respectively. In contrast, the secondary frequencies shown in
Fig. \ref{fig:sdds}(b) are not ordered with respect to the wave number
but still retain the block structure. As it will be shown later in
this section these secondary frequencies provide additional resonances
among the modes. Notice that only the modes
$m\in\{3,4,5,6,13,14,15,16,17,18\}$ have a secondary peak in the
frequency spectrum and thus are quasiperiodic. These modes are those
located contiguously at the boundaries of the block regions, see for
instance the modes $m=4,5,6$ in the case of the MRW at
$\Ray=1.3030\times 10^8$ in Fig. \ref{fig:sdds}(a). The other modes
$m\in\{1,2,7,8,9,10,11,19,20\}$ are purely periodic and lie in the
interior of the block regions.

Figure \ref{fig:sdds} can be easily compared with Table I of
\cite{Lin21}. In that study, for $\Ek=10^{-6}$ and $\Pra=0.001$, the
resonance conditions involving the azimuthal wave numbers $m_i=4$,
$m_j=1$ and $m_k=3$, were found. The associated frequencies were
$\omega_i/\Omega=0.25$, $\omega_j/\Omega=0.16$, and
$\omega_k/\Omega=0.09$ which are roughly one order of magnitude larger
than those presented in Fig. \ref{fig:sdds} for the low azimuthal wave
numbers $m<5$. This is not surprising since our Ekman number
($\Ek=3\times 10^{-7}$) is smaller.

\begin{table*}[t!]
\caption{Relations between the main frequencies $f_m$ of the different
  modes $(m,l)=(m,m)$ for the 2T MRW with $m_2=1$ azimuthal symmetry at
  $\Ray=1.3030\times 10^8$. For the modes with $m\in\{4,5,6\}$ the
spectrum has two peaks $f_m$ and $f^2_m$. These relations are
satisfied up to $(f_{m_i}-f_{m_j}-f_{m_k})/f_{m_i}<\epsilon_f$ with
$\epsilon_f=10^{-4}$. }
\label{table:res1}
\renewcommand{\arraystretch}{1.5}
\begin{tabular}{|l|l|l|l|l|l|l|l|l|}
\hline
 $m=3$                     & $m=4$                       &  $m=5$                     & $m=6$                     & $m=7$                       & $m=8$                  \\
\hline
$f_3=f_1+f_2$              & $f_4=f_1+f_3$             & $f_5=f_1+f_4$               & $f_6=f_1+f^2_5$            &  $f_7=f_1+f_6$              & $f_8=f_1+f_7$              \\
                          & $\hspace{4.mm}=2f_2$      & $\hspace{4.mm}=f_2+f_3$     & $\hspace{4.mm}=f_2+f^2_4$  &  $\hspace{4.mm}=f_2+f^2_5$  & $\hspace{4.mm}=f_2+f_6$  \\
                          &                           & $f^2_5=f_1+f^2_4$           & $f^2_6=f_1+f_5$            &  $\hspace{4.mm}=f_3+f^2_4$   & $\hspace{4.mm}=f_3+f^2_5$  \\
                          &                           &                            & $\hspace{4.mm}=f_2+f_4$    &                             &                             \\
                          &                           &                            & $\hspace{4.mm}=2f_3$       &                             &                             \\
\hline
\hline
$m=9$                       & $m=10$                        & $m=11$                         & $m=12$                     & $m=13$            & $m=14$                       \\
\hline
$f_9=f_1+f_8$               & $f_{10}=f_1+f_9$                & $f_{11}=f_1+f_{10}$             & $f_{12}=f_1+f_{11}$          & $f_{13}=f_1+f_{12}$          & $f_{14}=f_1+f_{13}$  \\
$\hspace{4.mm}=f_2+f_7$     & $\hspace{5.mm}=f_2+f_8$        & $\hspace{5.mm}=f_2+f_{9}$      & $\hspace{5.mm}=f_2+f_{10}$  & $\hspace{5.mm}=f_2+f_{11}$   & $\hspace{5.mm}=f_2+f_{12}$ \\
$\hspace{4.mm}=f_3+f_6$     & $\hspace{5.mm}=f_3+f_7$        & $\hspace{5.mm}=f_3+f_{8}$      & $\hspace{5.mm}=f_3+f_{9}$   & $\hspace{5.mm}=f_3+f_{10}$   & $\hspace{5.mm}=f_3+f_{11}$ \\
$\hspace{4.mm}=f_4+f^2_5$   & $\hspace{5.mm}=f_4+f_{6}$       & $\hspace{5.mm}=f_4+f_{7}$      & $\hspace{5.mm}=f_4+f_{8}$  & $\hspace{5.mm}=f_4+f_{9}$    & $\hspace{5.mm}=f_4+f_{10}$\\
$\hspace{4.mm}=f^2_4+f_5$   & $\hspace{5.mm}=f^2_4+f^{2}_{6}$ & $\hspace{5.mm}=f_5+f_{6}$       & $\hspace{5.mm}=f_5+f_{7}$  & $\hspace{5.mm}=f_5+f_{8}$   & $\hspace{5.mm}=f_5+f_{9}$ \\
                           &                                &  $\hspace{5.mm}=f^2_5+f^{2}_{6}$ &                           & $\hspace{5.mm}=f^2_6+f_{7}$ & $\hspace{5.mm}=f^2_6+f_{8}$ \\
\hline
\end{tabular}
\end{table*}

\begin{table*}[t!]
\caption{Relations between the main frequencies $f_m$ of the different
  modes $(m,l)=(m,m)$ for the 2T MRW with $m_2=1$ azimuthal symmetry at
  $\Ray=1.3919\times 10^8$. For the modes with $m\in\{3,4,5,6,13,14\}$ the
spectrum has two peaks $f_m$ and $f^2_m$. These relations are
satisfied up to
$(f_{m_i}-f_{m_j}-f_{m_k})/f_{m_i}<\epsilon_f$
with $\epsilon_f=10^{-4}$. }
\label{table:res2}
\renewcommand{\arraystretch}{1.5}
\begin{tabular}{|l|l|l|l|l|l|l|l|l|}
\hline
 $m=3$                    & $m=4$                         &  $m=5$                      & $m=6$                     & $m=7$                        & $m=8$                      \\
\hline
$f_3=f_1+f_2$              & $f_4=f_1+f_3$                 & $f_5=f_1+f^2_4$              & $f_6=f_1+f_5$             &  $f_7=f_1+f_6$             & $f_8=f_1+f_7$               \\
                          & $\hspace{4.mm}=2f_2$          & $\hspace{4.mm}=f_2+f^2_3$    & $\hspace{4.mm}=f_2+f^2_4$ &  $\hspace{4.mm}=f_2+f_5$   & $\hspace{4.mm}=f_2+f_6$   \\
                          &  $f^2_4=f_1+f^2_3$             & $f^2_5=f_1+f_4$              & $f^2_6=f_1+f^2_5$         &  $\hspace{4.mm}=f_3+f^2_4$ & $\hspace{4.mm}=f_3+f_5$ \\
                          &                               & $\hspace{4.mm}=f_2+f_3$      & $\hspace{4.mm}=f_2+f_4$  & $\hspace{4.mm}=f^2_3+f_4$  & $\hspace{4.mm}=f^2_3+f^2_5$ \\
                          &                               &                              &  $\hspace{4.mm}=2f_3$    &                           &                            \\
\hline
\hline
$m=9$                      & $m=10$                         & $m=11$                      & $m=12$                     & $m=13$                      & $m=14$                       \\
\hline
$f_9=f_1+f_8$               & $f_{10}=f_1+f_9$                & $f_{11}=f_1+f_{10}$           & $f_{12}=f_1+f_{11}$         & $f_{13}=f_1+f_{12}$           & $f_{14}=f_1+f_{13}$           \\
$\hspace{4.mm}=f_2+f_7$     & $\hspace{5.mm}=f_2+f_8$        & $\hspace{5.mm}=f_2+f_{9}$    & $\hspace{5.mm}=f_2+f_{10}$  & $\hspace{5.mm}=f_2+f_{11}$   & $\hspace{5.mm}=f_2+f_{12}$   \\
$\hspace{4.mm}=f_3+f_6$     & $\hspace{5.mm}=f_3+f_7$        & $\hspace{5.mm}=f_3+f_{8}$    & $\hspace{5.mm}=f_3+f_{9}$   & $\hspace{5.mm}=f_3+f_{10}$   & $\hspace{5.mm}=f_3+f_{11}$ \\
$\hspace{4.mm}=f^2_3+f^2_6$ & $\hspace{5.mm}=f_4+f_{6}$       & $\hspace{5.mm}=f_4+f_{7}$    & $\hspace{5.mm}=f_4+f_{8}$   & $\hspace{5.mm}=f_4+f_{9}$    & $\hspace{5.mm}=f_4+f_{10}$ \\
$\hspace{4.mm}=f_4+f_5$     & $\hspace{5.mm}=f^2_4+f^{2}_{6}$ & $\hspace{5.mm}=f^2_5+f_{6}$   & $\hspace{5.mm}=f^2_5+f_{7}$ & $\hspace{5.mm}=f^2_5+f_{8}$  & $\hspace{5.mm}=f^2_5+f_{9}$ \\
$\hspace{4.mm}=f^2_4+f^2_5$ &                                & $\hspace{5.mm}=f_5+f^{2}_{6}$ &                             & $\hspace{5.mm}=f^2_6+f_{7}$  & $\hspace{5.mm}=f^2_6+f_{8}$ \\
                           &                                &                              &                             & $f^2_{13}=f^2_3+f_{10}$       & $f^2_{14}=f^2_3+f_{11}$ \\
                           &                                &                              &                             & $\hspace{5.mm}=f^2_4+f_{9}$  & $\hspace{5.mm}=f^2_4+f_{10}$ \\
                           &                                &                              &                             & $\hspace{5.mm}=f_5+f_{8}$    & $\hspace{5.mm}=f_5+f_{9}$ \\
                           &                                &                              &                             & $\hspace{5.mm}=f_6+f_{7}$    & $\hspace{5.mm}=f_6+f_{8}$ \\
                           &                                &                              &                             &                             & $\hspace{5.mm}=2f_{7}$ \\
\hline
\end{tabular}
\end{table*}

\begin{figure*}[t!]
\begin{center}
  \includegraphics[scale=1.1]{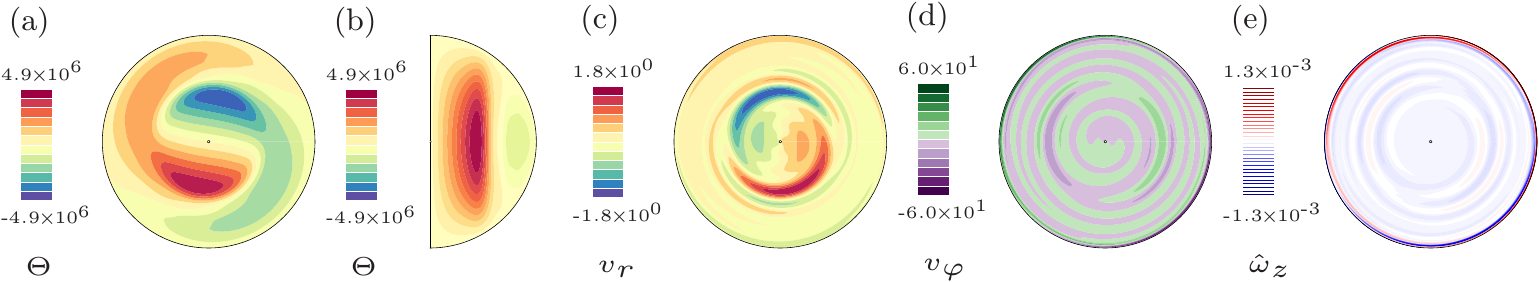}\\[3.mm]
  \includegraphics[scale=1.1]{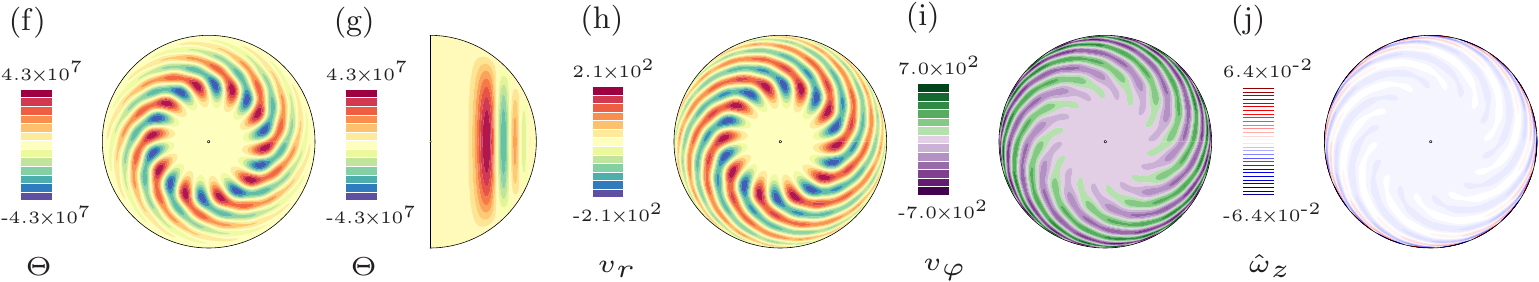}\\[3.mm]
  \includegraphics[scale=1.1]{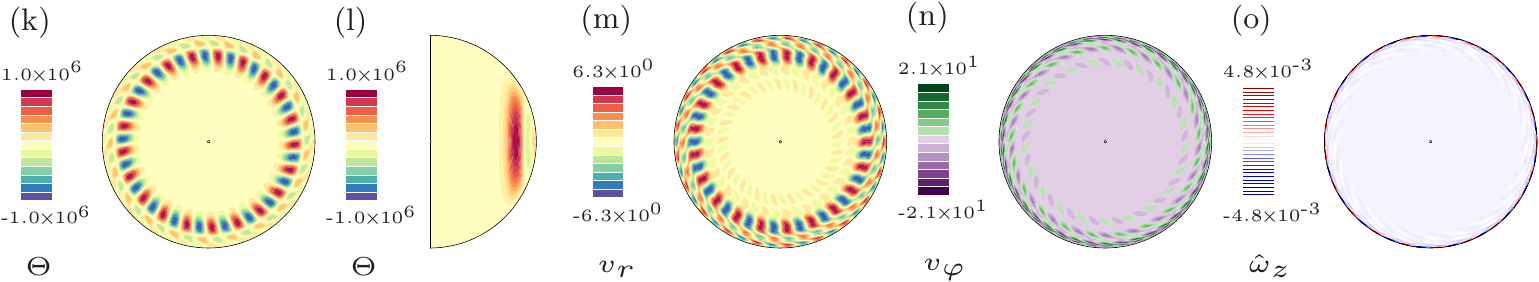}\\
\end{center}
\caption{Modulated rotating wave with azimuthal symmetry $m_2=1$ in
  the case of the set $P_3$ ($\Ek=3\times 10^{-7}$,$\Pra=0.003$) at
  $\Ray=1.3030\times 10^8$. The contour plots of the temperature
  perturbation $\Theta$ on an equatorial (a,f,k) and a meridional
  (b,g,l) section, and of the radial velocity $v_r$ (c,h,m), azimuthal
  velocity $v_{\varphi}$ (d,i,n), and vertical vorticity $\hat{w}_z$
  (e,j,o), normalized by the planetary vorticity $\hat{w}_z=w_z\Ek/2$,
  on equatorial sections, are displayed from left to right in each
  row.  From top to bottom only the $m=1$ (a-e), $m=10$ (f-j) and
  $m=21$ (k-o), respectively, azimuthal wave numbers, rather than all
  $m$'s, are considered for the contour plots.}
\label{fig:nl_cp_E3.d-7_m}   
\end{figure*}
\begin{figure*}[t!]
\begin{center}
  \includegraphics[scale=1.1]{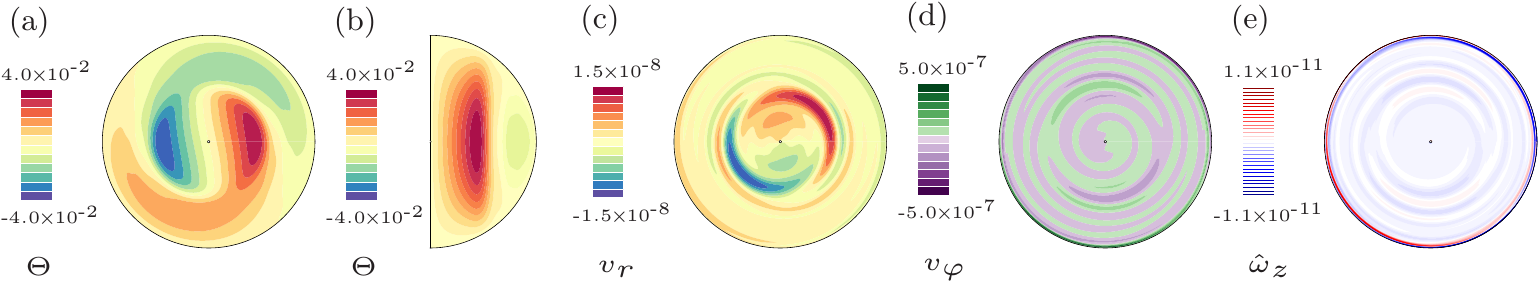}\\[2.mm]
  \includegraphics[scale=1.1]{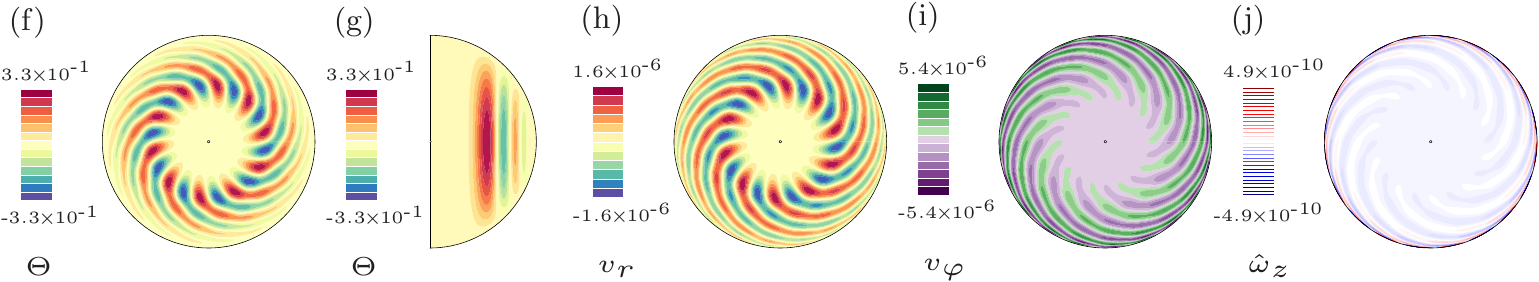}\\[2.mm]
  \includegraphics[scale=1.1]{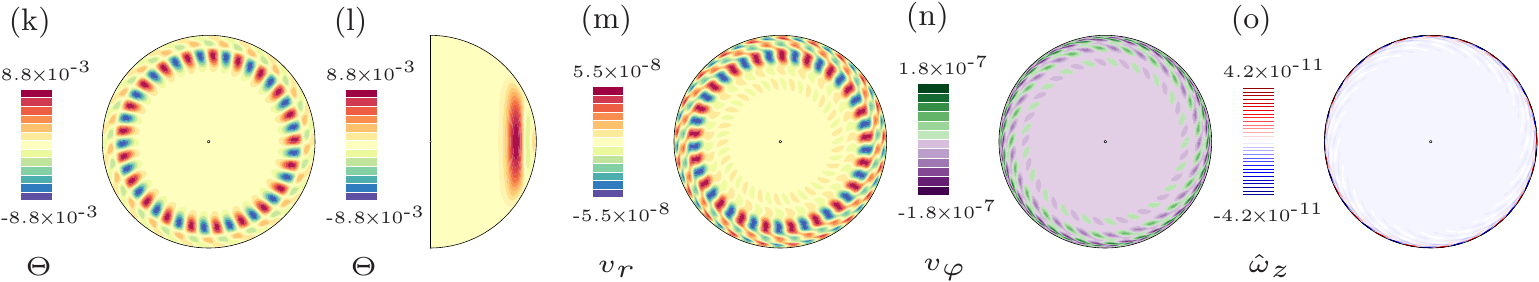}\\[2.mm]
\end{center}  
\caption{Leading eigenfunction, with azimuthal symmetry $m_1=1$, of a
  rotating wave with $m_0=11$, in the case of the set $P_3$
  ($\Ek=3\times 10^{-7}$,$\Pra=0.003$) at $\Ray=1.3011\times
  10^8$. The contour plots of the temperature perturbation $\Theta$ on
  an equatorial (a,f,k) and a meridional (b,g,l) section, and of the
  radial velocity $v_r$ (c,h,m), azimuthal velocity $v_{\varphi}$
  (d,i,n), and vertical vorticity $\hat{w}_z$ (e,j,o), normalized by
  the planetary vorticity $\hat{w}_z=w_z\Ek/2$, on equatorial
  sections, are displayed from left to right in each row. From top to
  bottom only the $m=1$ (a-e), $m=10$ (f-j) and $m=21$ (k-o),
  respectively, azimuthal wave numbers, rather than all $m$'s, are
  considered for the contour plots.}
\label{fig:eig_cp_E3.d-7_brm11_Ra2.45_m}   
\end{figure*}

\begin{figure*}[t!]
\begin{center}
  \includegraphics[scale=0.85]{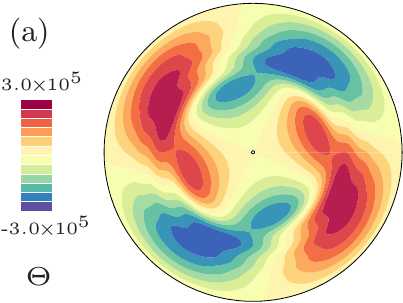}
  \includegraphics[scale=0.85]{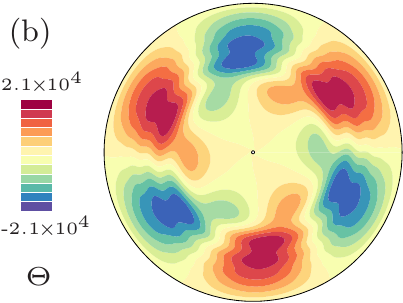}
  \includegraphics[scale=0.85]{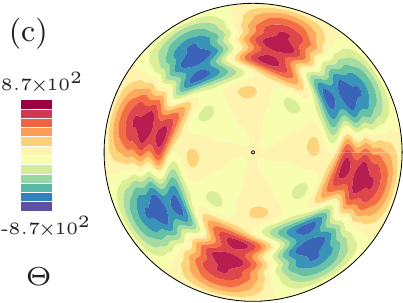}
  \includegraphics[scale=0.85]{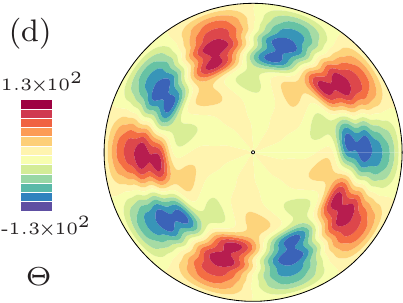}      
  \includegraphics[scale=0.85]{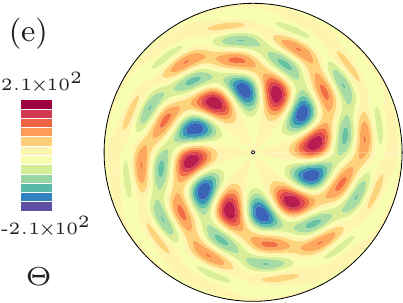}\\[2.mm]  
  \includegraphics[scale=0.85]{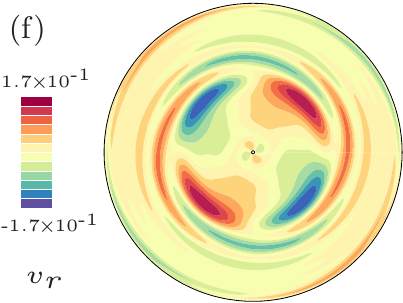}
  \includegraphics[scale=0.85]{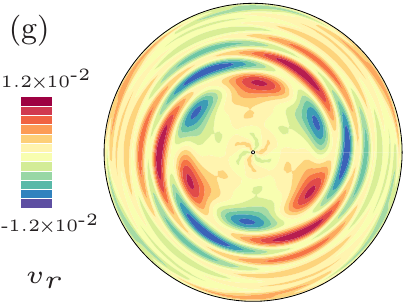}
  \includegraphics[scale=0.85]{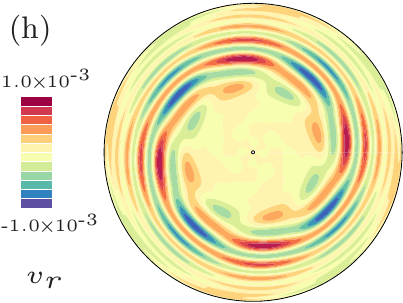}
  \includegraphics[scale=0.85]{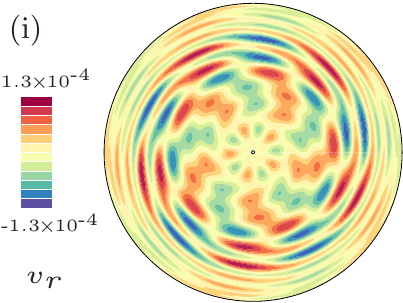}      
  \includegraphics[scale=0.85]{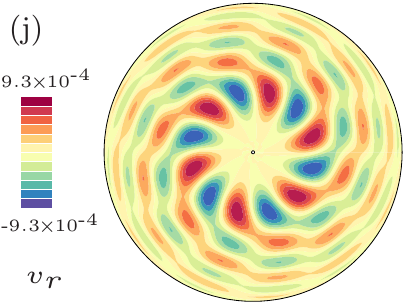}\\[2.mm]        
  \includegraphics[scale=0.85]{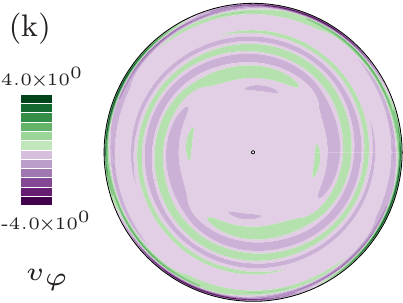}
  \includegraphics[scale=0.85]{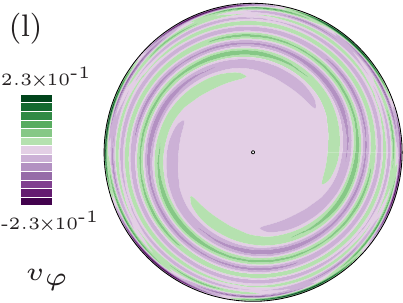}
  \includegraphics[scale=0.85]{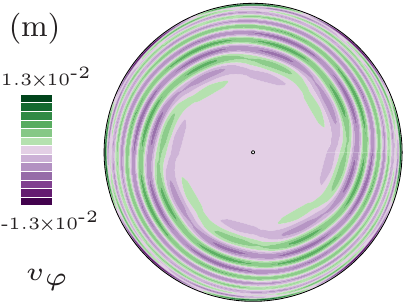}
  \includegraphics[scale=0.85]{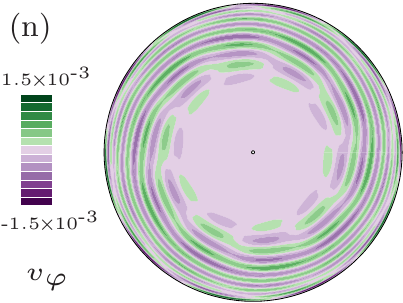}      
  \includegraphics[scale=0.85]{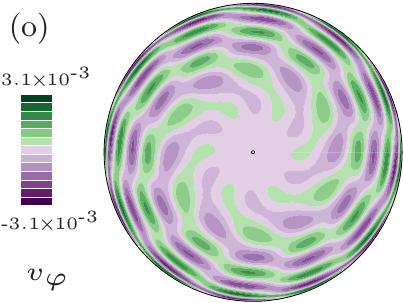}      
\end{center}  
\caption{Modulated rotating wave with azimuthal symmetry $m_2=1$ at in
  the case of the set $P_3$ ($\Ek=3\times 10^{-7}$,$\Pra=0.003$) at
  $\Ray=1.3030\times 10^8$. The contour plots of the temperature
  perturbation $\Theta$ (a-e), of the radial velocity $v_r$ (f-j), and
  of the azimuthal velocity $v_{\varphi}$ (k-o) on an equatorial
  section are displayed from top to bottom rows.  The $m=2,...,6$
  azimuthal components of the solution are displayed in (a,f,k),
  (b,g,l), (c,h,m), (d,i,n), and (e,j,o), respectively.}
\label{fig:nl_cp_E3.d-7_mb}   
\end{figure*}

\begin{figure*}[t!]
\begin{center}
  \includegraphics[scale=0.85]{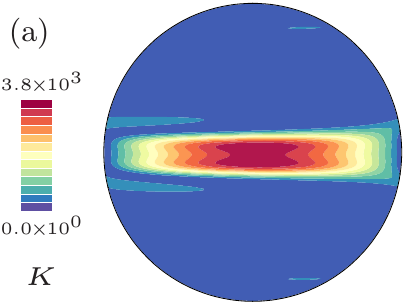}
  \includegraphics[scale=0.85]{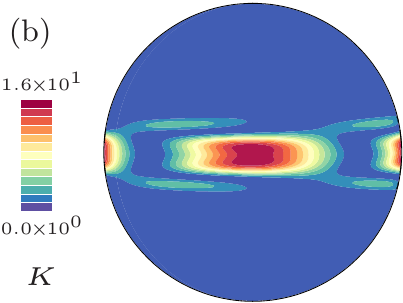}
  \includegraphics[scale=0.85]{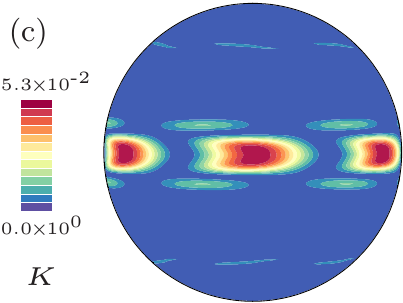}
  \includegraphics[scale=0.85]{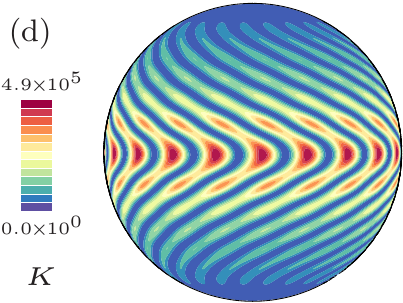}    
  \includegraphics[scale=0.85]{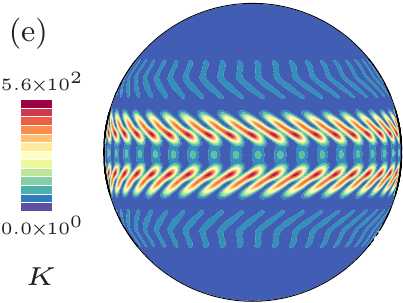}\\  
  \includegraphics[scale=0.85]{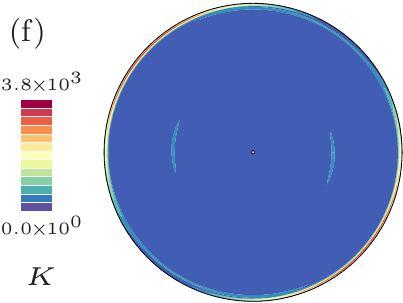}
  \includegraphics[scale=0.85]{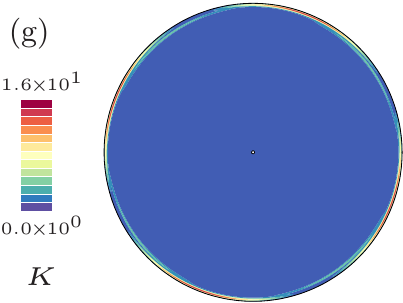}
  \includegraphics[scale=0.85]{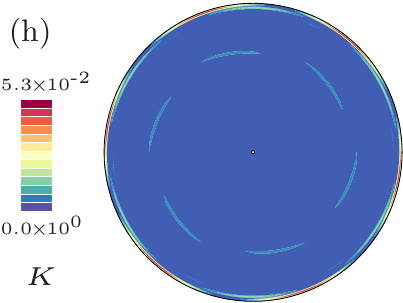}
  \includegraphics[scale=0.85]{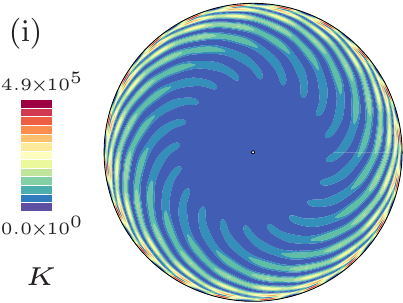}    
  \includegraphics[scale=0.85]{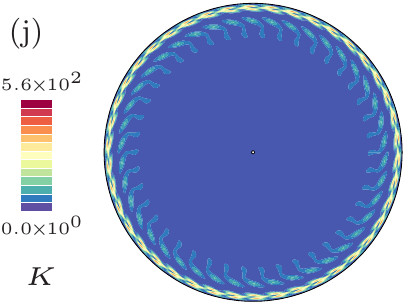}\\
  \includegraphics[scale=0.85]{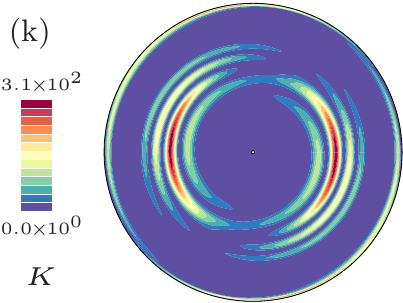}
  \includegraphics[scale=0.85]{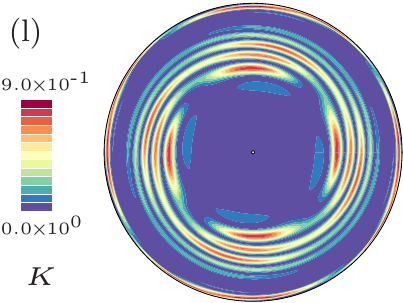}    
  \includegraphics[scale=0.85]{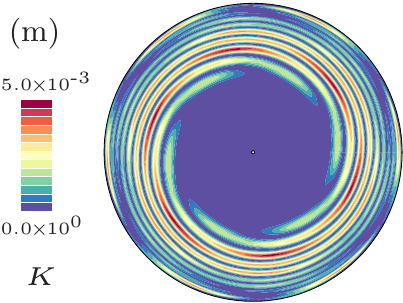}
  \includegraphics[scale=0.85]{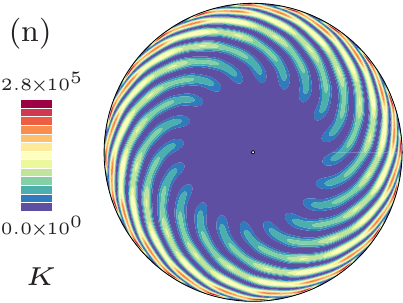}
  \includegraphics[scale=0.85]{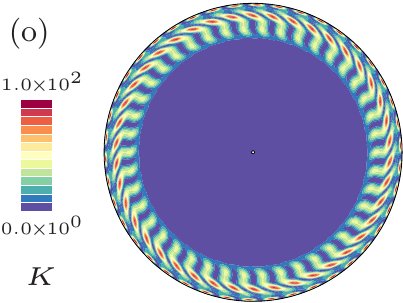}        
\end{center}  
\caption{Modulated rotating wave with azimuthal symmetry $m_2=1$ in
  the case of the set $P_3$ ($\Ek=3\times 10^{-7}$,$\Pra=0.003$) at
  $\Ray=1.3030\times 10^8$. The contour plots for kinetic energy $K$
  on a spherical section at $r\approx 0.99 r_o$ (a-e), on an
  equatorial section (f-j), and on a colatitudinal section at
  $\theta=75^{\circ}$ (k-o).  The $m=1,2,3,10,21$ azimuthal components
  of the solution are displayed in (a,f,k), (b,g,l), (c,h,m), (d,i,n),
  and (e,j,o), respectively.}
\label{fig:nl_cp_K_E3.d-7}   
\end{figure*}

The resonance conditions found for the azimuthal wave numbers $m\le
14$, corresponding to the two MRWs at $\Ray=1.3030\times 10^8$ and at
$\Ray=1.3919\times 10^8$, are listed in Table \ref{table:res1} and
Table \ref{table:res2}, respectively.  The conditions, relating the
largest peaks in the frequency spectrum $f_m$, are of two types. The
first type corresponds to relations involving only low wave numbers
$m\le 5$ whereas the second type involves one low wave number and two
moderate $6\le m\le 16$ wave numbers (see Table \ref{table:res1}). In
contrast, the relations for the second largest peak in the spectrum
($f^2_m$) can involve three moderate wave numbers (i.e
$f_{13}=f^2_6+f_{7}$) since for the second peak the azimuthal wave
number ordering is broken. For instance the modes $m\in\{4,5\}$ have
$f_m$ in the small range but $f^2_m$ in the moderate range while the
reverse occurs for $m=6$ (see Fig. \ref{fig:sdds}). The quasiperiodic
modes $m\in\{4,5\}$ are then dual in the sense that according to their
first peak $f_m$ they may be classified as slow and according to their
second peak $f^2_m$ the may be classified as moderate. The reverse
occurs for $m=6$ and similarly for the moderate and fast modes. The
situation for the MRW at $\Ray=1.3919\times 10^8$ is similar. In this
case the kinetic energy of the nondominant azimuthal wave numbers is
larger (see Fig. \ref{fig:ener_spec}(c)) so more triadic resonant
conditions due to nonlinear interactions are obtained (compare Table
\ref{table:res1} with Table \ref{table:res2}). At $\Ray=1.3919\times
10^8$ there are more dual quasiperiodic modes
$m\in\{3,4,5,6,13,14,15,16,17\}$ and the modes corresponding to small,
moderate, and large frequencies are now $m\le 4$, $5\le m\le 14$, and
$m\ge 15$.

Figure \ref{fig:nl_cp_E3.d-7_m} displays the flow patterns, on a
snapshot, corresponding to selected azimuthal modes with frequencies
on each of the blocks of Fig. \ref{fig:sdds}(a). Specifically, we
select the azimuthal wave numbers $m=1$, $m=10$, and $m=21$ which
correspond to slow, moderate, and fast purely periodic modes. The
contour plots of the temperature perturbation (on equatorial and
meridional sections), of the radial and azimuthal velocity (on an
equatorial section), and of the vertical vorticity (on an equatorial
section), are shown from left to right in each row (see figure
caption). We note that only a single mode for each type (slow,
moderate, or fast) is selected in Fig. \ref{fig:nl_cp_E3.d-7_m} since
the modes for each type have similar flow structure. Slow modes have
vortices of $\Theta$ and $v_r$ located close to the origin of the
sphere, with spiralling arms towards the outer boundary. In contrast,
the situation for the flow structures is reversed. They are mainly
attached to the outer sphere with spiralling arms towards the interior
of the sphere (see $v_{\varphi}$ and $\hat{w}_z$ equatorial
sections). For the moderate modes the vortices of $\Theta$ and $v_r$
are now located at a radial distance around $r_o/2$ so the spiralling
arms towards the outer boundary are smaller. The flow structures are
still mainly attached to the outer sphere but the spiralling arms now
extend up to a radial distance around $r_o/2$. In contrast to this,
the spiralling structures almost disappear in the case of the fast
modes which have the vortices of $\Theta$ and $v_r$ located close to
the outer boundary. The maximum flow velocities are not attached to
the outer sphere, although remain very close to it (see equatorial
section of $v_{\varphi}$).

The same contour plots as Fig. \ref{fig:nl_cp_E3.d-7_m} are displayed
in Fig. \ref{fig:eig_cp_E3.d-7_brm11_Ra2.45_m} corresponding to same
azimuthal wave number decomposition of the Floquet eigenfunction of
the RW with $m_0=11$ azimuthal symmetry at $\Ray=1.3011\times 10^8$
(already analyzed in Sec.\ref{sec:rw_stab} and displayed in
Fig. \ref{fig:cp_eig1}). The patterns are almost the same and make
evident the relation between the resonant modes and the Floquet
eigenfunctions. The eigenfunction, at $\Ray=1.3011\times 10^8$ close
to the bifurcation point giving rise the MRW of Fig.
\ref{fig:nl_cp_E3.d-7_m}, has dominant modes $m=1,10,12,21,23,..$,
i.e, $m=1$ and $m=11k\pm 1$, $k\in\mathbb{Z}$ and the other modes have
nearly zero velocity. When this spatial structure is coupled with the
$m_0=11$ azimuthal symmetry of the unstable RW the main modes are
$m=1$, $m=11k$, and $m=11k\pm 1$, $k\in\mathbb{Z}$ as exhibited by the
MRW in Fig. \ref{fig:sdds}(a) (also Fig. \ref{fig:ener_spec}). It is
interesting to note then that resonant modes arise due to the Hopf
bifurcation giving rise to the MRW and that the Floquet eigenfunctions
reveal the main structure of the resonant slow, moderate, and fast
modes.

To further investigate the flow topology of slow and moderate modes
Fig. \ref{fig:nl_cp_E3.d-7_mb} displays the equatorial sections of
$\Theta$, $v_r$, and $v_{\varphi}$ for the azimuthal wave numbers
$m=2,3,4,5$, and $m=6$. The former correspond to slow modes while the
latter is a moderate mode. The main characteristic of this figure is
that in the case of slow modes the spiraling arms form a polygonal
structure to bound the interior of the sphere (this is best seen on
the sections of $v_{\varphi}$). For the slow mode with $m=2$ the
pattern is a square, for $m=3$ is an hexagon, etc. We note that the
$m=2,3,4,5,6$ modes of Fig. \ref{fig:nl_cp_E3.d-7_mb} are
negligible in the azimuthal wave number decomposition of the leading
eigenfunction and thus are excited due to nonlinear interactions among
the modes of the RW ($m=11k$, $k\in\mathbb{Z}$) and those of the
eigenfunctions ($m=1$, $m=11k\pm 1$, $k\in\mathbb{Z}$). As the wave
number is increased (from $m=2$ up to $m=5$) the vortices of $\Theta$
and $v_r$ of the slow modes tend to be located farther away from the
interior and the spiraling arms of $v_r$ and $v_{\varphi}$ contain
more cells. The patterns of the moderate mode, $m=6$, are changed
significantly (compare with the slow mode $m=5$), especially for the
case of $\Theta$ and $v_r$.

The azimuthal and latitudinal topology of the flow close to the outer
sphere is displayed in the contour plots of the kinetic energy $K$ on
a spherical surface of Fig. \ref{fig:nl_cp_K_E3.d-7} (top row). In
this figure the slow ($m=1,2,3$), moderate ($m=10$), and fast ($m=21$)
modes are displayed from left to right. In the case of slow modes the
convective motions are restricted to a relatively narrow belt
surrounding the equator whereas for the moderate modes the convective
vortices spiral in the azimuthal as well as latitudinal
directions. For both types of modes the maximum value of $K$ is at the
equator. In contrast, for the fast modes motions are almost forbidden
at the equator but develop just above and below. The corresponding
colatitudinal sections at the equator and at colatitude
$\theta=75^{\circ}$ are displayed on the middle and bottom row,
respectively. The equatorial sections now clearly show that in the
case of the slow modes the motions are mainly attached to the outer
boundary. However the bimodal nature of the flow, exhibiting interior
polygonal structures of second order (notice the weak interior
vortices for $m=1$ and $m=3$ at the equatorial plane) can be
identified if the colatitudinal section does not intercept with the
main vortices, for instance at $\theta=75^{\circ}$ (see bottom row of
Fig. \ref{fig:nl_cp_K_E3.d-7}).

\section{Conclusions}
\label{sec:conc}

We have performed a numerical study of thermal convection in an
internally heated rotating sphere with very low Prandtl and Ekman
numbers, appropriate for the study of planetary fluid
cores. Concretely, three sets of parameters are considered
$P_1=(\Pr,\Ek)=(0.03,3\times 10^{-6})$, $P_2=(\Pr,\Ek)=(0.01,10^{-6})$
and $P_3=(\Pr,\Ek)=(0.003,3\times 10^{-7})$ which have already been
studied in \cite{KSVC17}. The focus of our investigation is on weakly
nonlinear flows (weak branch of \cite{KSVC17}) occurring near the
onset of convection, i.\,e. at weakly supercritical conditions
$\Rayt=\Ray/\Rayc-1\leq 1$. By means of continuation methods
(\cite{Kel77,DoTu00,SaNe16}) we have computed branches of rotating
waves (RWs), whose time dependence is described by a steady drift in
the azimuthal direction, bifurcating directly from the base state. The
stability analysis of RWs has evidenced that they are stable for all
the models $P_i$. Additional direct numerical simulations (DNS) allow
us to study secondary quasiperiodic flows (modulated rotating waves,
MRWs) by analyzing Poincar\'e sections, kinetic energy spectra, and
the time series of the flow and its individual modes.

The bifurcation diagrams of the Peclet number $\Pec$ of the RWs follow
the $\sqrt{\Ray-\Rayc}$ law for $\Rayt <3\times 10^{-2}$ since a Hopf
bifurcation breaks the axisymmetry of the conduction state
(\cite{EZK92}). In this interval the rotation frequencies $\omega$ of
the RWs remain nearly constant. For larger values of $\Rayt$ the
bifurcation diagrams become more complicated and can exhibit
saddle-node points (as found for the model $P_1$). In contrast to
\cite{KSVC17}, we have been able to compute the weak branch for the
model $P_3=(\Pr,\Ek)=(0.003,3\times 10^{-7})$. The use of continuation
methods helped us in this task since with DNS very long initial
transients, about $30$ diffusion or $5\times 10^4$ rotation time
units, are required before the nonlinear saturation of the
solution. While steadily drifting solutions have neither been found in
liquid gallium experiments nor numerical simulations of
\cite{HoSc17,ABGHV18,VHA21}, we demonstrate that they can be found
even with smaller $\Pra$ and $\Ek$. The existence of very long initial
transients may make it unfeasible to detect them using experiments and
require massive numerical simulations very close to the onset.

Our results show that for the lowest $\Ek$ and $\Pra$ considered (the
set $P_3$) the RWs are of multicellular type as described in
\cite{NGS08,Lin21} with azimuthal symmetry $m_0=12$ or $m_0=11$.  A
two-layer structure with some vortices of the kinetic energy ($K$)
located close to the outer sphere and others located in the bulk of
the fluid, displaying a polygonal pattern, is formed at the largest
supercritical conditions studied, $\Rayt=1$. The present systematic
computation of multicellular RWs complements the previous studies of
\cite{SGN13}, considering a small inner core and at $\Ek=10^{-4}$ and
$\Pra=0.1$, and \cite{GCW19}, in the case of a very thin shell
($\eta=0.9$) at $\Ek=10^{-4}$ and $\Pra=0.003$. The study of
\cite{SGN13} corresponds to the systematic computation of RWs of
spiralling type (e.\,g. \cite{Zha92}), and that of \cite{GCW19}
corresponds to RWs of polar type (described in \cite{GCW18}). In
agreement with \cite{SGN13,GCW19} RWs become unstable as a result of a
supercritical Hopf bifurcation. We have found that for the set $P_3$
the analysis of stability of RWs is numerically challenging. This is
because the eigenvalues are clustered near the unit circle, which
degrades the convergence of eigenvalue solver, and means that multiple
bifurcations take place near the onset (as in \cite{GCW19}).  The
analysis of the structure and symmetry $m_1$ of the eigenfunctions
(Floquet modes) allows us to predict MRWs with azimuthal symmetry
$m_2=1$.

The DNS presented here, starting from an unstable RW initial
condition, exhibit strongly oscillatory and very long initial
transients, about $30$ diffusion or $5\times 10^4$ rotation time
units, before a weakly oscillatory quasiperiodic flow (MRW) is
statistically saturated. This is because the perturbations grow very
slowly in the unstable directions, given by Floquet modes, which are
predicted by the stability analysis of RWs. Close to the bifurcation
point the azimuthal wave number structure is inherited from the
leading Floquet mode. The azimuthal wave number and time dependence of
the long initial transients and the saturated solution is
significantly different. Initial transients are characterized by
strong time dependence and a large energy component of low azimuthal
wave numbers $m<6$, whereas the kinetic energy spectra of the
saturated solution are nearly constant in time and have significant
peaks only for a reduced set of modes, including $m=1$.  In addition,
the time series of the temperature perturbation, at several points
inside the sphere, reveal two very different time scales, slow and
fast, associated to the interior ($r<r_o/5$) or the exterior
($r>r_o/2$) of the sphere, respectively. The former is characteristic
of low wave numbers (e.\,g. $m=1$) whereas the latter is
characteristic of moderate and large wave numbers (e.g. $m=11$).

As in \cite{Lin21} our DNS exhibit triadic resonances among different
equatorially symmetric modes characterized by the spherical harmonic
degree $l$ and order $m$, and in agreement with \cite{GGS21} the
solutions are MRWs. A characteristic block pattern with low, moderate,
and large resonant wave numbers described by small, moderate, and
large frequencies, respectively, is found in the frequency
spectra. For the MRW closest to the bifurcation point these modes are
$m\le 5$, $6\le m\le 16$, and $m\ge 17$. The modes having largest peaks
in the frequency spectrum are the non-vanishing components of the
Floquet mode, $m=1$ and $m=11k\pm 1$, $k\in\mathbb{Z}$, which includes
the $m=11$ mode of the parent RW.

The flow and temperature perturbation contour plots of the individual
modes $m=1,10,21$ forming the leading Floquet eigenfunction are almost
the same as the contour plots of the $m=1,10,21$ modes forming the
resonant flow (MRW). For the slow modes (such as $m=1$) convective
motions mainly occur close to the outer sphere (wall modes), on a
narrow band around the equator. However, weak regular and polygonal
structures (oval, square, hexagon) develop in the bulk of the fluid
(interior modes) so the flow topology is of bimodal nature. The flow
patterns of the $m=10$ moderate mode, although still attached to the
outer sphere (wall modes) and with a maximum amplitude vortex at the
equator, spiral to high latitudes and to the bulk of the fluid.  In
contrast, for the large mode $m=21$ the single vortex splits in two
which are located symmetricaly above and below the equator, a little away
from the outer boundary but without going deep into the
interior. Either moderate ($m=10$) or large ($m=21$) modes have single
mode structure since in this case there is no interior differentiated
pattern.

While the patterns of RWs can be described by a single mode predicted
by the linear stability analysis of the onset of convection, the
patterns of MRWs can be multimodal and can be predicted by the
stability analysis of RWs (periodic flows) and the computation of the
leading Floquet modes. According to \cite{Lin21} (see introduction)
the mechanism giving rise to the multimodal nature, i.\, e. to flows
from which dominant modes with different spatial localization can be
identified (\cite{HoSc17,ABGHV18,VHA21}), in the case of rotating
convection at low $\Pra$ still remains a puzzle. Our study
demonstrates that in this regime multimodal convection is generated by
a Hopf bifurcaton of RWs (weak branch). Moreover, we have found that
the specific spatial structure of the different spatially localized
modes is determined by the stability analysis (Floquet modes) of the
RWs.



\medskip
\section{Acknowledgements}

This project has received funding from the European Research Council
(ERC) under the European Union’s Horizon 2020 research and innovation
programme (grant agreement No 787544). The authors kindly thank
N. Schaeffer for his valuable comments.


\begin{thebibliography}{69}%
\makeatletter
\providecommand \@ifxundefined [1]{%
 \@ifx{#1\undefined}
}%
\providecommand \@ifnum [1]{%
 \ifnum #1\expandafter \@firstoftwo
 \else \expandafter \@secondoftwo
 \fi
}%
\providecommand \@ifx [1]{%
 \ifx #1\expandafter \@firstoftwo
 \else \expandafter \@secondoftwo
 \fi
}%
\providecommand \natexlab [1]{#1}%
\providecommand \enquote  [1]{``#1''}%
\providecommand \bibnamefont  [1]{#1}%
\providecommand \bibfnamefont [1]{#1}%
\providecommand \citenamefont [1]{#1}%
\providecommand \href@noop [0]{\@secondoftwo}%
\providecommand \href [0]{\begingroup \@sanitize@url \@href}%
\providecommand \@href[1]{\@@startlink{#1}\@@href}%
\providecommand \@@href[1]{\endgroup#1\@@endlink}%
\providecommand \@sanitize@url [0]{\catcode `\\12\catcode `\$12\catcode
  `\&12\catcode `\#12\catcode `\^12\catcode `\_12\catcode `\%12\relax}%
\providecommand \@@startlink[1]{}%
\providecommand \@@endlink[0]{}%
\providecommand \url  [0]{\begingroup\@sanitize@url \@url }%
\providecommand \@url [1]{\endgroup\@href {#1}{\urlprefix }}%
\providecommand \urlprefix  [0]{URL }%
\providecommand \Eprint [0]{\href }%
\providecommand \doibase [0]{http://dx.doi.org/}%
\providecommand \selectlanguage [0]{\@gobble}%
\providecommand \bibinfo  [0]{\@secondoftwo}%
\providecommand \bibfield  [0]{\@secondoftwo}%
\providecommand \translation [1]{[#1]}%
\providecommand \BibitemOpen [0]{}%
\providecommand \bibitemStop [0]{}%
\providecommand \bibitemNoStop [0]{.\EOS\space}%
\providecommand \EOS [0]{\spacefactor3000\relax}%
\providecommand \BibitemShut  [1]{\csname bibitem#1\endcsname}%
\let\auto@bib@innerbib\@empty
\bibitem [{\citenamefont {Glatzmaier}\ and\ \citenamefont
  {Roberts}(1995)}]{GlRo95a}%
  \BibitemOpen
  \bibfield  {author} {\bibinfo {author} {\bibfnamefont {G.A.}\ \bibnamefont
  {Glatzmaier}}\ and\ \bibinfo {author} {\bibfnamefont {P.H.}\ \bibnamefont
  {Roberts}},\ }\bibfield  {title} {\enquote {\bibinfo {title} {A
  three-dimensional self-consistent computer simulation of a geomagnetic field
  reversal},}\ }\href@noop {} {\bibfield  {journal} {\bibinfo  {journal}
  {Nature}\ }\textbf {\bibinfo {volume} {377}},\ \bibinfo {pages} {203--209}
  (\bibinfo {year} {1995})}\BibitemShut {NoStop}%
\bibitem [{\citenamefont {Schaeffer}\ \emph {et~al.}(2017)\citenamefont
  {Schaeffer}, \citenamefont {Jault}, \citenamefont {Nataf},\ and\
  \citenamefont {Fournier}}]{SJNF17}%
  \BibitemOpen
  \bibfield  {author} {\bibinfo {author} {\bibfnamefont {N.}~\bibnamefont
  {Schaeffer}}, \bibinfo {author} {\bibfnamefont {D.}~\bibnamefont {Jault}},
  \bibinfo {author} {\bibfnamefont {H.-C.}\ \bibnamefont {Nataf}}, \ and\
  \bibinfo {author} {\bibfnamefont {A.}~\bibnamefont {Fournier}},\ }\bibfield
  {title} {\enquote {\bibinfo {title} {{Turbulent geodynamo simulations: a leap
  towards Earth’s core}},}\ }\href {\doibase 10.1093/gji/ggx265} {\bibfield
  {journal} {\bibinfo  {journal} {Geophys. J. Int.}\ }\textbf {\bibinfo
  {volume} {211}},\ \bibinfo {pages} {1--29} (\bibinfo {year}
  {2017})}\BibitemShut {NoStop}%
\bibitem [{\citenamefont {Heimpel}\ \emph {et~al.}(2015)\citenamefont
  {Heimpel}, \citenamefont {Gastine},\ and\ \citenamefont {Wicht}}]{HGW15}%
  \BibitemOpen
  \bibfield  {author} {\bibinfo {author} {\bibfnamefont {M.}~\bibnamefont
  {Heimpel}}, \bibinfo {author} {\bibfnamefont {T.}~\bibnamefont {Gastine}}, \
  and\ \bibinfo {author} {\bibfnamefont {J.}~\bibnamefont {Wicht}},\ }\bibfield
   {title} {\enquote {\bibinfo {title} {Simulation of deep-seated zonal jets
  and shallow vortices in gas giant atmospheres},}\ }\href@noop {} {\bibfield
  {journal} {\bibinfo  {journal} {Nat. Geosci.}\ }\textbf {\bibinfo {volume}
  {9}},\ \bibinfo {pages} {19--23} (\bibinfo {year} {2015})}\BibitemShut
  {NoStop}%
\bibitem [{\citenamefont {Garcia}\ \emph {et~al.}(2020)\citenamefont {Garcia},
  \citenamefont {Chambers},\ and\ \citenamefont {Watts}}]{GCW20}%
  \BibitemOpen
  \bibfield  {author} {\bibinfo {author} {\bibfnamefont {F.}~\bibnamefont
  {Garcia}}, \bibinfo {author} {\bibfnamefont {F.~R.~N.}\ \bibnamefont
  {Chambers}}, \ and\ \bibinfo {author} {\bibfnamefont {A.~L.}\ \bibnamefont
  {Watts}},\ }\bibfield  {title} {\enquote {\bibinfo {title} {Deep model
  simulation of polar vortices in gas giant atmospheres},}\ }\href@noop {}
  {\bibfield  {journal} {\bibinfo  {journal} {Mon.\ Not.\ R.\ astr.\ Soc.}\
  }\textbf {\bibinfo {volume} {499}},\ \bibinfo {pages} {4698--4715} (\bibinfo
  {year} {2020})}\BibitemShut {NoStop}%
\bibitem [{\citenamefont {R{\"u}diger}(1989)}]{Rud89}%
  \BibitemOpen
  \bibfield  {author} {\bibinfo {author} {\bibfnamefont {G.}~\bibnamefont
  {R{\"u}diger}},\ }\href@noop {} {\emph {\bibinfo {title} {Differential
  Rotation and Stellar Convection: Sun and Solar-type Stars}}},\ Fluid
  mechanics of astrophysics and geophysics\ (\bibinfo  {publisher} {Gordon and
  Breach Science Publishers},\ \bibinfo {year} {1989})\BibitemShut {NoStop}%
\bibitem [{\citenamefont {{Brun}}\ \emph {et~al.}(2004)\citenamefont {{Brun}},
  \citenamefont {{Miesch}},\ and\ \citenamefont {{Toomre}}}]{BMT04}%
  \BibitemOpen
  \bibfield  {author} {\bibinfo {author} {\bibfnamefont {A.~S.}\ \bibnamefont
  {{Brun}}}, \bibinfo {author} {\bibfnamefont {M.~S.}\ \bibnamefont
  {{Miesch}}}, \ and\ \bibinfo {author} {\bibfnamefont {J.}~\bibnamefont
  {{Toomre}}},\ }\bibfield  {title} {\enquote {\bibinfo {title} {{Global-Scale
  Turbulent Convection and Magnetic Dynamo Action in the Solar Envelope}},}\
  }\href@noop {} {\bibfield  {journal} {\bibinfo  {journal} {Astrophys.\ J.}\
  }\textbf {\bibinfo {volume} {614}},\ \bibinfo {pages} {1073--1098} (\bibinfo
  {year} {2004})}\BibitemShut {NoStop}%
\bibitem [{\citenamefont {Jones}(2007)}]{Jon07}%
  \BibitemOpen
  \bibfield  {author} {\bibinfo {author} {\bibfnamefont {C.~A.}\ \bibnamefont
  {Jones}},\ }\bibfield  {title} {\enquote {\bibinfo {title} {Thermal and
  compositional convection in the outer core},}\ }\href@noop {} {\bibfield
  {journal} {\bibinfo  {journal} {Treat. Geophys.}\ }\textbf {\bibinfo {volume}
  {8}},\ \bibinfo {pages} {131--185} (\bibinfo {year} {2007})}\BibitemShut
  {NoStop}%
\bibitem [{\citenamefont {Gailitis}\ \emph {et~al.}(2002)\citenamefont
  {Gailitis}, \citenamefont {Lielausis}, \citenamefont {Platacis},
  \citenamefont {Gerbeth},\ and\ \citenamefont {Stefani}}]{GLPGS02}%
  \BibitemOpen
  \bibfield  {author} {\bibinfo {author} {\bibfnamefont {A.}~\bibnamefont
  {Gailitis}}, \bibinfo {author} {\bibfnamefont {O.}~\bibnamefont {Lielausis}},
  \bibinfo {author} {\bibfnamefont {E.}~\bibnamefont {Platacis}}, \bibinfo
  {author} {\bibfnamefont {G}~\bibnamefont {Gerbeth}}, \ and\ \bibinfo {author}
  {\bibfnamefont {F.}~\bibnamefont {Stefani}},\ }\bibfield  {title} {\enquote
  {\bibinfo {title} {Colloquium: {L}aboratory experiments on hydromagnetic
  dynamos},}\ }\href@noop {} {\bibfield  {journal} {\bibinfo  {journal} {Rev.\
  Mod. Phys.}\ }\textbf {\bibinfo {volume} {74}},\ \bibinfo {pages} {973--990}
  (\bibinfo {year} {2002})}\BibitemShut {NoStop}%
\bibitem [{\citenamefont {Moffatt}\ and\ \citenamefont {Dormy}(2019)}]{MoDo19}%
  \BibitemOpen
  \bibfield  {author} {\bibinfo {author} {\bibfnamefont {K.}~\bibnamefont
  {Moffatt}}\ and\ \bibinfo {author} {\bibfnamefont {E.}~\bibnamefont
  {Dormy}},\ }\href@noop {} {\emph {\bibinfo {title} {Self-Exciting Fluid
  Dynamos}}},\ Cambridge Texts in Applied Mathematics\ (\bibinfo  {publisher}
  {Cambridge University press},\ \bibinfo {year} {2019})\BibitemShut {NoStop}%
\bibitem [{\citenamefont {Dormy}\ and\ \citenamefont {Soward}(2007)}]{DoSo07}%
  \BibitemOpen
  \bibinfo {editor} {\bibfnamefont {E.}~\bibnamefont {Dormy}}\ and\ \bibinfo
  {editor} {\bibfnamefont {A.~M.}\ \bibnamefont {Soward}},\ eds.,\ \href@noop
  {} {\emph {\bibinfo {title} {{M}athematical {A}spects of {N}atural
  {D}ynamos}}},\ \bibinfo {series} {{T}he {F}luid {M}echanics of {A}strophysics
  and {G}eophysics}, Vol.~\bibinfo {volume} {13}\ (\bibinfo  {publisher}
  {Chapman \& Hall/CRC, {B}oca {R}aton, {FL}},\ \bibinfo {year}
  {2007})\BibitemShut {NoStop}%
\bibitem [{\citenamefont {Julien}\ \emph {et~al.}(2012)\citenamefont {Julien},
  \citenamefont {Rubio}, \citenamefont {Grooms},\ and\ \citenamefont
  {Knobloch}}]{JRGK12}%
  \BibitemOpen
  \bibfield  {author} {\bibinfo {author} {\bibfnamefont {K.}~\bibnamefont
  {Julien}}, \bibinfo {author} {\bibfnamefont {A.~M.}\ \bibnamefont {Rubio}},
  \bibinfo {author} {\bibfnamefont {I.}~\bibnamefont {Grooms}}, \ and\ \bibinfo
  {author} {\bibfnamefont {E.}~\bibnamefont {Knobloch}},\ }\bibfield  {title}
  {\enquote {\bibinfo {title} {Statistical and physical balances in low
  {R}ossby number {R}ayleigh–{B}\'enard convection},}\ }\href@noop {}
  {\bibfield  {journal} {\bibinfo  {journal} {Geophys. Astrophys. Fluid
  Dynamics}\ }\textbf {\bibinfo {volume} {106}},\ \bibinfo {pages} {392--428}
  (\bibinfo {year} {2012})}\BibitemShut {NoStop}%
\bibitem [{\citenamefont {Guervilly}\ \emph {et~al.}(2019)\citenamefont
  {Guervilly}, \citenamefont {Cardin},\ and\ \citenamefont
  {Schaeffer}}]{GCS19}%
  \BibitemOpen
  \bibfield  {author} {\bibinfo {author} {\bibfnamefont {C.}~\bibnamefont
  {Guervilly}}, \bibinfo {author} {\bibfnamefont {P.}~\bibnamefont {Cardin}}, \
  and\ \bibinfo {author} {\bibfnamefont {N.}~\bibnamefont {Schaeffer}},\
  }\bibfield  {title} {\enquote {\bibinfo {title} {Turbulent convective length
  scale in planetary cores},}\ }\href@noop {} {\bibfield  {journal} {\bibinfo
  {journal} {Nature}\ }\textbf {\bibinfo {volume} {570}},\ \bibinfo {pages}
  {368--371} (\bibinfo {year} {2019})}\BibitemShut {NoStop}%
\bibitem [{\citenamefont {Christensen}\ \emph {et~al.}(2001)\citenamefont
  {Christensen}, \citenamefont {Aubert}, \citenamefont {Cardin}, \citenamefont
  {Dormy}, \citenamefont {Gibbons}, \citenamefont {Glatzmaier}, \citenamefont
  {Grote}, \citenamefont {Honkura}, \citenamefont {Jones}, \citenamefont
  {Kono}, \citenamefont {Matsushima}, \citenamefont {Sakuraba}, \citenamefont
  {Takahashi}, \citenamefont {Tilgner}, \citenamefont {Wicht},\ and\
  \citenamefont {Zhang}}]{CACDGGGHJKMSTTWZ01}%
  \BibitemOpen
  \bibfield  {author} {\bibinfo {author} {\bibfnamefont {U.R.}\ \bibnamefont
  {Christensen}}, \bibinfo {author} {\bibfnamefont {J.}~\bibnamefont {Aubert}},
  \bibinfo {author} {\bibfnamefont {P.}~\bibnamefont {Cardin}}, \bibinfo
  {author} {\bibfnamefont {E.}~\bibnamefont {Dormy}}, \bibinfo {author}
  {\bibfnamefont {S.}~\bibnamefont {Gibbons}}, \bibinfo {author} {\bibfnamefont
  {G.A.}\ \bibnamefont {Glatzmaier}}, \bibinfo {author} {\bibfnamefont
  {E.}~\bibnamefont {Grote}}, \bibinfo {author} {\bibfnamefont
  {Y.}~\bibnamefont {Honkura}}, \bibinfo {author} {\bibfnamefont
  {C.}~\bibnamefont {Jones}}, \bibinfo {author} {\bibfnamefont
  {M.}~\bibnamefont {Kono}}, \bibinfo {author} {\bibfnamefont {M.}~\bibnamefont
  {Matsushima}}, \bibinfo {author} {\bibfnamefont {A.}~\bibnamefont
  {Sakuraba}}, \bibinfo {author} {\bibfnamefont {F.}~\bibnamefont {Takahashi}},
  \bibinfo {author} {\bibfnamefont {A.}~\bibnamefont {Tilgner}}, \bibinfo
  {author} {\bibfnamefont {J.}~\bibnamefont {Wicht}}, \ and\ \bibinfo {author}
  {\bibfnamefont {K.}~\bibnamefont {Zhang}},\ }\bibfield  {title} {\enquote
  {\bibinfo {title} {A numerical dynamo benchmark},}\ }\href@noop {} {\bibfield
   {journal} {\bibinfo  {journal} {Phys.\ Earth\ Planet.\ Inter.}\ }\textbf
  {\bibinfo {volume} {128}},\ \bibinfo {pages} {25--34} (\bibinfo {year}
  {2001})}\BibitemShut {NoStop}%
\bibitem [{\citenamefont {Marti}\ \emph {et~al.}(2014)\citenamefont {Marti},
  \citenamefont {Schaeffer}, \citenamefont {Hollerbach}, \citenamefont
  {Cébron}, \citenamefont {Nore}, \citenamefont {Luddens}, \citenamefont
  {Guermond}, \citenamefont {Aubert}, \citenamefont {Takehiro}, \citenamefont
  {Sasaki}, \citenamefont {Hayashi}, \citenamefont {Simitev}, \citenamefont
  {Busse}, \citenamefont {Vantieghem},\ and\ \citenamefont
  {Jackson}}]{Mar_et_al15}%
  \BibitemOpen
  \bibfield  {author} {\bibinfo {author} {\bibfnamefont {P.}~\bibnamefont
  {Marti}}, \bibinfo {author} {\bibfnamefont {N.}~\bibnamefont {Schaeffer}},
  \bibinfo {author} {\bibfnamefont {R.}~\bibnamefont {Hollerbach}}, \bibinfo
  {author} {\bibfnamefont {D.}~\bibnamefont {Cébron}}, \bibinfo {author}
  {\bibfnamefont {C.}~\bibnamefont {Nore}}, \bibinfo {author} {\bibfnamefont
  {F.}~\bibnamefont {Luddens}}, \bibinfo {author} {\bibfnamefont {J.-L.}\
  \bibnamefont {Guermond}}, \bibinfo {author} {\bibfnamefont {J.}~\bibnamefont
  {Aubert}}, \bibinfo {author} {\bibfnamefont {S.}~\bibnamefont {Takehiro}},
  \bibinfo {author} {\bibfnamefont {Y.}~\bibnamefont {Sasaki}}, \bibinfo
  {author} {\bibfnamefont {Y.-Y.}\ \bibnamefont {Hayashi}}, \bibinfo {author}
  {\bibfnamefont {R.}~\bibnamefont {Simitev}}, \bibinfo {author} {\bibfnamefont
  {F.}~\bibnamefont {Busse}}, \bibinfo {author} {\bibfnamefont
  {S.}~\bibnamefont {Vantieghem}}, \ and\ \bibinfo {author} {\bibfnamefont
  {A.}~\bibnamefont {Jackson}},\ }\bibfield  {title} {\enquote {\bibinfo
  {title} {{Full sphere hydrodynamic and dynamo benchmarks}},}\ }\href@noop {}
  {\bibfield  {journal} {\bibinfo  {journal} {Geophys. J. Int.}\ }\textbf
  {\bibinfo {volume} {197}},\ \bibinfo {pages} {119--134} (\bibinfo {year}
  {2014})}\BibitemShut {NoStop}%
\bibitem [{\citenamefont {Chandrasekhar}(1981)}]{Cha81}%
  \BibitemOpen
  \bibfield  {author} {\bibinfo {author} {\bibfnamefont {S.}~\bibnamefont
  {Chandrasekhar}},\ }\href@noop {} {\emph {\bibinfo {title} {Hydrodynamic and
  Hydromagnetic Stability}}}\ (\bibinfo  {publisher} {Dover publications, inc.
  New York},\ \bibinfo {year} {1981})\BibitemShut {NoStop}%
\bibitem [{\citenamefont {Rand}(1982)}]{Ran82}%
  \BibitemOpen
  \bibfield  {author} {\bibinfo {author} {\bibfnamefont {D.}~\bibnamefont
  {Rand}},\ }\bibfield  {title} {\enquote {\bibinfo {title} {Dynamics and
  symmetry. {P}redictions for modulated waves in rotating fluids},}\
  }\href@noop {} {\bibfield  {journal} {\bibinfo  {journal} {Arch.\ Ration.\
  Mech.\ Anal.}\ }\textbf {\bibinfo {volume} {79}},\ \bibinfo {pages} {1--37}
  (\bibinfo {year} {1982})}\BibitemShut {NoStop}%
\bibitem [{\citenamefont {Golubitsky}\ \emph {et~al.}(2000)\citenamefont
  {Golubitsky}, \citenamefont {LeBlanc},\ and\ \citenamefont
  {Melbourne}}]{GLM00}%
  \BibitemOpen
  \bibfield  {author} {\bibinfo {author} {\bibfnamefont {M.}~\bibnamefont
  {Golubitsky}}, \bibinfo {author} {\bibfnamefont {V.~G.}\ \bibnamefont
  {LeBlanc}}, \ and\ \bibinfo {author} {\bibfnamefont {I.}~\bibnamefont
  {Melbourne}},\ }\bibfield  {title} {\enquote {\bibinfo {title} {Hopf
  bifurcation from rotating waves and patterns in physical space},}\
  }\href@noop {} {\bibfield  {journal} {\bibinfo  {journal} {J.\,Nonlinear
  Sci.}\ }\textbf {\bibinfo {volume} {10}},\ \bibinfo {pages} {69--101}
  (\bibinfo {year} {2000})}\BibitemShut {NoStop}%
\bibitem [{\citenamefont {Zhang}(1992)}]{Zha92}%
  \BibitemOpen
  \bibfield  {author} {\bibinfo {author} {\bibfnamefont {K.}~\bibnamefont
  {Zhang}},\ }\bibfield  {title} {\enquote {\bibinfo {title} {Spiralling
  columnar convection in rapidly rotating spherical fluid shells},}\
  }\href@noop {} {\bibfield  {journal} {\bibinfo  {journal} {J.\,Fluid Mech.}\
  }\textbf {\bibinfo {volume} {236}},\ \bibinfo {pages} {535--556} (\bibinfo
  {year} {1992})}\BibitemShut {NoStop}%
\bibitem [{\citenamefont {Dormy}\ \emph {et~al.}(2004)\citenamefont {Dormy},
  \citenamefont {Soward}, \citenamefont {Jones}, \citenamefont {Jault},\ and\
  \citenamefont {Cardin}}]{DSJJC04}%
  \BibitemOpen
  \bibfield  {author} {\bibinfo {author} {\bibfnamefont {E.}~\bibnamefont
  {Dormy}}, \bibinfo {author} {\bibfnamefont {A.~M.}\ \bibnamefont {Soward}},
  \bibinfo {author} {\bibfnamefont {C.~A.}\ \bibnamefont {Jones}}, \bibinfo
  {author} {\bibfnamefont {D.}~\bibnamefont {Jault}}, \ and\ \bibinfo {author}
  {\bibfnamefont {P.}~\bibnamefont {Cardin}},\ }\bibfield  {title} {\enquote
  {\bibinfo {title} {The onset of thermal convection in rotating spherical
  shells},}\ }\href@noop {} {\bibfield  {journal} {\bibinfo  {journal}
  {J.\,Fluid Mech.}\ }\textbf {\bibinfo {volume} {501}},\ \bibinfo {pages}
  {43--70} (\bibinfo {year} {2004})}\BibitemShut {NoStop}%
\bibitem [{\citenamefont {Zhang}(1993)}]{Zha93}%
  \BibitemOpen
  \bibfield  {author} {\bibinfo {author} {\bibfnamefont {K.}~\bibnamefont
  {Zhang}},\ }\bibfield  {title} {\enquote {\bibinfo {title} {On equatorially
  trapped boundary inertial waves},}\ }\href@noop {} {\bibfield  {journal}
  {\bibinfo  {journal} {J.\,Fluid Mech.}\ }\textbf {\bibinfo {volume} {248}},\
  \bibinfo {pages} {203--217} (\bibinfo {year} {1993})}\BibitemShut {NoStop}%
\bibitem [{\citenamefont {{\relax Net}}\ \emph {et~al.}(2008)\citenamefont
  {{\relax Net}}, \citenamefont {Garcia},\ and\ \citenamefont
  {S\'anchez}}]{NGS08}%
  \BibitemOpen
  \bibfield  {author} {\bibinfo {author} {\bibfnamefont {M.}~\bibnamefont
  {{\relax Net}}}, \bibinfo {author} {\bibfnamefont {F.}~\bibnamefont
  {Garcia}}, \ and\ \bibinfo {author} {\bibfnamefont {J.}~\bibnamefont
  {S\'anchez}},\ }\bibfield  {title} {\enquote {\bibinfo {title} {On the onset
  of low-{P}randtl-number convection in rotating spherical shells: non-slip
  boundary conditions},}\ }\href@noop {} {\bibfield  {journal} {\bibinfo
  {journal} {J.\,Fluid Mech.}\ }\textbf {\bibinfo {volume} {601}},\ \bibinfo
  {pages} {317--337} (\bibinfo {year} {2008})}\BibitemShut {NoStop}%
\bibitem [{\citenamefont {Garcia}\ \emph {et~al.}(2008)\citenamefont {Garcia},
  \citenamefont {S\'anchez},\ and\ \citenamefont {{\relax Net}}}]{GSN08}%
  \BibitemOpen
  \bibfield  {author} {\bibinfo {author} {\bibfnamefont {F.}~\bibnamefont
  {Garcia}}, \bibinfo {author} {\bibfnamefont {J.}~\bibnamefont {S\'anchez}}, \
  and\ \bibinfo {author} {\bibfnamefont {M.}~\bibnamefont {{\relax Net}}},\
  }\bibfield  {title} {\enquote {\bibinfo {title} {Antisymmetric polar modes of
  thermal convection in rotating spherical fluid shells at high {T}aylor
  numbers},}\ }\href@noop {} {\bibfield  {journal} {\bibinfo  {journal} {Phys.\
  Rev.\ Lett.}\ }\textbf {\bibinfo {volume} {101}},\ \bibinfo {pages}
  {194501--(1--4)} (\bibinfo {year} {2008})}\BibitemShut {NoStop}%
\bibitem [{\citenamefont {Garcia}\ \emph {et~al.}(2018)\citenamefont {Garcia},
  \citenamefont {Chambers},\ and\ \citenamefont {Watts}}]{GCW18}%
  \BibitemOpen
  \bibfield  {author} {\bibinfo {author} {\bibfnamefont {F.}~\bibnamefont
  {Garcia}}, \bibinfo {author} {\bibfnamefont {F.~R.~N.}\ \bibnamefont
  {Chambers}}, \ and\ \bibinfo {author} {\bibfnamefont {A.~L.}\ \bibnamefont
  {Watts}},\ }\bibfield  {title} {\enquote {\bibinfo {title} {The onset of low
  {P}randtl number thermal convection in thin spherical shells},}\ }\href@noop
  {} {\bibfield  {journal} {\bibinfo  {journal} {Phys.\ Rev.\,Fluids}\ }\textbf
  {\bibinfo {volume} {3}},\ \bibinfo {pages} {024801} (\bibinfo {year}
  {2018})}\BibitemShut {NoStop}%
\bibitem [{\citenamefont {S\'anchez}\ \emph {et~al.}(2016)\citenamefont
  {S\'anchez}, \citenamefont {Garcia},\ and\ \citenamefont {Net}}]{SGN16b}%
  \BibitemOpen
  \bibfield  {author} {\bibinfo {author} {\bibfnamefont {J.}~\bibnamefont
  {S\'anchez}}, \bibinfo {author} {\bibfnamefont {F.}~\bibnamefont {Garcia}}, \
  and\ \bibinfo {author} {\bibfnamefont {M.}~\bibnamefont {Net}},\ }\bibfield
  {title} {\enquote {\bibinfo {title} {Critical torsional modes of convection
  in rotating fluid spheres at high taylor numbers},}\ }\href {\doibase
  10.1017/jfm.2016.52} {\bibfield  {journal} {\bibinfo  {journal} {J.\,Fluid
  Mech.}\ }\textbf {\bibinfo {volume} {791}} (\bibinfo {year} {2016}),\
  10.1017/jfm.2016.52}\BibitemShut {NoStop}%
\bibitem [{\citenamefont {Zhang}\ \emph {et~al.}(2017)\citenamefont {Zhang},
  \citenamefont {Lam},\ and\ \citenamefont {Kong}}]{ZLK17}%
  \BibitemOpen
  \bibfield  {author} {\bibinfo {author} {\bibfnamefont {K.}~\bibnamefont
  {Zhang}}, \bibinfo {author} {\bibfnamefont {K.}~\bibnamefont {Lam}}, \ and\
  \bibinfo {author} {\bibfnamefont {D.}~\bibnamefont {Kong}},\ }\bibfield
  {title} {\enquote {\bibinfo {title} {Asymptotic theory for torsional
  convection in rotating fluid spheres},}\ }\href {\doibase 10.1017/jfm.2017.9}
  {\bibfield  {journal} {\bibinfo  {journal} {Journal of Fluid Mechanics}\
  }\textbf {\bibinfo {volume} {813}} (\bibinfo {year} {2017}),\
  10.1017/jfm.2017.9}\BibitemShut {NoStop}%
\bibitem [{\citenamefont {Ardes}\ \emph {et~al.}(1997)\citenamefont {Ardes},
  \citenamefont {Busse},\ and\ \citenamefont {Wicht}}]{ABW97}%
  \BibitemOpen
  \bibfield  {author} {\bibinfo {author} {\bibfnamefont {M.}~\bibnamefont
  {Ardes}}, \bibinfo {author} {\bibfnamefont {F.~H.}\ \bibnamefont {Busse}}, \
  and\ \bibinfo {author} {\bibfnamefont {J.}~\bibnamefont {Wicht}},\ }\bibfield
   {title} {\enquote {\bibinfo {title} {Thermal convection in rotating
  spherical shells},}\ }\href@noop {} {\bibfield  {journal} {\bibinfo
  {journal} {Phys.\ Earth\ Planet.\ Inter.}\ }\textbf {\bibinfo {volume}
  {99}},\ \bibinfo {pages} {55--67} (\bibinfo {year} {1997})}\BibitemShut
  {NoStop}%
\bibitem [{\citenamefont {Simitev}\ and\ \citenamefont {Busse}(2003)}]{SiBu03}%
  \BibitemOpen
  \bibfield  {author} {\bibinfo {author} {\bibfnamefont {R.}~\bibnamefont
  {Simitev}}\ and\ \bibinfo {author} {\bibfnamefont {F.~H.}\ \bibnamefont
  {Busse}},\ }\bibfield  {title} {\enquote {\bibinfo {title} {Patterns of
  convection in rotating spherical shells},}\ }\href@noop {} {\bibfield
  {journal} {\bibinfo  {journal} {New J. Phys}\ }\textbf {\bibinfo {volume}
  {5}},\ \bibinfo {pages} {97.1--97.20} (\bibinfo {year} {2003})}\BibitemShut
  {NoStop}%
\bibitem [{\citenamefont {Oruba}\ and\ \citenamefont {Dormy}(2014)}]{OrDo14}%
  \BibitemOpen
  \bibfield  {author} {\bibinfo {author} {\bibfnamefont {L.}~\bibnamefont
  {Oruba}}\ and\ \bibinfo {author} {\bibfnamefont {E.}~\bibnamefont {Dormy}},\
  }\bibfield  {title} {\enquote {\bibinfo {title} {Predictive scaling laws for
  spherical rotating dynamos},}\ }\href@noop {} {\bibfield  {journal} {\bibinfo
   {journal} {Geophys. J. Int.}\ }\textbf {\bibinfo {volume} {198}},\ \bibinfo
  {pages} {828--847} (\bibinfo {year} {2014})}\BibitemShut {NoStop}%
\bibitem [{\citenamefont {{Gastine}}\ \emph {et~al.}(2016)\citenamefont
  {{Gastine}}, \citenamefont {{Wicht}},\ and\ \citenamefont
  {{Aubert}}}]{GWA16}%
  \BibitemOpen
  \bibfield  {author} {\bibinfo {author} {\bibfnamefont {T.}~\bibnamefont
  {{Gastine}}}, \bibinfo {author} {\bibfnamefont {J.}~\bibnamefont {{Wicht}}},
  \ and\ \bibinfo {author} {\bibfnamefont {J.}~\bibnamefont {{Aubert}}},\
  }\bibfield  {title} {\enquote {\bibinfo {title} {{Scaling regimes in
  spherical shell rotating convection}},}\ }\href@noop {} {\bibfield  {journal}
  {\bibinfo  {journal} {J.\,Fluid Mech.}\ }\textbf {\bibinfo {volume} {808}},\
  \bibinfo {pages} {690--732} (\bibinfo {year} {2016})}\BibitemShut {NoStop}%
\bibitem [{\citenamefont {Garcia}\ \emph {et~al.}(2015)\citenamefont {Garcia},
  \citenamefont {S{\'a}nchez}, \citenamefont {Dormy},\ and\ \citenamefont
  {{\relax Net}}}]{GSDN15}%
  \BibitemOpen
  \bibfield  {author} {\bibinfo {author} {\bibfnamefont {F.}~\bibnamefont
  {Garcia}}, \bibinfo {author} {\bibfnamefont {J.}~\bibnamefont {S{\'a}nchez}},
  \bibinfo {author} {\bibfnamefont {E.}~\bibnamefont {Dormy}}, \ and\ \bibinfo
  {author} {\bibfnamefont {M.}~\bibnamefont {{\relax Net}}},\ }\bibfield
  {title} {\enquote {\bibinfo {title} {Oscillatory convection in rotating
  spherical shells: Low {P}randtl number and non-slip boundary conditions.}}\
  }\href@noop {} {\bibfield  {journal} {\bibinfo  {journal} {SIAM J.\,Appl.\
  Dynam.\ Systems}\ }\textbf {\bibinfo {volume} {14}},\ \bibinfo {pages}
  {1787--1807} (\bibinfo {year} {2015})}\BibitemShut {NoStop}%
\bibitem [{\citenamefont {Horn}\ and\ \citenamefont {Schmid}(2017)}]{HoSc17}%
  \BibitemOpen
  \bibfield  {author} {\bibinfo {author} {\bibfnamefont {S.}~\bibnamefont
  {Horn}}\ and\ \bibinfo {author} {\bibfnamefont {P.~J.}\ \bibnamefont
  {Schmid}},\ }\bibfield  {title} {\enquote {\bibinfo {title} {Prograde,
  retrograde, and oscillatory modes in rotating {R}ayleigh-{B}\'enard
  convection},}\ }\href@noop {} {\bibfield  {journal} {\bibinfo  {journal}
  {J.\,Fluid Mech.}\ }\textbf {\bibinfo {volume} {831}},\ \bibinfo {pages}
  {182--211} (\bibinfo {year} {2017})}\BibitemShut {NoStop}%
\bibitem [{\citenamefont {{Kaplan}}\ \emph {et~al.}(2017)\citenamefont
  {{Kaplan}}, \citenamefont {{Schaeffer}}, \citenamefont {{Vidal}},\ and\
  \citenamefont {{Cardin}}}]{KSVC17}%
  \BibitemOpen
  \bibfield  {author} {\bibinfo {author} {\bibfnamefont {E.~J.}\ \bibnamefont
  {{Kaplan}}}, \bibinfo {author} {\bibfnamefont {N.}~\bibnamefont
  {{Schaeffer}}}, \bibinfo {author} {\bibfnamefont {J.}~\bibnamefont
  {{Vidal}}}, \ and\ \bibinfo {author} {\bibfnamefont {P.}~\bibnamefont
  {{Cardin}}},\ }\bibfield  {title} {\enquote {\bibinfo {title} {{Subcritical
  Thermal Convection of Liquid Metals in a Rapidly Rotating Sphere}},}\
  }\href@noop {} {\bibfield  {journal} {\bibinfo  {journal} {Phys.\ Rev.\
  Lett.}\ }\textbf {\bibinfo {volume} {119}},\ \bibinfo {pages} {094501}
  (\bibinfo {year} {2017})}\BibitemShut {NoStop}%
\bibitem [{\citenamefont {Lam}\ \emph {et~al.}(2018)\citenamefont {Lam},
  \citenamefont {Kong},\ and\ \citenamefont {Zhang}}]{LKZ18}%
  \BibitemOpen
  \bibfield  {author} {\bibinfo {author} {\bibfnamefont {K.}~\bibnamefont
  {Lam}}, \bibinfo {author} {\bibfnamefont {D.}~\bibnamefont {Kong}}, \ and\
  \bibinfo {author} {\bibfnamefont {K.}~\bibnamefont {Zhang}},\ }\bibfield
  {title} {\enquote {\bibinfo {title} {Nonlinear thermal inertial waves in
  rotating fluid spheres},}\ }\href@noop {} {\bibfield  {journal} {\bibinfo
  {journal} {Geophys. Astrophys. Fluid Dynamics}\ }\textbf {\bibinfo {volume}
  {112}},\ \bibinfo {pages} {357--374} (\bibinfo {year} {2018})}\BibitemShut
  {NoStop}%
\bibitem [{\citenamefont {Aurnou}\ \emph {et~al.}(2018)\citenamefont {Aurnou},
  \citenamefont {Bertin}, \citenamefont {Grannan}, \citenamefont {Horn},\ and\
  \citenamefont {Vogt}}]{ABGHV18}%
  \BibitemOpen
  \bibfield  {author} {\bibinfo {author} {\bibfnamefont {J.~M.}\ \bibnamefont
  {Aurnou}}, \bibinfo {author} {\bibfnamefont {V.}~\bibnamefont {Bertin}},
  \bibinfo {author} {\bibfnamefont {A.~M.}\ \bibnamefont {Grannan}}, \bibinfo
  {author} {\bibfnamefont {S.}~\bibnamefont {Horn}}, \ and\ \bibinfo {author}
  {\bibfnamefont {T.}~\bibnamefont {Vogt}},\ }\bibfield  {title} {\enquote
  {\bibinfo {title} {Rotating thermal convection in liquid gallium: multi-modal
  flow, absent steady columns},}\ }\href@noop {} {\bibfield  {journal}
  {\bibinfo  {journal} {J.\,Fluid Mech.}\ }\textbf {\bibinfo {volume} {846}},\
  \bibinfo {pages} {846--876} (\bibinfo {year} {2018})}\BibitemShut {NoStop}%
\bibitem [{\citenamefont {Garcia}\ \emph {et~al.}(2019)\citenamefont {Garcia},
  \citenamefont {Chambers},\ and\ \citenamefont {Watts}}]{GCW19}%
  \BibitemOpen
  \bibfield  {author} {\bibinfo {author} {\bibfnamefont {F.}~\bibnamefont
  {Garcia}}, \bibinfo {author} {\bibfnamefont {F.~R.~N.}\ \bibnamefont
  {Chambers}}, \ and\ \bibinfo {author} {\bibfnamefont {A.~L.}\ \bibnamefont
  {Watts}},\ }\bibfield  {title} {\enquote {\bibinfo {title} {Polar waves and
  chaotic flows in thin rotating spherical shells},}\ }\href@noop {} {\bibfield
   {journal} {\bibinfo  {journal} {Phys. Rev. Fluids}\ }\textbf {\bibinfo
  {volume} {4}},\ \bibinfo {pages} {074802} (\bibinfo {year}
  {2019})}\BibitemShut {NoStop}%
\bibitem [{\citenamefont {Lin}(2021)}]{Lin21}%
  \BibitemOpen
  \bibfield  {author} {\bibinfo {author} {\bibfnamefont {Y.}~\bibnamefont
  {Lin}},\ }\bibfield  {title} {\enquote {\bibinfo {title} {Triadic resonances
  driven by thermal convection in a rotating sphere},}\ }\href@noop {}
  {\bibfield  {journal} {\bibinfo  {journal} {J.\,Fluid Mech.}\ }\textbf
  {\bibinfo {volume} {909}},\ \bibinfo {pages} {R3} (\bibinfo {year}
  {2021})}\BibitemShut {NoStop}%
\bibitem [{\citenamefont {Massaguer}(1991)}]{Mas91}%
  \BibitemOpen
  \bibfield  {author} {\bibinfo {author} {\bibfnamefont {J.~M.}\ \bibnamefont
  {Massaguer}},\ }\bibfield  {title} {\enquote {\bibinfo {title} {Stellar
  convection as a low {P}randtl number flow},}\ }in\ \href@noop {} {\emph
  {\bibinfo {booktitle} {The Sun and Cool Stars: activity, magnetism, dynamos:
  Proceedings of Colloquium No. 130 of the International Astronomical Union
  Held in Helsinki, Finland, 17--20 July 1990}}},\ \bibinfo {editor} {edited
  by\ \bibinfo {editor} {\bibfnamefont {I.}~\bibnamefont {Tuominen}}, \bibinfo
  {editor} {\bibfnamefont {D.}~\bibnamefont {Moss}}, \ and\ \bibinfo {editor}
  {\bibfnamefont {G.}~\bibnamefont {R{\"u}diger}}}\ (\bibinfo  {publisher}
  {Springer Berlin Heidelberg},\ \bibinfo {year} {1991})\ pp.\ \bibinfo {pages}
  {57--61}\BibitemShut {NoStop}%
\bibitem [{\citenamefont {Vogt}\ \emph {et~al.}(2021)\citenamefont {Vogt},
  \citenamefont {Horn},\ and\ \citenamefont {Aurnou}}]{VHA21}%
  \BibitemOpen
  \bibfield  {author} {\bibinfo {author} {\bibfnamefont {T.}~\bibnamefont
  {Vogt}}, \bibinfo {author} {\bibfnamefont {S.}~\bibnamefont {Horn}}, \ and\
  \bibinfo {author} {\bibfnamefont {J.~M.}\ \bibnamefont {Aurnou}},\ }\bibfield
   {title} {\enquote {\bibinfo {title} {Oscillatory thermal–inertial flows in
  liquid metal rotating convection},}\ }\href {\doibase 10.1017/jfm.2020.976}
  {\bibfield  {journal} {\bibinfo  {journal} {J.\,Fluid Mech.}\ }\textbf
  {\bibinfo {volume} {911}},\ \bibinfo {pages} {A5} (\bibinfo {year}
  {2021})}\BibitemShut {NoStop}%
\bibitem [{\citenamefont {Ecke}\ \emph {et~al.}(1992)\citenamefont {Ecke},
  \citenamefont {Zhong},\ and\ \citenamefont {Knobloch}}]{EZK92}%
  \BibitemOpen
  \bibfield  {author} {\bibinfo {author} {\bibfnamefont {R.~E.}\ \bibnamefont
  {Ecke}}, \bibinfo {author} {\bibfnamefont {F.}~\bibnamefont {Zhong}}, \ and\
  \bibinfo {author} {\bibfnamefont {E.}~\bibnamefont {Knobloch}},\ }\bibfield
  {title} {\enquote {\bibinfo {title} {Hopf bifurcation with broken reflection
  symmetry in rotating {R}ayleigh-{B}\'enard convection},}\ }\href@noop {}
  {\bibfield  {journal} {\bibinfo  {journal} {Europhys.\ Lett.}\ }\textbf
  {\bibinfo {volume} {19}},\ \bibinfo {pages} {177--182} (\bibinfo {year}
  {1992})}\BibitemShut {NoStop}%
\bibitem [{\citenamefont {S\'anchez~Umbr\'{\i}a}\ and\ \citenamefont
  {Net}(2019)}]{SaNe19}%
  \BibitemOpen
  \bibfield  {author} {\bibinfo {author} {\bibfnamefont {J.}~\bibnamefont
  {S\'anchez~Umbr\'{\i}a}}\ and\ \bibinfo {author} {\bibfnamefont
  {M.}~\bibnamefont {Net}},\ }\bibfield  {title} {\enquote {\bibinfo {title}
  {Torsional solutions of convection in rotating fluid spheres},}\ }\href
  {\doibase 10.1103/PhysRevFluids.4.013501} {\bibfield  {journal} {\bibinfo
  {journal} {Phys.\ Rev.\,Fluids}\ }\textbf {\bibinfo {volume} {4}},\ \bibinfo
  {pages} {013501} (\bibinfo {year} {2019})}\BibitemShut {NoStop}%
\bibitem [{\citenamefont {Keller}(1977)}]{Kel77}%
  \BibitemOpen
  \bibfield  {author} {\bibinfo {author} {\bibfnamefont {H.~B.}\ \bibnamefont
  {Keller}},\ }\bibfield  {title} {\enquote {\bibinfo {title} {Numerical
  solution of bifurcation and nonlinear eigenvalue problems},}\ }in\ \href@noop
  {} {\emph {\bibinfo {booktitle} {Applications of {B}ifurcation {T}heory}}},\
  \bibinfo {editor} {edited by\ \bibinfo {editor} {\bibfnamefont {P.~H.}\
  \bibnamefont {Rabinowitz}}}\ (\bibinfo  {publisher} {Academic Press, New
  York},\ \bibinfo {year} {1977})\ pp.\ \bibinfo {pages} {359--384}\BibitemShut
  {NoStop}%
\bibitem [{\citenamefont {Doedel}\ and\ \citenamefont
  {Tuckerman}(2000)}]{DoTu00}%
  \BibitemOpen
  \bibinfo {editor} {\bibfnamefont {E.}~\bibnamefont {Doedel}}\ and\ \bibinfo
  {editor} {\bibfnamefont {L.~S.}\ \bibnamefont {Tuckerman}},\ eds.,\
  \href@noop {} {\emph {\bibinfo {title} {Numerical {M}ethods for {B}ifurcation
  {P}roblems and {L}arge-{S}cale {D}ynamical {S}ystems}}},\ \bibinfo {series}
  {{IMA} Volumes in Mathematics and its Applications}, Vol.\ \bibinfo {volume}
  {119}\ (\bibinfo  {publisher} {Springer--{V}erlag, Berlin},\ \bibinfo {year}
  {2000})\BibitemShut {NoStop}%
\bibitem [{\citenamefont {S{\'a}nchez}\ and\ \citenamefont {{\relax
  Net}}(2016)}]{SaNe16}%
  \BibitemOpen
  \bibfield  {author} {\bibinfo {author} {\bibfnamefont {J.}~\bibnamefont
  {S{\'a}nchez}}\ and\ \bibinfo {author} {\bibfnamefont {M.}~\bibnamefont
  {{\relax Net}}},\ }\bibfield  {title} {\enquote {\bibinfo {title} {Numerical
  continuation methods for large-scale dissipative dynamical systems},}\
  }\href@noop {} {\bibfield  {journal} {\bibinfo  {journal} {Eur. Phys. J.
  Spec. Top.}\ }\textbf {\bibinfo {volume} {225}},\ \bibinfo {pages}
  {2465--2486} (\bibinfo {year} {2016})}\BibitemShut {NoStop}%
\bibitem [{\citenamefont {Garcia}\ \emph {et~al.}(2016)\citenamefont {Garcia},
  \citenamefont {{\relax Net}},\ and\ \citenamefont {S\'anchez}}]{GNS16}%
  \BibitemOpen
  \bibfield  {author} {\bibinfo {author} {\bibfnamefont {F.}~\bibnamefont
  {Garcia}}, \bibinfo {author} {\bibfnamefont {M.}~\bibnamefont {{\relax
  Net}}}, \ and\ \bibinfo {author} {\bibfnamefont {J.}~\bibnamefont
  {S\'anchez}},\ }\bibfield  {title} {\enquote {\bibinfo {title} {Continuation
  and stability of convective modulated rotating waves in spherical shells},}\
  }\href@noop {} {\bibfield  {journal} {\bibinfo  {journal} {Phys.\ Rev.\,E}\
  }\textbf {\bibinfo {volume} {93}},\ \bibinfo {pages} {013119} (\bibinfo
  {year} {2016})}\BibitemShut {NoStop}%
\bibitem [{\citenamefont {Garcia}\ \emph
  {et~al.}(2021{\natexlab{a}})\citenamefont {Garcia}, \citenamefont
  {Giesecke},\ and\ \citenamefont {Stefani}}]{GGS21}%
  \BibitemOpen
  \bibfield  {author} {\bibinfo {author} {\bibfnamefont {F.}~\bibnamefont
  {Garcia}}, \bibinfo {author} {\bibfnamefont {A.}~\bibnamefont {Giesecke}}, \
  and\ \bibinfo {author} {\bibfnamefont {F.}~\bibnamefont {Stefani}},\
  }\bibfield  {title} {\enquote {\bibinfo {title} {Modulated rotating waves and
  triadic resonances in spherical fluid systems: The case of magnetized
  spherical couette flow},}\ }\href@noop {} {\bibfield  {journal} {\bibinfo
  {journal} {Phys.\ Fluids}\ }\textbf {\bibinfo {volume} {33}},\ \bibinfo
  {pages} {044105} (\bibinfo {year} {2021}{\natexlab{a}})}\BibitemShut
  {NoStop}%
\bibitem [{\citenamefont {Garcia}\ \emph {et~al.}(2010)\citenamefont {Garcia},
  \citenamefont {{\relax Net}}, \citenamefont {Garc{\'\i}a-Archilla},\ and\
  \citenamefont {S\'anchez}}]{GNGS10}%
  \BibitemOpen
  \bibfield  {author} {\bibinfo {author} {\bibfnamefont {F.}~\bibnamefont
  {Garcia}}, \bibinfo {author} {\bibfnamefont {M.}~\bibnamefont {{\relax
  Net}}}, \bibinfo {author} {\bibfnamefont {B.}~\bibnamefont
  {Garc{\'\i}a-Archilla}}, \ and\ \bibinfo {author} {\bibfnamefont
  {J.}~\bibnamefont {S\'anchez}},\ }\bibfield  {title} {\enquote {\bibinfo
  {title} {A comparison of high-order time integrators for thermal convection
  in rotating spherical shells},}\ }\href@noop {} {\bibfield  {journal}
  {\bibinfo  {journal} {J.\,Comput.\ Phys.}\ }\textbf {\bibinfo {volume}
  {229}},\ \bibinfo {pages} {7997--8010} (\bibinfo {year} {2010})}\BibitemShut
  {NoStop}%
\bibitem [{\citenamefont {S{\'a}nchez}\ \emph {et~al.}(2016)\citenamefont
  {S{\'a}nchez}, \citenamefont {Garcia},\ and\ \citenamefont {{\relax
  Net}}}]{SGN16}%
  \BibitemOpen
  \bibfield  {author} {\bibinfo {author} {\bibfnamefont {J.}~\bibnamefont
  {S{\'a}nchez}}, \bibinfo {author} {\bibfnamefont {F.}~\bibnamefont {Garcia}},
  \ and\ \bibinfo {author} {\bibfnamefont {M.}~\bibnamefont {{\relax Net}}},\
  }\bibfield  {title} {\enquote {\bibinfo {title} {Radial collocation methods
  for the onset of convection in rotating spheres},}\ }\href@noop {} {\bibfield
   {journal} {\bibinfo  {journal} {J.\,Comput.\ Phys.}\ }\textbf {\bibinfo
  {volume} {308}},\ \bibinfo {pages} {273--288} (\bibinfo {year}
  {2016})}\BibitemShut {NoStop}%
\bibitem [{\citenamefont {Orszag}(1970)}]{Ors70}%
  \BibitemOpen
  \bibfield  {author} {\bibinfo {author} {\bibfnamefont {S.~A.}\ \bibnamefont
  {Orszag}},\ }\bibfield  {title} {\enquote {\bibinfo {title} {Transform method
  for calculation of vector-coupled sums: {A}pplication to the spectral form of
  the vorticity equation},}\ }\href@noop {} {\bibfield  {journal} {\bibinfo
  {journal} {J. Atmos. Sci.}\ }\textbf {\bibinfo {volume} {27}},\ \bibinfo
  {pages} {890--895} (\bibinfo {year} {1970})}\BibitemShut {NoStop}%
\bibitem [{\citenamefont {Frigo}\ and\ \citenamefont {Johnson}(2005)}]{FrJo05}%
  \BibitemOpen
  \bibfield  {author} {\bibinfo {author} {\bibfnamefont {Matteo}\ \bibnamefont
  {Frigo}}\ and\ \bibinfo {author} {\bibfnamefont {Steven~G.}\ \bibnamefont
  {Johnson}},\ }\bibfield  {title} {\enquote {\bibinfo {title} {The design and
  implementation of {FFTW3}},}\ }\href@noop {} {\bibfield  {journal} {\bibinfo
  {journal} {Proceedings of the IEEE}\ }\textbf {\bibinfo {volume} {93}},\
  \bibinfo {pages} {216--231} (\bibinfo {year} {2005})},\ \bibinfo {note}
  {special issue on "Program Generation, Optimization, and Platform
  Adaptation"}\BibitemShut {NoStop}%
\bibitem [{\citenamefont {Goto}\ and\ \citenamefont {van~de
  Geijn}(2008)}]{GoGe08}%
  \BibitemOpen
  \bibfield  {author} {\bibinfo {author} {\bibfnamefont {Kazushige}\
  \bibnamefont {Goto}}\ and\ \bibinfo {author} {\bibfnamefont {Robert~A.}\
  \bibnamefont {van~de Geijn}},\ }\bibfield  {title} {\enquote {\bibinfo
  {title} {Anatomy of high-performance matrix multiplication},}\ }\href@noop {}
  {\bibfield  {journal} {\bibinfo  {journal} {ACM Trans. Math. Softw.}\
  }\textbf {\bibinfo {volume} {34}},\ \bibinfo {pages} {1--25} (\bibinfo {year}
  {2008})}\BibitemShut {NoStop}%
\bibitem [{\citenamefont {Crawford}\ and\ \citenamefont
  {Knobloch}(1991)}]{CrKn91}%
  \BibitemOpen
  \bibfield  {author} {\bibinfo {author} {\bibfnamefont {J.~D.}\ \bibnamefont
  {Crawford}}\ and\ \bibinfo {author} {\bibfnamefont {E.}~\bibnamefont
  {Knobloch}},\ }\bibfield  {title} {\enquote {\bibinfo {title} {Symmetry and
  symmetry-breaking bifurcations in fluid dynamics},}\ }\href@noop {}
  {\bibfield  {journal} {\bibinfo  {journal} {Annu.\ Rev.\ Fluid Mech.}\
  }\textbf {\bibinfo {volume} {23}},\ \bibinfo {pages} {341--387} (\bibinfo
  {year} {1991})}\BibitemShut {NoStop}%
\bibitem [{\citenamefont {{Coughlin}}\ and\ \citenamefont
  {{Marcus}}(1992)}]{CoMa92}%
  \BibitemOpen
  \bibfield  {author} {\bibinfo {author} {\bibfnamefont {K.~T.}\ \bibnamefont
  {{Coughlin}}}\ and\ \bibinfo {author} {\bibfnamefont {P.~S.}\ \bibnamefont
  {{Marcus}}},\ }\bibfield  {title} {\enquote {\bibinfo {title} {{Modulated
  waves in {T}aylor-{C}ouette flow. {P}art 1. {A}nalysis}},}\ }\href@noop {}
  {\bibfield  {journal} {\bibinfo  {journal} {J.\,Fluid Mech.}\ }\textbf
  {\bibinfo {volume} {234}},\ \bibinfo {pages} {1--18} (\bibinfo {year}
  {1992})}\BibitemShut {NoStop}%
\bibitem [{\citenamefont {Kawahara}\ \emph {et~al.}(2012)\citenamefont
  {Kawahara}, \citenamefont {Uhlmann},\ and\ \citenamefont {van Veen}}]{KUL12}%
  \BibitemOpen
  \bibfield  {author} {\bibinfo {author} {\bibfnamefont {G.}~\bibnamefont
  {Kawahara}}, \bibinfo {author} {\bibfnamefont {M.}~\bibnamefont {Uhlmann}}, \
  and\ \bibinfo {author} {\bibfnamefont {L.}~\bibnamefont {van Veen}},\
  }\bibfield  {title} {\enquote {\bibinfo {title} {The signficance of simple
  invariant solutions in turbulent flows},}\ }\href@noop {} {\bibfield
  {journal} {\bibinfo  {journal} {Arch.\ Ration.\ Mech.\ Anal.}\ }\textbf
  {\bibinfo {volume} {44}},\ \bibinfo {pages} {203--225} (\bibinfo {year}
  {2012})}\BibitemShut {NoStop}%
\bibitem [{\citenamefont {Hof~et al.}(2004)}]{Hof_et_al04}%
  \BibitemOpen
  \bibfield  {author} {\bibinfo {author} {\bibfnamefont {B.}~\bibnamefont
  {Hof~et al.}},\ }\bibfield  {title} {\enquote {\bibinfo {title} {Experimental
  observation of nonlinear traveling waves in turbulent pipe flow},}\
  }\href@noop {} {\bibfield  {journal} {\bibinfo  {journal} {Science}\ }\textbf
  {\bibinfo {volume} {305}},\ \bibinfo {pages} {1594--1598} (\bibinfo {year}
  {2004})}\BibitemShut {NoStop}%
\bibitem [{\citenamefont {Doedel}(1986)}]{Doe86}%
  \BibitemOpen
  \bibfield  {author} {\bibinfo {author} {\bibfnamefont {E.}~\bibnamefont
  {Doedel}},\ }\href@noop {} {\emph {\bibinfo {title} {{AUTO}: Software for
  continuation and bifurcation problems in ordinary differential equations}}},\
  \bibinfo {organization} {Report {A}pplied {M}athematics, {C}alifornia
  {I}nstitute of {T}echnology},\ \bibinfo {address} {Pasadena, USA} (\bibinfo
  {year} {1986})\BibitemShut {NoStop}%
\bibitem [{\citenamefont {S{\'a}nchez}\ \emph {et~al.}(2004)\citenamefont
  {S{\'a}nchez}, \citenamefont {{\relax Net}}, \citenamefont
  {Garc{\'\i}a-Archilla},\ and\ \citenamefont {Sim\'o}}]{SNGS04b}%
  \BibitemOpen
  \bibfield  {author} {\bibinfo {author} {\bibfnamefont {J.}~\bibnamefont
  {S{\'a}nchez}}, \bibinfo {author} {\bibfnamefont {M.}~\bibnamefont {{\relax
  Net}}}, \bibinfo {author} {\bibfnamefont {B.}~\bibnamefont
  {Garc{\'\i}a-Archilla}}, \ and\ \bibinfo {author} {\bibfnamefont
  {C.}~\bibnamefont {Sim\'o}},\ }\bibfield  {title} {\enquote {\bibinfo {title}
  {{Newton-Krylov} continuation of periodic orbits for {Navier-Stokes}
  flows},}\ }\href@noop {} {\bibfield  {journal} {\bibinfo  {journal}
  {J.\,Comput.\ Phys.}\ }\textbf {\bibinfo {volume} {201}},\ \bibinfo {pages}
  {13--33} (\bibinfo {year} {2004})}\BibitemShut {NoStop}%
\bibitem [{\citenamefont {S{\'a}nchez}\ \emph {et~al.}(2013)\citenamefont
  {S{\'a}nchez}, \citenamefont {Garcia},\ and\ \citenamefont {{\relax
  Net}}}]{SGN13}%
  \BibitemOpen
  \bibfield  {author} {\bibinfo {author} {\bibfnamefont {J.}~\bibnamefont
  {S{\'a}nchez}}, \bibinfo {author} {\bibfnamefont {F.}~\bibnamefont {Garcia}},
  \ and\ \bibinfo {author} {\bibfnamefont {M.}~\bibnamefont {{\relax Net}}},\
  }\bibfield  {title} {\enquote {\bibinfo {title} {Computation of azimuthal
  waves and their stability in thermal convection in rotating spherical shells
  with application to the study of a double-{H}opf bifurcation},}\ }\href@noop
  {} {\bibfield  {journal} {\bibinfo  {journal} {Phys.\ Rev.\,E}\ }\textbf
  {\bibinfo {volume} {87}},\ \bibinfo {pages} {033014/ 1--11} (\bibinfo {year}
  {2013})}\BibitemShut {NoStop}%
\bibitem [{\citenamefont {Feudel}\ \emph {et~al.}(2013)\citenamefont {Feudel},
  \citenamefont {Seehafer}, \citenamefont {Tuckerman},\ and\ \citenamefont
  {Gellert}}]{FSTG13}%
  \BibitemOpen
  \bibfield  {author} {\bibinfo {author} {\bibfnamefont {F.}~\bibnamefont
  {Feudel}}, \bibinfo {author} {\bibfnamefont {N.}~\bibnamefont {Seehafer}},
  \bibinfo {author} {\bibfnamefont {L.~S.}\ \bibnamefont {Tuckerman}}, \ and\
  \bibinfo {author} {\bibfnamefont {M.}~\bibnamefont {Gellert}},\ }\bibfield
  {title} {\enquote {\bibinfo {title} {Multistability in rotating spherical
  shell convection},}\ }\href@noop {} {\bibfield  {journal} {\bibinfo
  {journal} {Phys.\ Rev.\,E}\ }\textbf {\bibinfo {volume} {87}},\ \bibinfo
  {pages} {023021--1--023021--8} (\bibinfo {year} {2013})}\BibitemShut
  {NoStop}%
\bibitem [{\citenamefont {Feudel}\ \emph {et~al.}(2015)\citenamefont {Feudel},
  \citenamefont {Tuckerman}, \citenamefont {Gellert},\ and\ \citenamefont
  {Seehafer}}]{FTGS15}%
  \BibitemOpen
  \bibfield  {author} {\bibinfo {author} {\bibfnamefont {F.}~\bibnamefont
  {Feudel}}, \bibinfo {author} {\bibfnamefont {L.~S.}\ \bibnamefont
  {Tuckerman}}, \bibinfo {author} {\bibfnamefont {M.}~\bibnamefont {Gellert}},
  \ and\ \bibinfo {author} {\bibfnamefont {N.}~\bibnamefont {Seehafer}},\
  }\bibfield  {title} {\enquote {\bibinfo {title} {Bifurcations of rotating
  waves in rotating spherical shell convection},}\ }\href@noop {} {\bibfield
  {journal} {\bibinfo  {journal} {Phys.\ Rev.\,E}\ }\textbf {\bibinfo {volume}
  {92}},\ \bibinfo {pages} {053015} (\bibinfo {year} {2015})}\BibitemShut
  {NoStop}%
\bibitem [{\citenamefont {Tuckerman}\ \emph {et~al.}(2019)\citenamefont
  {Tuckerman}, \citenamefont {Langham},\ and\ \citenamefont {Willis}}]{TLW19}%
  \BibitemOpen
  \bibfield  {author} {\bibinfo {author} {\bibfnamefont {L.~S.}\ \bibnamefont
  {Tuckerman}}, \bibinfo {author} {\bibfnamefont {J.}~\bibnamefont {Langham}},
  \ and\ \bibinfo {author} {\bibfnamefont {A.}~\bibnamefont {Willis}},\
  }\enquote {\bibinfo {title} {Order-of-magnitude speedup for steady states and
  traveling waves via stokes preconditioning in channelflow and
  openpipeflow},}\ in\ \href@noop {} {\emph {\bibinfo {booktitle}
  {Computational Modelling of Bifurcations and Instabilities in Fluid
  Dynamics}}}\ (\bibinfo  {publisher} {Springer International, Cham,
  Switzerland},\ \bibinfo {year} {2019})\ pp.\ \bibinfo {pages}
  {3--31}\BibitemShut {NoStop}%
\bibitem [{\citenamefont {Jordan}\ and\ \citenamefont {Smith}(2007)}]{JoSm07}%
  \BibitemOpen
  \bibfield  {author} {\bibinfo {author} {\bibfnamefont {D.}~\bibnamefont
  {Jordan}}\ and\ \bibinfo {author} {\bibfnamefont {P.}~\bibnamefont {Smith}},\
  }\href@noop {} {\emph {\bibinfo {title} {Nonlinear Ordinary Differential
  Equations : An Introduction for Scientists and Engineers}}},\ \bibinfo
  {series} {Oxford Texts in Applied and Engineering Mathematics}, Vol.~\bibinfo
  {volume} {10}\ (\bibinfo  {publisher} {Oxford University Press},\ \bibinfo
  {year} {2007})\BibitemShut {NoStop}%
\bibitem [{\citenamefont {Lehoucq}\ \emph {et~al.}(1998)\citenamefont
  {Lehoucq}, \citenamefont {Sorensen},\ and\ \citenamefont {Yang}}]{LSY98}%
  \BibitemOpen
  \bibfield  {author} {\bibinfo {author} {\bibfnamefont {R.~B.}\ \bibnamefont
  {Lehoucq}}, \bibinfo {author} {\bibfnamefont {D.~C.}\ \bibnamefont
  {Sorensen}}, \ and\ \bibinfo {author} {\bibfnamefont {C.}~\bibnamefont
  {Yang}},\ }\href@noop {} {\emph {\bibinfo {title} {{ARPACK} User's Guide:
  {S}olution of {L}arge-{S}cale {E}igenvalue {P}roblems with {I}mplicitly
  {R}estarted {A}rnoldi {M}ethods}}}\ (\bibinfo  {publisher} {SIAM},\ \bibinfo
  {year} {1998})\BibitemShut {NoStop}%
\bibitem [{\citenamefont {Tuckerman}(2015)}]{Tuc15}%
  \BibitemOpen
  \bibfield  {author} {\bibinfo {author} {\bibfnamefont {L.~S.}\ \bibnamefont
  {Tuckerman}},\ }\bibfield  {title} {\enquote {\bibinfo {title} {Laplacian
  preconditioning for the inverse {A}rnoldi method},}\ }\href@noop {}
  {\bibfield  {journal} {\bibinfo  {journal} {Commun. Comput. Phys.}\ }\textbf
  {\bibinfo {volume} {18}},\ \bibinfo {pages} {1336--1351} (\bibinfo {year}
  {2015})}\BibitemShut {NoStop}%
\bibitem [{\citenamefont {Saad}(1992)}]{Saa92}%
  \BibitemOpen
  \bibfield  {author} {\bibinfo {author} {\bibfnamefont {Y.}~\bibnamefont
  {Saad}},\ }\href@noop {} {\emph {\bibinfo {title} {Numerical {M}ethods for
  {L}arge {E}igenvalue {P}roblems}}}\ (\bibinfo  {publisher} {Manchester
  {U}niversity {P}ress},\ \bibinfo {address} {Manchester},\ \bibinfo {year}
  {1992})\BibitemShut {NoStop}%
\bibitem [{\citenamefont {Garcia}\ \emph {et~al.}(2014)\citenamefont {Garcia},
  \citenamefont {Dormy}, \citenamefont {S\'anchez},\ and\ \citenamefont
  {{\relax Net}}}]{GDSN14}%
  \BibitemOpen
  \bibfield  {author} {\bibinfo {author} {\bibfnamefont {F.}~\bibnamefont
  {Garcia}}, \bibinfo {author} {\bibfnamefont {E.}~\bibnamefont {Dormy}},
  \bibinfo {author} {\bibfnamefont {J.}~\bibnamefont {S\'anchez}}, \ and\
  \bibinfo {author} {\bibfnamefont {M.}~\bibnamefont {{\relax Net}}},\
  }\bibfield  {title} {\enquote {\bibinfo {title} {Two computational approaches
  for the simulation of fluid problems in rotating spherical shells},}\ }in\
  \href@noop {} {\emph {\bibinfo {booktitle} {Proc. of the 5th International
  Conference on Computational Methods - ICCM2014. Cambridge, England}}},\
  Vol.~\bibinfo {volume} {1},\ \bibinfo {editor} {edited by\ \bibinfo {editor}
  {\bibfnamefont {G.~R.}\ \bibnamefont {Liu}}\ and\ \bibinfo {editor}
  {\bibfnamefont {Z.~W.}\ \bibnamefont {Guan}}}\ (\bibinfo {year}
  {2014})\BibitemShut {NoStop}%
\bibitem [{\citenamefont {Marcus}\ and\ \citenamefont
  {Tuckerman}(1987)}]{MaTu87a}%
  \BibitemOpen
  \bibfield  {author} {\bibinfo {author} {\bibfnamefont {P.~S.}\ \bibnamefont
  {Marcus}}\ and\ \bibinfo {author} {\bibfnamefont {L.~S.}\ \bibnamefont
  {Tuckerman}},\ }\bibfield  {title} {\enquote {\bibinfo {title} {Simulation of
  flow between concentric rotating spheres. {P}art 1. {S}teady states},}\
  }\href@noop {} {\bibfield  {journal} {\bibinfo  {journal} {J.\,Fluid Mech.}\
  }\textbf {\bibinfo {volume} {185}},\ \bibinfo {pages} {1--30} (\bibinfo
  {year} {1987})}\BibitemShut {NoStop}%
\bibitem [{\citenamefont {Barik}\ \emph {et~al.}(2018)\citenamefont {Barik},
  \citenamefont {Triana}, \citenamefont {Hoff},\ and\ \citenamefont
  {Wicht}}]{BTHW18}%
  \BibitemOpen
  \bibfield  {author} {\bibinfo {author} {\bibfnamefont {A.}~\bibnamefont
  {Barik}}, \bibinfo {author} {\bibfnamefont {S.~A.}\ \bibnamefont {Triana}},
  \bibinfo {author} {\bibfnamefont {M.}~\bibnamefont {Hoff}}, \ and\ \bibinfo
  {author} {\bibfnamefont {J.}~\bibnamefont {Wicht}},\ }\bibfield  {title}
  {\enquote {\bibinfo {title} {Triadic resonances in the wide-gap spherical
  {C}ouette system},}\ }\href@noop {} {\bibfield  {journal} {\bibinfo
  {journal} {J.\,Fluid Mech.}\ }\textbf {\bibinfo {volume} {843}},\ \bibinfo
  {pages} {211--243} (\bibinfo {year} {2018})}\BibitemShut {NoStop}%
\bibitem [{\citenamefont {Laskar}(1993)}]{Las93}%
  \BibitemOpen
  \bibfield  {author} {\bibinfo {author} {\bibfnamefont {J.}~\bibnamefont
  {Laskar}},\ }\bibfield  {title} {\enquote {\bibinfo {title} {Frequency
  analysis of a dynamical system},}\ }\href@noop {} {\bibfield  {journal}
  {\bibinfo  {journal} {Celestial Mech. Dyn. Astron.}\ }\textbf {\bibinfo
  {volume} {56}},\ \bibinfo {pages} {191--196} (\bibinfo {year}
  {1993})}\BibitemShut {NoStop}%
\bibitem [{\citenamefont {Garcia}\ \emph
  {et~al.}(2021{\natexlab{b}})\citenamefont {Garcia}, \citenamefont
  {Seilmayer}, \citenamefont {Giesecke},\ and\ \citenamefont
  {Stefani}}]{GSGS21}%
  \BibitemOpen
  \bibfield  {author} {\bibinfo {author} {\bibfnamefont {F.}~\bibnamefont
  {Garcia}}, \bibinfo {author} {\bibfnamefont {M.}~\bibnamefont {Seilmayer}},
  \bibinfo {author} {\bibfnamefont {A.}~\bibnamefont {Giesecke}}, \ and\
  \bibinfo {author} {\bibfnamefont {F.}~\bibnamefont {Stefani}},\ }\bibfield
  {title} {\enquote {\bibinfo {title} {Long term time dependent frequency
  analysis of chaotic waves in the weakly magnetised spherical {C}ouette
  system},}\ }\href@noop {} {\bibfield  {journal} {\bibinfo  {journal}
  {Physica\ D}\ }\textbf {\bibinfo {volume} {418}},\ \bibinfo {pages} {132836}
  (\bibinfo {year} {2021}{\natexlab{b}})}\BibitemShut {NoStop}%
\end{thebibliography}
\end{document}